\documentclass[twocolumn, reprint, amsmath,amssymb, superscriptaddress,floatfix]{revtex4-2}

\usepackage{graphicx}
\usepackage{dcolumn}
\usepackage{bm}
\usepackage{mathtools}

\usepackage{color}

\def\*#1{\mathbf{#1}}
\def\^#1{\boldsymbol{#1}}
\def\br#1{\left[ #1 \right]}
\def\pr#1{\left( #1\right)}
\def\brc#1{\left\{ #1\right\}}
\def\ang#1{\left\langle #1\right\rangle}
\def\abs#1{\lvert #1\rvert}
\newcommand{\iu}{\mathrm{i}\mkern1mu}
\DeclareMathOperator*{\argmax}{arg\,max}

\makeatletter
\newcommand{\vast}{\bBigg@{3}}
\newcommand{\Vast}{\bBigg@{4}}
\makeatother

\usepackage[resetlabels,labeled]{multibib} 
\newcites{supp}{SupplementaryReferences}

\begin{document}

\preprint{APS/123-QED}

\title{Nonequilibrium thermodynamics of the asymmetric Sherrington-Kirkpatrick model}%

\author{Miguel Aguilera}
   \email{sci@maguilera.net}
   \affiliation{ BCAM -- Basque Center for Applied Mathematics.
        Bilbao, Spain}
   \affiliation{IKERBASQUE, Basque Foundation for Science. Bilbao, Spain }
   \affiliation{School of Engineering and Informatics, University of Sussex, Falmer, Brighton, United Kingdom}%
\author{Masanao Igarashi}
    \affiliation{Graduate School of Engineering, Hokkaido University,
        Sapporo, Japan}
\author{Hideaki Shimazaki}
    \affiliation{Graduate School of Informatics, Kyoto University, Kyoto, Japan}
    \affiliation{Center for Human Nature, Artificial Intelligence, and Neuroscience (CHAIN), Hokkaido University, Sapporo, Japan}

\date{\today}


\begin{abstract}
Most natural systems operate far from equilibrium, displaying time-asymmetric, irreversible dynamics characterized by a positive entropy production while exchanging energy and matter with the environment. Although stochastic thermodynamics underpins the irreversible dynamics of small systems, the nonequilibrium thermodynamics of larger, more complex systems remains unexplored. Here, we investigate the asymmetric Sherrington-Kirkpatrick model with synchronous and asynchronous updates as a prototypical example of large-scale nonequilibrium processes. Using a path integral method, we calculate a generating functional over trajectories, obtaining exact solutions of the order parameters, path entropy, and steady-state entropy production of infinitely large networks. Entropy production peaks at critical order-disorder phase transitions, but is significantly larger for quasi-deterministic disordered dynamics. Consequently, entropy production can increase under distinct scenarios, requiring multiple thermodynamic quantities to describe the system accurately. These results contribute to developing an exact analytical theory of the nonequilibrium thermodynamics of large-scale physical and biological systems and their phase transitions.

\end{abstract}

\maketitle


\section*{Introduction}

While isolated systems tend toward thermodynamic equilibrium, many physical, chemical, and biological processes operate far from equilibrium. Such nonequilibrium systems -- from molecules to organisms and machines -- persist by exchanging matter and energy with their surroundings, being causally driven by time-varying external stimuli or by their past states (e.g., the adaptive action of sensor and effector interfaces \cite{wiener_newtonian_1961}).
Nonequilibrium processes inherently break time reversal symmetry, describing spatial and temporal patterns with a definite past-future order, and being thus  strikingly different from the reversible dynamics found at thermodynamic equilibrium. 
Understanding these dissipative processes -- from chemical reactions to neural dynamics or flocks of birds -- brings critical insights into the self-organization of open systems  \cite{kondepudi2014modern}. 
Although these ideas have attracted the interest of disparate fields, from evolutionary dynamics \cite{england2013statistical} to neuroscience \cite{lynn2021broken,perl2021nonequilibrium,delafuente2022temporal,deco2021deep}, little is known about the thermodynamic description of nonequilibrium systems comprising many interacting particles. While stochastic thermodynamics has been greatly influential in the study of small systems with appreciable fluctuations \cite{seifert_stochastic_2012}, the thermodynamics of large-scale nonequilibrium systems and their phase transitions has attracted attention only very recently \cite{herpich2018collective,sune2019out,herpich2020stochastic}.

When the elements of a system are numerous, characterizing its nonequilibrium states is challenging due to the expansion of its state space. Inspired by the success of the equilibrium Ising model in investigating disordered systems in the thermodynamic limit, we study the nonequilibrium thermodynamics of a stochastic, kinetic Ising model with both synchronous and asynchronous updates. 
The Ising model is a cornerstone of statistical mechanics, originally conceived as a model describing phase transitions in magnetic materials \cite{nishimori_statistical_2001}. A natural extension of the model introducing Markovian dynamics either in discrete or continuous time is the kinetic Ising model, a prototypical model of both equilibrium and nonequilibrium systems such as recurrent neural networks \cite{roudi2015multi} or genetic regulatory networks \cite{lang2014epigenetic}. 
With time-independent parameters and symmetric couplings (under synchronous or asynchronous updates in the absence of lagged self-couplings), the kinetic Ising model results in an equilibrium process exhibiting a variety of complex phenomena, including ordered (ferromagnetic), disordered (paramagnetic), and quenched disordered states (known as spin glasses). The celebrated Sherrington–Kirkpatrick (SK) model, characterized by quenched random couplings resulting in a spin-glass phase \cite{sherrington1975solvable}, can be solved using the replica mean-field method \cite{parisi1980sequence}. A kinetic version of this symmetric-coupling model has been represented as a bipartite network, also solved using the replica trick \cite{brunetti_asymmetric_1992}.

The kinetics of equilibrium Ising systems are indistinguishable when observed in a forward or backward direction in time, i.e., they are invariant under the reversal of the arrow of time. This time-symmetry breaks down under time-varying external fields or asymmetric couplings comprising history-dependent, non-conservative forces. Such time-asymmetric processes violate detailed balance, leading to nonequilibrium dynamics yielding a positive entropy production \cite{schnakenberg_network_1976,gaspard2004time,seifert_stochastic_2012,itoUnifiedFrameworkEntropy2020}. In the latter case of asymmetric couplings with constant fields, the system may relax towards a steady state known as a nonequilibrium steady state after some time. Time-asymmetric trajectories in steady state are linked with entropy change of the heat baths under `local detailed balance' for a system coupled to equilibrium reservoirs or heat baths \cite{jarzynskiHamiltonianDerivationDetailed2000,maes2021local}, suggesting that steady-state entropy production is critical for unveiling the interaction of out-of-equilibrium systems with their environments. Yet, unlike its equilibrium counterpart, the properties of irreversible Ising dynamics remain unclear due to the lack of theoretical underpinnings for its entropy production.

Here, we study the kinetics of the SK model with asymmetric connections under synchronous and asynchronous updates as a prototypical model of nonlinear and nonequilibrium processes. As the model does not have a free energy defined in classical terms, we resort to a dynamical equivalent in the form of a generating functional. We apply a path integral approach to obtain exact solutions on its statistical moments and nonequilibrium thermodynamic properties. Unlike the replica method, the generating functional for fully asymmetric couplings has exact solutions in the thermodynamic limit without additional assumptions like analytic continuation and replica symmetry breaking \cite{crisanti2018path}.
Previous studies using this method \cite{crisanti1987dynamics, crisanti1988dynamics} have shown that the asymmetric kinetic Ising model with asynchronous updates does not have a spin glass phase. In this manuscript, we will extend the generating functional path integral method to confirm this result in both cases of synchronous and asynchronous updates and further retrieve an exact solution of the entropy production of the system.

One of the open questions in empirical studies is whether an increase in entropy production observed in specific nonequilibrium systems under investigation is linked with the critical properties of  systems approaching continuous phase transitions \cite{perl2021nonequilibrium,delafuente2022temporal}. Entropy production is not necessarily maximized under such conditions and can display a continuous change \cite{falasco2018information}, or discontinuities in its derivative \cite{nguyen2018phase}. However, a number of simple nonequilibrium systems maximize their entropy production at a critical point. Examples are the entropy production of an Ising model with an oscillatory field and a mean-field majority vote model \cite{zhangCriticalBehaviorEntropy2016a,noaEntropyProductionTool2019a,crochik2005entropy}. It is therefore important to investigate the case of the kinetic Ising system as a general model of physical and biological networks.
We previously showed that the entropy production of the stationary asymmetric SK model with finite size takes a maximum around a critical point by applying mean-field approximations preserving fluctuations in the system \cite{aguilera2021unifying}.
However, this result (and the aforementioned references) relies on approximations and numerical simulations. Therefore, the assumption that entropy production is maximized near continuous phase transitions has not yet been ratified by exact solutions of spin models in the thermodynamic limit.
In this study, we show analytically that the entropy production is locally maximized at critical phase transition points, representing a potentially useful phase transition correlate for systems without a globally defined free energy or heat capacity.
Nevertheless, we also show that entropy production can take larger values for largely heterogeneous couplings in low-temperature regimes exhibiting disordered but nearly deterministic dynamics. Thus, entropy production must be examined carefully, as its increase does not necessarily indicate that the system is approaching a critical state. Instead, combining the entropy rate and entropy production yields a more precise picture of the irreversible processes.

The paper is organized as follows. First, we introduce maximum entropy Markov processes, their entropy production, and a generating functional method used to compute the system's moments and the entropy production in both discrete and continuous time. Next, we describe the asymmetric SK model with synchronous and asynchronous updates and a path integral method calculating the configurational average of the generating functional. This yields an exact solution of the entropy production, magnetization, and correlations in an infinite system. We employ our theoretical results to draw phase diagrams of the order parameters and entropy production for synchronous and asynchronous dynamics with and without randomly sampled external fields. The theoretical predictions are then corroborated by numerical simulation. We also examine the critical line of nonequilibrium phase transitions, the temporal structure of the dynamics, and their relations to the entropy production. Finally, we conclude the paper by discussing the implications of our results for the study of biological systems.

\section*{Results}


\subsection*{Maximum entropy Markov chains}

The principle of maximum entropy is a foundation of equilibrium statistical mechanics \cite{jaynes2003probability}. The principle has been later generalized for treating time-dependent phenomena, as the principle of maximum caliber or maximum path entropy \cite{jaynes1985macroscopic, presse2013principles}.
Under consistency requirements preserving causal interactions, the maximum caliber principle yields a Markov process \cite{ge2012markov}. 
To see this, we start with a discrete-time stochastic process with $N$ discrete-state elements defined at time $u$ as $\*s_u = \{s_{1,u}, \ldots, s_{N,u}\}$ for discrete-time trajectories of length $t+1$ defined by a path probability $p(\*s_{0:t})$. 
Later, we will show this discrete-time formulation can be generalized to an equivalent continuous-time formulation under appropriate assumptions.

Path entropy is defined as
\begin{align}
    S_{0:t} =&- \sum_{\*s_{0:t}} p(\*s_{0:t}) \log p(\*s_{0:t}).
    \label{eq:path-entropy}
\end{align}
Maximizing Eq.~\ref{eq:path-entropy}, subject to constraints, yields the least structured distribution $p(\*s_{0:t})$ consistent with observations \cite{livesey1985maximum}.
In causal network models, entropy maximization has to be constrained with a set of temporal consistency requirements \cite{ge2012markov}, as was first established by \cite{kolomogoroff2013grundbegriffe}.
Specifically, for any positive integer $u\,\,(\leq t)$, we impose 
\begin{align}
    \sum_{\*s_{u}} p_{0:u}(\*s_{0:u})=& p_{0:u-1}(\*s_{0:{u-1}}),
    \label{eq:consistency-condition}
\end{align}
where $p_{0:u}(\*s_{0:u})$ is given by
\begin{align}
    p_{0:u}(\*s_{0:u}) =&\argmax_{p(\*s_{0:u})} S_{0:u}.
\end{align}
That is, we impose consistency between the marginal distribution for the maximum entropy path $\*s_{0:u-1}$ in $p_{0:u}(\*s_{0:u})$ and the maximum entropy distribution of path $\*s_{0:u-1}$, $p_{0:u-1}(\*s_{0:u-1})$. This constrains path distribution dependencies between consecutive states. We will drop the subscript in the path probability when not needed.

Maximizing Eq.~\ref{eq:path-entropy} with constraints $f_n(\*s_{u},\*s_{u-1}) = C_{n,u}$ (where $C_{n,u}$ is a constant for the $n$-th constraint at time $u$), an initial distribution $p(\*s_{0})$, and  Eq.~\ref{eq:consistency-condition} results in a Markovian process (cf. \cite{ge2012markov})
\begin{align}
    p(\*s_{0:t}) =& p(\*s_{0})\prod_{u=1}^t p(\*s_{u}\,|\,\*s_{u-1})
    \nonumber\\ \propto&  p(\*s_{0}) \prod_{u=1}^t \exp\br{\sum_n \lambda_n f_n(\*s_{u},\*s_{u-1}) }.
    \label{eq:maxcal-solution}
\end{align}
The path entropy can be then decomposed into
\begin{align}
    S_{0:t} =& -\sum_{\*s_{0:t}} p(\*s_{0:t}) \pr{\sum_u \log p(\*s_{u}\,|\,\*s_{u-1}) +\log p(\*s_{0}) }
    \nonumber\\ =& \sum_u S_{u|u-1} + S_{0},
    \label{eq:path-entropy-II}
\end{align}
where $S_0$ is the entropy of the initial distribution and $S_{u|u-1}$ is a conditional entropy, defined as
\begin{align}
	S_{u|u-1} 
	= &- \sum_{\*s_u,\*s_{u-1}}p(\*s_{u},\*s_{u-1})
	\log p(\*s_u\,|\,\*s_{u-1}),
	    \label{eq:conditional-entropy}
\end{align}
which, at the steady state described in the following, corresponds to the Kolmogorov–Sinai entropy or entropy rate, $\lim_{t\to\infty} \frac{1}{t} S_{0:t}$.

\subsection*{Nonequilibrium steady state}

A Markov chain converges to a unique stationary distribution if the system is irreducible (all states are accessible from any state in finite time) and aperiodic (the greatest common divisor of the number of steps for returning to the same state with non-zero probability is one) \cite{freedman2017convergence}. 
We can confirm that these requirements are satisfied by Eq.~\ref{eq:maxcal-solution} with finite transition probabilities, thus warranting the existence of a steady-state distribution $\pi(\*s_u)$, which can be either in or out of thermodynamic equilibrium, as explained in the following.

For a discrete-time  Markov chain, the evolution of the state probability distribution follows a master equation:
\begin{align}
    p_{u}(\*s_u)=&\sum_{\*s_{u-1}} p(\*s_u\,|\,\*s_{u-1})p_{u-1}(\*s_{u-1})
     \nonumber\\ =& p_{u-1}(\*s_{u}) + \sum_{\*s_{u-1}}  j_{\*s_{u-1}\to \*s_u}^u.
     \label{eq:master-equation}
\end{align}
Here $p_v(\*s_u)$ is a marginal probability distribution of a state $\*s_{u}$ calculated for the distribution at time $v$. For simplicity, we will omit the subscript and write $p(\*s_u)$ when $v=u$. $j_{\*s_{u-1}\to \*s_u}^u$ are the system's probability fluxes:
\begin{align}
   j_{\*s_{u-1}\to \*s_u}^u \equiv p(\*s_u\,|\,\*s_{u-1})p(\*s_{u-1})
     -p(\*s_{u-1}\,|\,\*s_u)p_{u-1}(\*s_u).
    \label{eq:transition_rate}
\end{align}
In the limit of small probability fluxes, the system can be described by an equivalent continuous-time process:
\begin{align}
    \frac{dp(\*s,t)}{dt} =& \sum_{\*s'} j_{\*s'\to \*s}(t) 
    \\j_{\*s'\to \*s}(t) \equiv& w(\*s\,|\,\*s')p(\*s',t) - w(\*s'\,|\,\*s)p(\*s,t),
    \label{eq:continuous-time-master-equation}
\end{align}
where $t$ refers to the continuous time and $w(\*s\,|\,\*s')$ are transition rates. 

The system is stationary or in a steady state if the sum of all  probability fluxes is zero for all $\*s_u$, i.e., $\sum_{\*s_{u},\*s_{u-1}}  j_{\*s_{u-1}\to \*s_u}^u=0 $.
In addition, this will be an equilibrium steady state if $j_{\*s_{u-1}\to \*s_u}^u=0$ for all pairs $\*s_{u-1},\*s_u$, resulting in the detailed balance condition
\begin{align}
p(\*s_u\,|\,\*s_{u-1})\pi(\*s_{u-1})=p(\*s_{u-1}\,|\,\*s_u)\pi(\*s_u), 
\label{eq:detailed_balance}
\end{align}
where $\pi(\*s_u)$ is the steady-state distribution. When detailed balance is broken under the stationary condition, i.e., some $j_{\*s_{u-1}\to \*s_u}^u\neq 0$  but their sum is equal to zero, the stationary system is in a nonequilibrium steady state.

\subsection*{Steady-state entropy production}
Stochastic thermodynamics describes a link between the time-irreversible stochastic trajectories with surroundings in the form of heat (entropy) dissipation.  
As the system evolves, it experiences an entropy change $\sigma_u^\mathrm{sys}$:
\begin{align}
    \sigma_u^\mathrm{sys} =& S_u - S_{u-1} = \sum_{\*s_{u},\*s_{u-1}} p(\*s_{u},\*s_{u-1}) \log \frac{ p(\*s_{u-1})}{p(\*s_{u})}. 
\end{align}
Nonequilibrium systems maintain irreversible dynamics by continuously dissipating heat (entropy) to their environments. Under local detailed balance \cite{jarzynskiHamiltonianDerivationDetailed2000,van2015ensemble,maes2021local} in a system coupled to a heat bath, the entropy change results from subtracting the entropy dissipated to the heat bath $\sigma_{u}^\mathrm{bath}$ from the (total) entropy production $\sigma_{u}$: 
\begin{align}
     \sigma_{u}^\mathrm{sys} = \sigma_u - \sigma_{u}^\mathrm{bath},
     \label{eq:decompositon}
\end{align}
where the entropy change of the heat bath is given as
\begin{align}
    \sigma_{u}^\mathrm{bath} =  \sum_{\*s_{u},\*s_{u-1}} p(\*s_{u},\*s_{u-1}) \log \frac{p(\*s_{u}\,|\,\*s_{u-1})}{p(\*s_{u-1}\,|\,\*s_{u})}. 
    \label{eq:bath_entropy_change}
\end{align}
Here $p(\*s_{u-1}\,|\,\*s_{u})$ is a transition probability (from Eq.~\ref{eq:maxcal-solution}) but evaluated by the reverse trajectory \cite{jarzynskiHamiltonianDerivationDetailed2000,crooks1998nonequilibrium}, that is, we define it using the transition function at time $u$, but switch $\*s_{u}$ and $\*s_{u-1}$. This equation relates the system’s time asymmetry with the entropy change of the reservoir.

The entropy production $\sigma_u$ at time $u$ is then given as
\begin{align}
	\sigma_{u} 
	=& \sum_{\*s_u,\*s_{u-1}}p(\*s_u,\*s_{u-1})
	\log \frac{p(\*s_u\,|\,\*s_{u-1}) p(\*s_{u-1})}{p(\*s_{u-1}\,|\,\*s_{u})p(\*s_{u})},
	\label{eq:entropy_production}
\end{align}
which is the Kullback-Leibler divergence between the forward and backward trajectories \cite{schnakenberg_network_1976,seifert_stochastic_2012,esposito2010three,itoUnifiedFrameworkEntropy2020}.
Due to the non-negativity of the divergence, the entropy production is non-negative, $\sigma_{u} \geq 0$. This entropy production vanishes if the probability of forward trajectories is identical to a posterior of past states given the future state \cite{itoUnifiedFrameworkEntropy2020}, i.e., when the process loses time-asymmetry in prediction and postdiction \cite{igarashi2022entropy}. 

Alternatively, the dissipation function \cite{Evans2002TheFT,seifert_stochastic_2012,yang2020unified} quantifies the difference between incoming and outgoing fluxes in Eq.~\ref{eq:transition_rate}:
\begin{align}
	\widetilde \sigma_{u} 
	=& \sum_{\*s_u,\*s_{u-1}}p(\*s_u,\*s_{u-1})
	\log \frac{p(\*s_u\,|\,\*s_{u-1}) p(\*s_{u-1})}{p(\*s_{u-1}\,|\,\*s_{u})p_{u-1}(\*s_{u})},
	\label{eq:dissipation_function}
\end{align}
which directly assesses a violation of the detailed balance. The entropy production $\sigma_{u}$ and dissipation function $\widetilde \sigma_{u}$ are equivalent under steady-state conditions. Furthermore, both quantities become equivalent in the continuous-time limit  \cite{seifert_stochastic_2012,yang2020unified,igarashi2022entropy} and converge to the entropy production rate \cite{van2015ensemble}:
\begin{align}
    \frac{d \sigma(t)}{dt} =& \frac{1}{2}\sum_{\*s,\*s'} j_{\*s'\to \*s}(t)  \log \frac{w(\*s\,|\,\*s') p(\*s',t)}{w(\*s'\,|\,\*s)p(\*s,t)}.
    \label{eq:entropy-production-rate}
\end{align}

In a steady state, the entropy production is caused by dissipation only and becomes equivalent to the \emph{house-keeping} entropy production caused by the non-conservative forces under a steady state \cite{hatano2001steady,dechant2022geometric}. Both $\sigma_{u}$ and $\widetilde \sigma_{u}$ result in:
\begin{align}
    \sigma_u 
    =& \widetilde \sigma_{u}  =\sigma_{u}^\mathrm{bath} 
    = - S_{u|u-1} + S_{u|u-1}^r.
    \label{eq:entropy_productionII}
\end{align}
Here  $S_{u|u-1}^r$ is the entropy of the time-reversed conditional distribution:
\begin{align}
	S_{u|u-1}^r 
	&\equiv - \sum_{\*s_u,\*s_{u-1}}p(\*s_u,\*s_{u-1}) 
	\log p(\*s_{u-1}\,|\,\*s_u).
	    \label{eq:reverse_conditional-entropy}
\end{align}
In this paper, we study the steady-state entropy production in Eq.~\ref{eq:entropy_productionII}, which is critical for evaluating the interaction of the nonequilibrium processes with their environment.

\subsection*{Generating functional}

Consider a maximum caliber path probability (Eq.~\ref{eq:maxcal-solution}) 
\begin{align}
    p(\*s_{0:t}) =& \prod_{u=1}^{t} p(\*s_u\,|\,\*s_{u-1}) p(\*s_0),
\end{align}
For simplicity, we will assume $p_0(\*s)=\prod_i \delta\br{s_i,s_{i,0}}$ -- the initial distribution is a Kronecker delta with a unique initial state -- and ignore the term. However, the following steps are general for any $p_0(\*s_0)$.

In equilibrium systems, the partition function retrieves its statistical moments. A nonequilibrium equivalent function is a generating functional or dynamical partition function. To obtain not only the statistical properties averaged over trajectories, but also the forward/time-reversed conditional entropies (Eqs.~\ref{eq:conditional-entropy}, \ref{eq:reverse_conditional-entropy}), we define the following generating functional:
\begin{align}
    Z_t(\*g) =&
    \sum_{\*s_{0:t}} p(\*s_{0:t})  \exp \Bigg[  \Gamma(\*g,\*s_{0:t}) \Bigg],
    \label{eq:SK-partition-function}
    \\\Gamma(\*g,\*s_{0:t}) =&  \sum_{i,u} g_{i,u} s_{i,u}
       - \sum_{u}  g_{u}^{S} \epsilon(\*s_{u}\,|\,\*s_{u-1})
     \nonumber\\ & -  \sum_{u}  g_{u}^{S^r}   \epsilon(\*s_{u-1}\,|\,\*s_{u}),
\end{align}
where  $\epsilon(\*s_{u}\,|\,\*s_{u-1}) \equiv - \log  p(\*s_u\,|\,\*s_{u-1})$.
In the limit $t\to\infty$, the logarithm of the generating functional converges to the large deviation function \cite{touchette2009large,touchette1large,touchette2018introduction},
\begin{equation}
    \varphi(\*g) =	\lim_{t\to\infty} \frac{1}{t} \log Z_t(\*g),
\end{equation}
which plays the role of a free-energy function for nonequilibrium trajectories \cite{lecomte_thermodynamic_2007}.
The vector $\*g$ is composed of parameters $g_{i,u}$ ($i=1,\ldots,N$, $u=1,\ldots,t$) and $g_{u}^{S}$, $g_{u}^{S^r}$ ($u=1,\ldots,t$) 
retrieving the system's statistical properties.
The parameters $g_{i,u}$ recover the statistical moments of the systems like the average rates and correlations:
\begin{align}
	\lim_{\*g\to\*0} \frac{\partial Z_t(\*g)}{\partial g_{i,u}} =&  \lim_{\*g\to\*0} \ang{ s_{i,u}}_{\*g} = \ang{ s_{i,u}}, \label{eq:m_iu}
	\\  
	\lim_{\*g\to\*0} \frac{\partial^2 Z_t(\*g)}{\partial g_{i,u} \partial g_{j,v}} =&   \lim_{\*g\to\*0} \ang{s_{i,u} s_{j,v}}_{\*g} = \ang{s_{i,u} s_{j,v}}, \label{eq:R_ij_uv}
\end{align}
where angle brackets are defined as
\begin{align}
    \ang{f(\*s_{0:t})}_{\*g} =& \sum_{\*s_{0:t}} f(\*s_{0:t}) \exp\Bigg[ \sum_{i,u} g_{i,u} s_{i,u}\Bigg] p(\*s_{0:t}),
    \\  \ang{f(\*s_{0:t})} =& \sum_{\*s_{0:t}} f(\*s_{0:t})  p(\*s_{0:t}).
\end{align}

In addition, $g_{u}^{S}$, $g_{u}^{S^r}$ retrieve the conditional and time-reversed conditional entropy terms, $S_{u|u-1}, S_{u|u-1}^r$:
\begin{align}
	S_{u|u-1} =& - \lim_{\*g\to\*0} \frac{\partial Z_t(\*g)}{\partial g_{u}^{S}}
	\nonumber\\ =& \lim_{\*g\to\*0}  \ang{\epsilon(\*s_{u}\,|\,\*s_{u-1}) }_{\*g}
	\label{eq:conditional_entropy_from_partition_func}
	=  \ang{ \epsilon(\*s_{u}\,|\,\*s_{u-1})},
	\\S_{u|u-1}^r =&  -\lim_{\*g\to\*0} \frac{\partial Z_t(\*g)}{\partial g_{u}^{S^r}}
	\nonumber\\ =& \lim_{\*g\to\*0} \ang{\epsilon(\*s_{u-1}\,|\,\*s_{u}) }_{\*g}
	=  \ang{\epsilon(\*s_{u-1}\,|\,\*s_{u})},
	\label{eq:reverse_conditional_entropy_from_partition_func}
\end{align}
and thus the steady-state entropy production (Eq.~\ref{eq:entropy_productionII}):
\begin{align}
	\sigma_{u} =& \lim_{\*g\to\*0} \pr{\frac{\partial Z_t(\*g)}{\partial g_{u}^{S}} - \frac{\partial Z_t(\*g)}{\partial g_{u}^{S^r}}}.
    \label{eq:entropy_production_from_partition_func}
\end{align}

\subsection*{Synchronous and asynchronous, asymmetric Sherrington-Kirkpatrick model}
\label{sec:model-description}
\begin{figure*}
\begin{center}
\includegraphics[width=14cm]{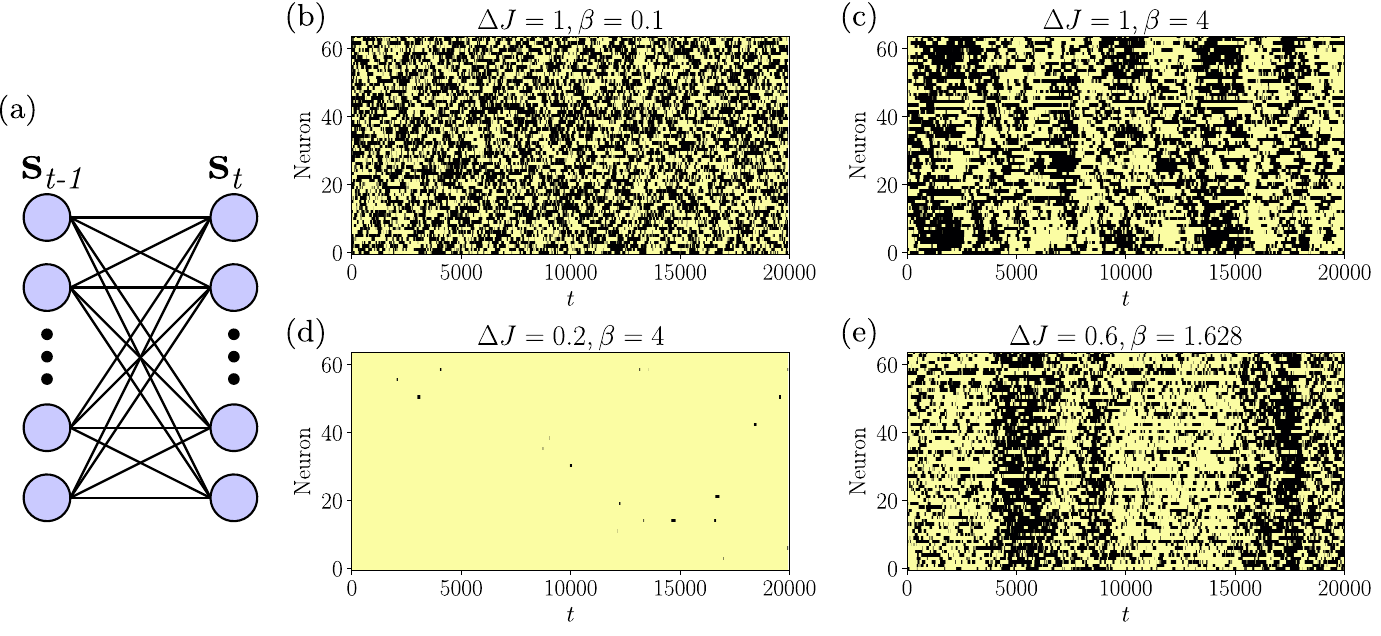} 
\end{center}
\caption{\textbf{Asymmetric kinetic SK model}. (a) The asymmetric kinetic Ising model describes a Markov chain where states at time $\*s_u$ depend on pairwise couplings to states $\*s_{u-1}$. (b-e) This model shows disordered dynamics for large coupling variance both at high and low temperatures (b and c), ordered dynamics for low temperatures and a low coupling variance (d), and critical dynamics at the phase transition (e).} 
\label{fig:raster-plots}
\end{figure*}

We consider $N$ interacting elements $\*s_u$ (spins or neurons), taking each element $i$ at time $u$ a binary state $s_{i,u}=\{-1,1\}$. Constraints take the form of delayed pairwise couplings (i.e., $f_{i,j}(\*s_u,\*s_{u-1})= s_{i,u}s_{j,u-1}$ in Eq.~\ref{eq:maxcal-solution}). This results in the dynamics:
\begin{align}
    p(\*s_{u}\,|\,\*{s}_{u-1})=& \prod_i \frac{\exp\br{ \beta s_{i,u} h_{i,u}}}{2 \cosh\br{\beta h_{i,u}}},
 \label{eq:Ising}
    \\ h_{i,u}=& H_{i,u}+\sum_j J_{ij} s_{j,u-1},
 \label{eq:Ising-field}
\end{align}
where $\beta$ is the inverse temperature. The system's state at time $u$ depends on the previous time-step (Fig.~\ref{fig:raster-plots}(a)).

The equation above is a general formulation of a kinetic Ising model with time-dependent fields $H_{i,u}$. The dynamics can include both synchronous and asynchronous Ising systems by introducing a set of independent Bernoulli random variables: $\tau_{i,u}={0,1}$ with probabilities $1-\alpha$ and $\alpha$ (i.e., $\tau_{i,u}\sim \mathrm{Bernoulli}\pr{\alpha}$) and making $H_{i,u}$ stochastic processes:
\begin{align}
    H_{i,u} = \Theta_{i,u} + (1-\tau_{i,u}) K s_{i,u-1}.
    \label{eq:asynchronous-fields}
\end{align}
Note that in the limit of $K\to\infty$, the state $s_{i,u}$ is tightly coupled to the previous state $s_{i,u-1}$. Therefore, the state changes only if $\tau_{i,u} = 1$. We have the following transition probability in the $K\to\infty$ limit:
\begin{align}
    p(\*s_{u}\,|\,\*{s}_{u-1})=& \prod_i \big(  \tau_{i,u} w(s_{i,u}\,|\,\*{s}_{u-1}) 
    \nonumber\\ &
    + (1-\tau_{i,u}) \delta\br{s_{i,u},s_{i,u-1}} 
    \big),
 \label{eq:Ising2}
\end{align}
with transition rate
\begin{align}
    w(s_{i,u}\,|\,\*{s}_{u-1}) = \frac{\exp\br{ \beta s_{i,u} h_{i,u}^1}}{2 \cosh\br{\beta h_{i,u}^1}},
\end{align}
where $h_{i,u}^1 = \Theta_{i,u}+\sum_j J_{ij} s_{j,u-1}$ and $\Theta_{i,u}$ is an external field.
With $\alpha=1$ we have a kinetic Ising system under parallel or synchronous updates. In the limit $\alpha \to 0$, we have in turn a kinetic Ising system with asynchronous updates (i.e., at most one spin is updated each time step), converging to a continuous-time master equation. 

The generating functional of the kinetic Ising system (Eq.~\ref{eq:SK-partition-function}) is defined by the functions
\begin{align}
	\epsilon(\*s_{u}\,|\,\*s_{u-1})  =& -\sum_i \pr{\beta  s_{i,u} h_{i,u} +  \log\br{ 2 \cosh \br{\beta h_{i,u}}}},
	\label{eq:epsilon}
	 \\ \epsilon(\*s_{u-1}\,|\,\*s_{u})  =&  -\sum_i \pr{ \beta  s_{i,u-1} h_{i,u}^r +  \log\br{2 \cosh \br{\beta h_{i,u}^r}}},
	 	\label{eq:epsilon_r}
\end{align}
where 
$h_{i,u}^r= H_{i,u}+\sum_j J_{ij} s_{j,u} = h_{i,u+1} +H_{i,u}-H_{i,u+1}$.

The equilibrium Ising model with symmetric random Gaussian couplings is referred to as the SK model. In the fully-asymmetric SK model, the couplings $J_{ij}$ are quenched independent variables, each following a Gaussian distribution 
\begin{equation}
    p(J_{ij}) = \frac{1}{\sqrt{2\pi \Delta J^2/N}} \exp\Bigg[ \frac{-1}{2\Delta J^2/N} \left(J_{ij} - \frac{J_0}{N}\right)^2 \Bigg],
    \label{eq:Gaussian-coulings}
\end{equation}
with mean $J_0/N$ and  variance $\Delta J^2/N$ scaled by $N$. 

The asymmetric SK model shows a variety of population dynamics. Fig.~\ref{fig:raster-plots} shows exemplary dynamics under asynchronous updates without fields ($\Theta_{i,u}=0$). It shows disordered dynamics for large coupling variance $\Delta J^2$ both at high and low temperatures (i.e. low and large $\beta$, Fig.~\ref{fig:raster-plots} (b,c)), ordered dynamics for low temperatures and low  $\Delta J^2$ (Fig.~\ref{fig:raster-plots}(d)), and critical dynamics at the phase transition (Fig.~\ref{fig:raster-plots}(e)).

\subsection*{Solution of the asymmetric Sherrington-Kirkpatrick model}
\label{sec:model-solution}

The solution of the kinetic version of the SK model with asymmetric and quenched couplings can be obtained by computing the generating functional averaged over the couplings (referred to as the configurational average):
\begin{equation}
    [Z_t(\*g)]_{\*J} = \int  d\*J  Z_t(\*g)\prod_{i,j}p(J_{ij}).
\end{equation}
This integral cannot be solved directly because of the $\log\br{ 2\cosh \br{\cdot}}$ terms in Eqs.~\ref{eq:epsilon} and \ref{eq:epsilon_r}, which depend nonlinearly on $J_{ij}$. A path integral method \cite{hertz2016path} to find a solution introduces a delta integral representing $\beta h_{i,u}$ with auxiliary variables $\theta_{i,u} = \beta (H_{i,u} + \sum_j J_{ij} s_{j,u-1})$ as well as $\beta h^{r}_{i,u}$  with an auxiliary variable $\vartheta_{i,u} = \theta_{i,u+1} + \beta( H_{i,u}-H_{i,u+1})$. Let $\^\theta=\{\theta_{i,u}\}$ (note $u=1,\ldots,t+1$) and $\^\vartheta=\{\vartheta_{i,u}\}$ ($u=0,\ldots,t$) denote a set of the auxiliary variables. Using conjugate variables $\^{\hat\theta} = \{\hat \theta_{i,u}\}$ to represent the delta function in the integral form, the configurational average is written as
\begin{align}
     [Z_t(\*g)]_{\*J} =& \frac{1}{\pr{2\pi}^{N(t+1)}}\int d\^\theta d\^{\hat\theta} d\*J
    \pr{\prod_{i,j} p(J_{ij})}
    \nonumber \\ 
    & \cdot \sum_{\*s_{1:t}} \exp \Bigg[ \sum_{i,u} ( s_{i,u}\theta_{i,u} - \log\br{2 \cosh\theta_{i,u}}) 
   \nonumber \\ &   +\sum_{i,u} \iu  \hat\theta_{i,u} (\theta_{i,u} - \beta H_{i,u} - \beta \sum_j J_{ij} s_{j,u-1}) 
   \nonumber \\ &   +  \Gamma(\*g,\*s_{0:t},\^\theta,\^\vartheta)
   \Bigg],
   \label{eq:configurational_average_SK}
\end{align}
with
\begin{align}
    \MoveEqLeft { \Gamma(\*g,\*s_{0:t},\^\theta,\^\vartheta) = \sum_{i,u} \Gamma_{i,u}(\*g,\*s_{0:t},\theta_{i,u},\vartheta_{i,u})} 
    \nonumber\\  \quad  & = \sum_{i,u} \Big(
    g_{i,u} s_{i,u} 
    + g_{u}^{S}   \pr{ s_{i,u} \theta_{i,u}  - \log\br{2 \cosh\br{\theta_{i,u}} }}
     \nonumber\\ &    \phantom{=}+  g_{u}^{S^r}   \pr{   s_{i,u-1}  \vartheta_{i,u}- \log\br{ 2 \cosh\br{\vartheta_{i,u}}  }}\Big).
\end{align}
Note that the summation of $\hat{\theta}_{i,u}$ terms is performed over $u=1,\dots,t+1$ to retrieve the fields of both the forward and backward trajectories.

The integral over $J_{ij}$ can be now performed directly over linear exponential terms (see Supplementary Note~1).
After integration, Eq.~\ref{eq:configurational_average_SK} incorporates quadruple-wise interactions among spins $\*s_{0:t}$ and conjugate variables $\^{\hat \theta}$ (Eq.~S1.10),
similar to replica interactions in the equilibrium SK model \cite{nishimori_statistical_2001}. These interactions are simplified by introducing Gaussian integrals and a saddle-point approximation in the thermodynamic limit (Eq.~S1.26).
The saddle-point solution can be simplified by introducing four types of order parameters (Eq.~S1.31).
In fully-asymmetric networks, two of these order parameters are found to be zero at $\*g=\*0$, yielding a solution in terms of the order parameters $m_u$ and $q_{u,v}$ (see Eq.~S1.48, S1.49): 
\begin{align}
	m_{u} =& \frac{1}{N}\sum_i \br{ \ang{s_{i,u}} }_{\*J},
	\label{eq:order_parameter_m}
	\\q_{u,v} = & \frac{1}{N}\sum_i\br{ \ang{s_{i,u} s_{i,v}}}_{\*J}.
	\label{eq:order_parameter_q}
\end{align}
Finally, conjugate variables $\^{\hat \theta}$ in the saddle-point solution can be substituted with a multivariate Gaussian integral (Eq.~S1.57),
leading to a factorized generating functional
\begin{align}
     [Z_t(\*g)]_{\*J}
	=&    
	\prod_i \sum_{\*s_{i,1:t}} \int d\^{\xi} p(\^{\xi}) \exp\Bigg[
	\sum_{u} s_{i,u} \overline h_{i,u}(\xi_{u}) 
	 \nonumber\\ &+ \sum_{u}  \beta  s_{i,u-1}\widetilde h_{i,u-1}-  \sum_{u} \log  2 \cosh\br{ \beta\overline h_{i,u}(\xi_{u})}
	 \nonumber\\ & + \sum_u \Gamma_{i,u}(\*g, \*s_{0:t},\beta{\overline h_{i,u}}({\xi_u}), \beta{\overline h^r_{i,u}}({\xi_{u+1}})) 
	 \Bigg],
	 \label{eq:[Z_t]_J}
\end{align}
where the stochastic elements $\^{\xi}=(\xi_{1},\ldots,\xi_{t+1})$  affecting each spin $i$  follow a multivariate normal distribution $p(\^{\xi})=\mathcal{N}(\*0,\*q)$ with $\*q$ defining $q_{u-1,v-1}$ as the covariance of each pair $\xi_u,\xi_v$ for $u,v \in 1,\ldots,t+1$. Here, at $\*g=\*0$, spin interactions are effectively substituted by same-spin temporal couplings in mean effective fields
\begin{align}
	\overline h_{i,u}( \xi_{u}) =& 
	 H_{i,u} + J_0 m_{u-1} 
	 + \Delta J  \xi_{u},
	 \\ \overline h^r_{i,u}(\xi_{u+1})=& 
	H_{i,u} + J_0 m_{u}  + \Delta J \xi_{u+1},
	\\ \widetilde h_{i,u-1} =& 0.
\end{align}
Applying Eqs.~\ref{eq:m_iu} and \ref{eq:R_ij_uv} to the configurational average in Eq.~\ref{eq:[Z_t]_J}, we obtain the order parameters $m_u$ and $q_{u,v}$:
\begin{align}
	m_{u} =& \frac{1}{N}\sum_i \int \mathrm{D}z \tanh\br{\beta\overline h_{i,u}(z) },
	\label{eq:def_order_parameter_m}
	\\  q_{u,v} =&\frac{1}{N}\sum_i \int \mathrm{D}xy^{(q_{u-1,v-1}) }
	\tanh\br{\beta\overline h_{i,u}(x) } 
	\nonumber\\ &\cdot\tanh\br{\beta\overline h_{i,v}(y) },
	\label{eq:def_order_parameter_q}
\end{align}
where the Gaussian stochastic terms are simplified to
\begin{align}
	\mathrm{D}z =&  \frac{1}{\sqrt{2\pi}} \exp\br{-\frac{1}{2}z^2},
	\\ \mathrm{D}xy^{(q_{u,v} )} =& \frac{1}{2\pi\sqrt{1-q_{u,v} ^2}}\exp\br{\frac{-x^{2}-y^{2}+2q_{u,v}  xy}{2(1-q_{u,v} ^2)}}.
\end{align}
Note that, in contrast with the equilibrium SK model, $m_u$ is independent of $q_{u,v}$, resulting in the lack of a spin-glass phase as suggested by previous studies  \cite{crisanti1988dynamics}. 

The configurational average of Eqs.~\ref{eq:conditional_entropy_from_partition_func} and \ref{eq:reverse_conditional_entropy_from_partition_func} results in the following conditional entropy and time-reversed conditional entropy
\begin{align}
	\br{S_{u|u-1} }_{\*J} =& \sum_i   \int  -\mathrm{D}z \Big( 
	\beta \pr{H_{i,u} + J_0 m_{u-1}} \tanh\br{\beta\overline h_{i,u}(z)}
	\nonumber\\  &+\beta^2 \Delta J^2 \pr{1-\tanh^2\br{\beta\overline h_{i,u}(z)}}
	\nonumber\\ & - \log\br{2 \cosh \br{\beta\overline h_{i,u}(z)}} \Big),
    \\
	\br{S_{u|u-1}^r }_{\*J} =&   \sum_i   \int  - \mathrm{D}z \Big( 
    \beta( H_{i,u} +  J_0 m_{u}) \tanh\br{\beta \overline h_{i,u-1}(z)}
	\nonumber\\  &+
	\beta^2 \Delta J^2  q_{u,u-2} \pr{1-\tanh^2\br{\beta \overline h_{i,u-1}(z)}}
	\nonumber\\ & -\log\br{2 \cosh \br{\beta\overline h^r_{i,u}(z)} } \Big).
\end{align}

Up to this point, our results are general for time-dependent fields $H_{i,u}$, covering synchronous and asynchronous updates by Eq.~\ref{eq:asynchronous-fields}. We obtain the results for the synchronous SK model by setting $\alpha=1$ or, equivalently, $H_{i,u}=\Theta_{i,u}$. For time-independent fields ($\Theta_{i,u}=\Theta_{i}$), the system converges to a steady state determined by the solution of the self-consistent equations given by Eqs.~\ref{eq:def_order_parameter_m} and \ref{eq:def_order_parameter_q}. 
Finally, using Eq.~\ref{eq:entropy_production_from_partition_func}, the steady-state entropy production under the synchronous updates is obtained as
\begin{align}
	\br{\sigma_{u}}_{\*J} =&  \beta^2 \Delta J^2 (1-q_{u,u-2}) 
	\sum_i \int  \mathrm{D}z
	(1-\nonumber \\ &\tanh^2\br{\beta \pr{\Theta_{i} + J_0 m_{u-1}  + \Delta J  z}}),
    \label{eq:sigma_mean-field}
\end{align}
with 
$m_{u-1}$ and $q_{u,u-2}$ given by their steady-state values (i.e., independent of $u$). Note that for the synchronous system the steady-state solution of $q_{u,v}$ is the same for all $u,v$. In the following, we will use $m$ and $q$ to represent these steady-state solutions. 

To calculate the steady-state solutions for the asynchronous SK model, we calculate the generating functional $\br{Z_t(\*g)}_{\*J,\^\tau}$ that is additionally averaged over the independent random variables $\tau_{i,u}$ in Eq.~\ref{eq:asynchronous-fields}. We show in Supplementary Note~2 that the resulting order parameters in continuous-time $m(t)$ and $q(t',t)$ are subject to the following dynamical equations:
\begin{align}
	\frac{d m(t)}{d t} =& \frac{1}{N}\sum_i  \int \mathrm{D}z \tanh\br{\beta h_{i}^{\ast}(z,t)} - m(t).
	\label{eq:async_order_parameter_m_cont}
	\\\frac{d q(t',t)}{d t} =& q^{1}(t',t) - q(t',t) ,
	 \label{eq:async_order_parameter_q_cont1}
	 \\ \frac{d q^{1}(t',t)}{dt'}=& 
	 \frac{1}{N}\sum_i   \int \mathrm{D}xy^{(q(t',t))} \tanh\br{\beta h_{i}^{\ast}(x,t')} 
	 \nonumber \\ &\cdot \tanh\br{\beta h_{i}^{\ast}(y,t)} - q^{1}(t',t)
	 \label{eq:async_order_parameter_q_cont2}
\end{align}
with $h_{i}^{\ast}(z,t) = \Theta_i + J_0 m(t) + \Delta J z$. Here $q^{1}(t',t)$ is the spin correlation conditioned on spins being updated at time $t$. The steady-state solutions of $m(t)$ and $q(t',t)$ (assuming $t' \gg t$) converge to the same steady-state values $m$ and $q$ found for the synchronous SK model (see Supplementary Note~2).

In the continuous-time limit, the steady-state entropy production converges to a steady-state entropy production rate (Eq.~\ref{eq:entropy-production-rate}) given by:
\begin{align}
	\br{\frac{d\sigma(t)}{dt}}_{\*J,\^\tau} =& \lim_{\alpha \to 0} \beta^2 \Delta J^2 (1-q(t+\alpha,t-\alpha) ) \nonumber\\
    & \cdot \sum_i \int  \mathrm{D}z(1-\tanh^2\br{\beta h_{i}^{\ast}(z,t)}).
    \label{eq:sigma-rate_mean-field}
\end{align}
We note that the delayed-self correlation $q(t',t)$ is discontinuous at $t'=t$ (i.e., $\lim_{\alpha \to 0} q(t+\alpha,t-\alpha)\neq q(t,t)=1$, see Fig.~S1), warranting that the entropy production rate can be non-zero for appropriate parameters.


Given the analytical solutions of the system, we will now study the phase diagram of the SK model. In contrast with the naive replica-symmetric solution of the equilibrium SK model, the equations above are exact in the model with asymmetric couplings in the thermodynamic limit under both synchronous and asynchronous updates.

\subsection*{The SK model without external fields}

\begin{figure}
\begin{center}
\includegraphics[width=5.5cm]{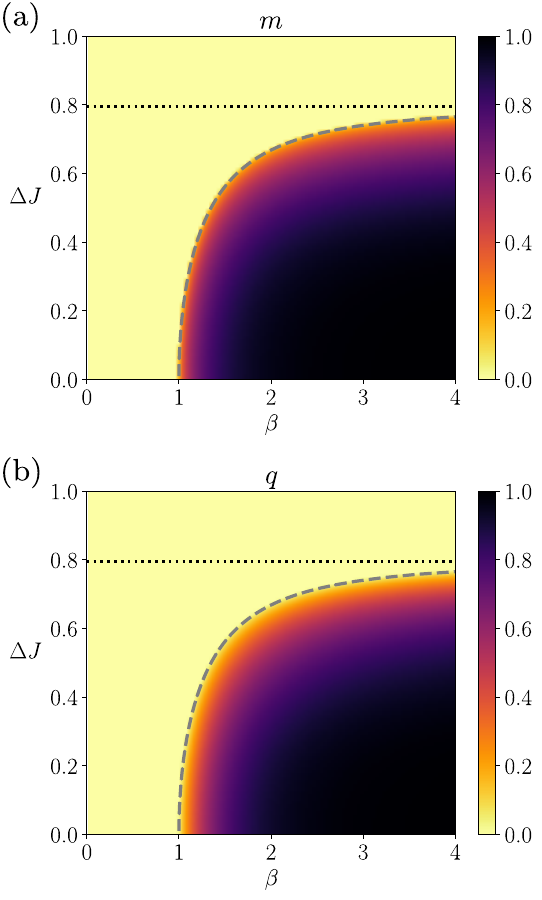}
\end{center}
\caption{\textbf{Order parameters of the asymmetric SK model with zero fields under synchronous and asynchronous updates}. The average magnetization $m$ and the average delayed self-coupling $q$ are shown in the phase diagram of the inverse temperature $\beta$ and coupling heterogeneity $\Delta J$ using a model with fixed parameters $J_0=1$, $\Delta H=0$. The dashed line represents the critical line separating ordered and disordered phases. The dotted line represents the critical value at zero temperature ($\beta\to\infty$).
} 
\label{fig:order-parameters-DeltaJ}
\end{figure}

Fig.~\ref{fig:order-parameters-DeltaJ}(a) and \ref{fig:order-parameters-DeltaJ}(b) display the phase diagram of the steady-state order parameters, $m$ and $q$, for both synchronous and asynchronous updates, respectively derived from Eqs.~\ref{eq:def_order_parameter_m} and \ref{eq:def_order_parameter_q} as a function of the inverse temperature $\beta$ and the width of the coupling distribution $\Delta J$, when the external fields are fixed at zeros ($\Theta_{i,u}=0$) and the mean coupling is $J_0 = 1$. In this setting, the inverse temperature $\beta$ controls the magnitude of the couplings. The phase diagram shows two distinct regions, one in which the order parameters are fixed at zero (zero magnetization and zero self-correlations, $m=0$ and $q=0$) -- indicating disordered states -- and the other in which the order parameters become positive ($m>0$ and $q>0$) -- indicating ordered states. Therefore, the system exhibits a nonequilibrium analogue of the paramagnetic-ferromagnetic (disorder-order) phase transition controlled by the parameters, $\beta$ and $\Delta J$. The dashed line in each panel shows the critical values of $\Delta J$ as a function of $\beta$, which is obtained by solving the following equation (see Supplementary Note~3),
\begin{align}
    \frac{1}{ \beta J_0} = \int \mathrm{D}z \pr{1-\tanh^2\br{\beta\pr{ \Delta J z }} }.
\end{align}
The solution will be denoted as $\Delta J^c(\beta)$.
As studied in Supplementary Note~3, this critical phase transition corresponds to the mean-field universality class, as in the order-disorder phase transition of the equilibrium SK model (note that the spin-glass phase has different exponents \cite{oppermann2008universality}).

Note that, depending on the coupling variance $\Delta J$, the dynamics do or do not undergo the nonequilibrium phase transition  by varying the inverse temperature $\beta$. The critical $\Delta J^c(\beta)$ at $\beta \rightarrow \infty$ is given as $\Delta J^c(\infty)=0.79501$ (dotted horizontal line). If the distribution is narrower than the critical value $\Delta J^c(\infty)$, the process undergoes the phase transition by changing $\beta$. If the distribution is wider than the critical value, the order parameters are fixed at zeros ($m=0$, $q=0$) for any $\beta$.  Note that, for $\beta\to\infty$ (zero temperature), the activation function approaches the threshold nonlinearity given by the sign function; therefore, the process becomes deterministic. That is, for the large values of $\beta$, the process approaches deterministic dynamics yielding either ordered or disordered states for smaller or larger $\Delta J$, respectively. We remark that the disordered state with $m=0$ and $q=0$ at high $\beta$ (low temperature) does not indicate the spin-glass phase as expected for the equilibrium Ising system (see Supplementary Note~4). 
We confirmed the non-existence of a spin-glass phase for the asymmetric kinetic SK model by finding that the system decays exponentially in this region (Supplementary Note~5).

\begin{figure*}[ht]
\begin{center}
\includegraphics[width=15cm]{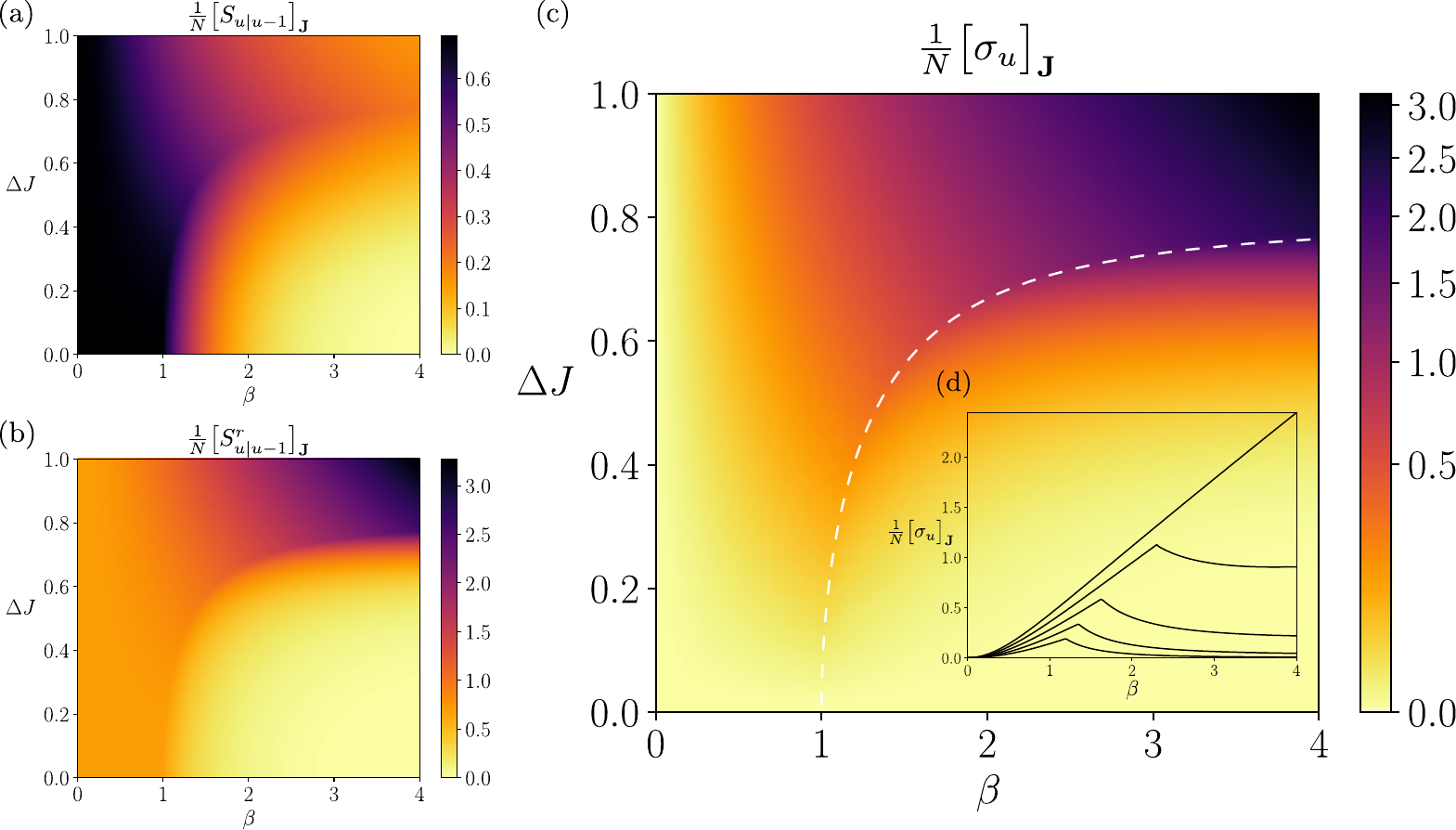} 
\end{center}
\caption{\textbf{Steady-state entropy rate and entropy production of the asymmetric SK model under synchronous updates}. (a) The phase diagram of the conditional entropy $\br{S_{u|u-1} }_{\*J}$ (equivalent to the entropy rate) as a function of the inverse temperature $\beta$ and the coupling heterogeneity $\Delta J$. (b) The conditional entropy of the reverse dynamics $\br{S_{u|u-1}^r }_{\*J}$. (c) The entropy production at a steady state. The white dashed line is a critical line for the nonequilibrium phase transitions. (d, inset) The horizontal sections of the entropy production ($\Delta J=0.4, 0.5, 0.6, 0.7$, and  $0.7950$), showing that it peaks at the critical line. All figures are based on a model with fixed parameters $H_{i,u}=0$ and $J_0=1$.
} 
\label{fig:entropy-DeltaJ}
\end{figure*}

The reduction in uncertainty at higher $\beta$ is indicated by the reduction of the conditional entropy (the path entropy) $S_{u|u-1}$ by increasing $\beta$ (Fig.~\ref{fig:entropy-DeltaJ}(a)). This figure additionally shows that the conditional entropy decreases slowly with increasing $\beta$ along the critical line of the phase transitions. This means that strong couplings and diverse patterns co-exist along the critical line. In contrast, the time-reversed conditional entropy $S_{u|u-1}^r$ (Fig.~\ref{fig:entropy-DeltaJ}(b)) displays opposite dependency on $\beta$ for the broader or narrower coupling distributions. Time-reversed conditional entropy quantifies how surprising the reverse process is under the forward model. With coupling distributions narrower than the critical value $\Delta J^c(\infty)$, the time-reversed conditional entropy diminishes by increasing $\beta$, indicating that the reverse processes takes place with increasingly high probabilities. This is because the spin state is fixated at all up or down under the ferromagnetic-like state for all time, losing temporal asymmetry. In contrast, the reverse process becomes less likely to happen as the dynamics becomes more deterministic by increasing $\beta$ yet remains disordered as long as the coupling distribution is broader than $\Delta J^c(\infty)$. This distinct behavior between the conditional entropy and its time-reversed version found at the wider coupling distributions and high inverse temperatures yields the strong time-asymmetry in this regime. 

The entropy production under the steady-state condition quantifies the difference between the conditional and time-reversed conditional entropy. Fig.~\ref{fig:entropy-DeltaJ}(c) displays the phase diagram of the entropy production for the synchronous Ising model (the asynchronous Ising model has a similar behavior but different scaling, see Supplementary Note~2 or Fig.~\ref{fig:simulation}). The entropy production is maximized at the high $\beta$ under the broader coupling distributions, where we find a significant difference between these two conditional entropies. Namely, strong time-asymmetry appears when the dynamics are disordered, nearly deterministic processes. Note that the entropy production increases with $\beta$ if the coupling distribution is wider than $\Delta J^c(\infty)$. In contrast, the entropy production is locally maximized at the critical point (white dashed line) with the  coupling distribution being narrower than $\Delta J^c(\infty)$ (see also Fig.~\ref{fig:entropy-DeltaJ}(d)). For the narrowly distributed couplings, the process exhibits a paramagnetic-like (randomized or disordered) phase at smaller $\beta$ and a ferromagnetic-like (ordered) phase at higher $\beta$ (Fig.~\ref{fig:order-parameters-DeltaJ}), neither of which can exhibit adequately asymmetric dynamics in time. Time-asymmetry appears between the ordered and disordered phases, namely at the critical point. As a consequence, the steady-state entropy production can be a measure of the criticality in this regime. 
However, more importantly, the magnitude of the entropy production is far more significant in the regime of large $\Delta J$ and $\beta$ than near the critical states, due to the strong time-asymmetry caused by the combination of disordered and quasi-deterministic dynamics.

To verify our theoretical predictions for the order parameters and steady-state entropy production, we compared them with the values computed from the sample trajectories by numerically simulating the kinetic SK models (see Supplementary Note~6 for the details). We constructed the kinetic Ising model with parameters $\Theta_{i,u} = 0$ and randomly generated $J_{ij}$ with $\Delta J=0.5$ and $J_0 = 1$ while changing the inverse temperature $\beta$. We ran simulations of the model for $t=128$ time steps and repeated the simulation $10^6$ times at each $\beta$. We computed the mean activation rate $\frac{1}{N}\sum_i \br{\ang{s_{i,u}}}_{\*J,\^\tau}$, the average delayed self-correlations $\frac{1}{N}\sum_i\br{\ang{s_{i,u}s_{i,u-1}}}_{\*J,\^\tau}$, and the normalized entropy production and entropy production rates $\frac{1}{N}\br{\sigma}_{\*J},\frac{1}{N}\br{\frac{d\sigma}{dt}}_{\*J,\^\tau}$ from trajectory and parameter sampling. We used the values at the last time step ($u=t$), where we confirmed that the statistics approached their steady-state values. 

\begin{figure*}[ht]
\begin{center}
\includegraphics[width=18cm]{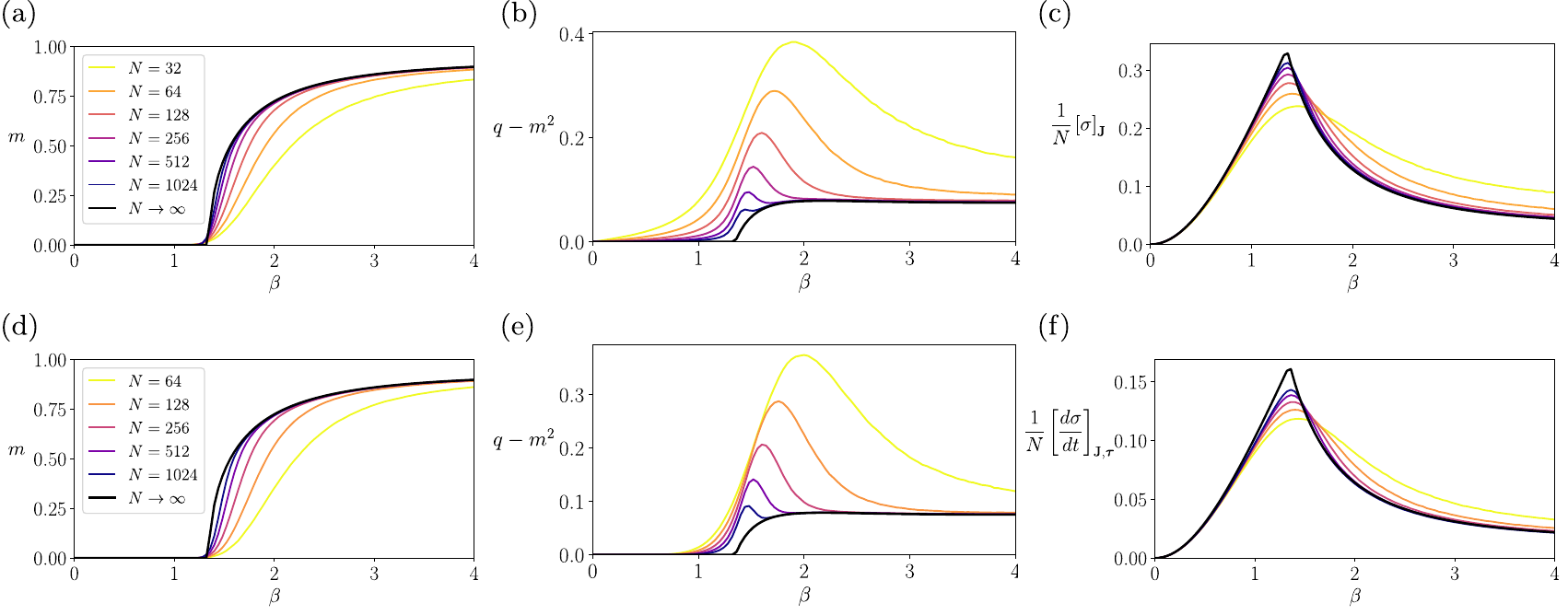}   
\end{center}
\caption{\textbf{Verification of the exact mean-field solutions by simulating the kinetic Ising systems with synchronous and asynchronous updates}. We repeated $400,000$ simulations of systems under synchronous (top) and asynchronous (bottom) updates with $\Theta_{i,u}=0$ and $\Delta J =0.5$:
(a,d) Sample estimates of the mean activation rate $\frac{1}{N}\sum_i \br{\ang{s_{i,u}}}_{\*J,\^\tau}$ compared with the theoretical order parameter $m$.   
(b,e) Sample estimates of the average delayed self-covariances $\frac{1}{N}\sum_i\br{\ang{s_{i,u}s_{i,u-d}}}_{\*J,\^\tau} - \frac{1}{N}\sum_i\br{\ang{s_{i,u}}}_{\*J,\^\tau}\frac{1}{N}\sum_i\br{\ang{s_{i,u-d}}}_{\*J,\^\tau}$  ($d=1$ for the synchronous system and $d=10N$ for the asynchronous one) computed from samples compared with the theoretical order parameter $q-m^2$. 
(c,f) Sample estimates of the entropy production and entropy production rate (Eqs.~S6.6 and S6.7
) compared with its mean-field value at the thermodynamic limit $\frac{1}{N}\br{\sigma}_{\*J},\frac{1}{N}\br{\frac{d\sigma}{dt}}_{\*J,\^\tau}$ (Eq.~\ref{eq:sigma_mean-field},\ref{eq:sigma-rate_mean-field}). 
} 
\label{fig:simulation}
\end{figure*}

Fig.~\ref{fig:simulation} compares the theoretical order parameter $m$ and $q$ with the mean activation rate and delayed self-correlations computed from the simulated trajectories for system size $N=32, \dots, 1024$ under synchronous (Fig.~\ref{fig:simulation}(a, b)) and asynchronous (Fig.~\ref{fig:simulation}(d, e)) updates. The simulated values approach the theoretical prediction as the size increases, albeit the convergence speed slows down near the critical temperature as it is expected. Similarly, we confirm in Fig.~\ref{fig:simulation}(c,f) that the entropy production from the sample trajectories for synchronous and asynchronous systems converges to the mean-field value at the thermodynamic limit as we increase the system size. Note that entropy production for synchronous updates differs from the entropy rate in the asynchronous update in continuous-time limit due to different values for  the delayed correlation term $q$ in Eqs.~\ref{eq:sigma_mean-field} and \ref{eq:sigma-rate_mean-field}. These results corroborate our theoretical predictions that the steady-state entropy production peaks at the critical nonequilibrium phase transitions. We further verified by simulations with $\Delta J=1$ that the steady-state entropy production increases when the significantly heterogeneous systems approach the quasi-deterministic regime (Fig.~S4).

\begin{figure}[b]
\begin{center}
\includegraphics[width=8.4cm]{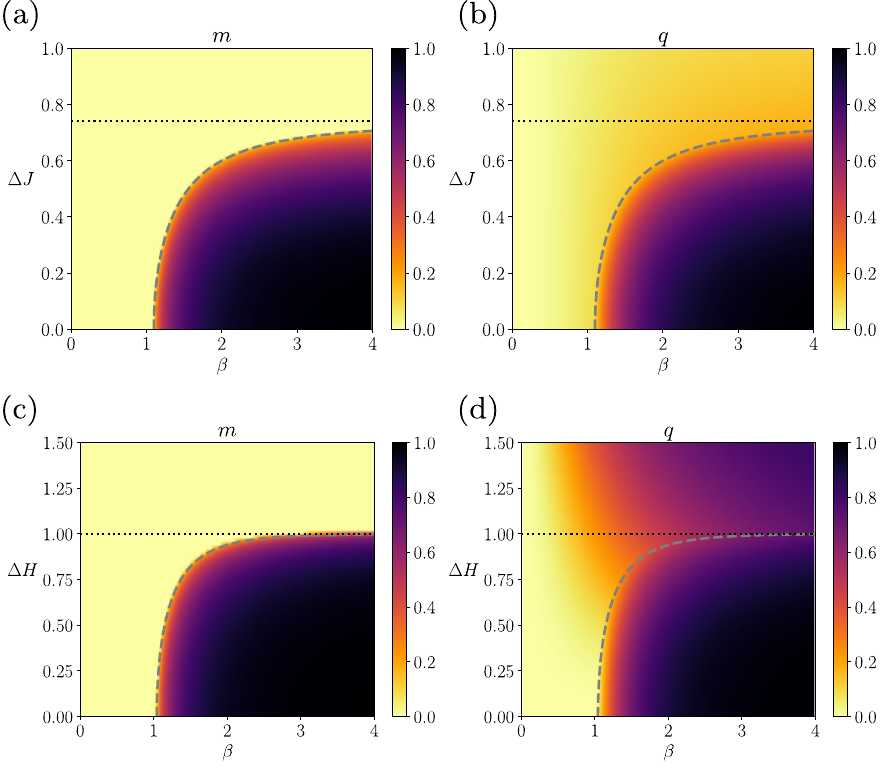}   
\end{center}
\caption{\textbf{Order parameters of the asymmetric SK model with heterogeneous fields}. (a, b) The average magnetization $m$ and average delayed self-coupling $q$ are shown as a function of $\Delta J$ and $\beta$. Fixed parameters are $J_0=1$, $\Delta H=0.5$. The dashed line represents the critical line separating ordered and disordered phases. The horizontal dotted line represents the critical $\Delta J$ at zero temperature ($\beta\to\infty$). (c, d) The phase diagram of order parameters as a function of $\Delta H$ and $\beta$. Fixed parameters are $J_0=1$ and $\Delta J=0.2$ and variable $\Delta H$. The dashed line is a critical $\Delta H$ at zero temperature.
} 
\label{fig:order-parameters-DeltaH}
\end{figure}


Finally, knowing the order parameters of the system under the configurational average, we can investigate the structure of the patterns emerging from the dynamics of a sufficiently large but finite system under certain conditions. We calculate the probability $\Omega^{(n)}$ of observing a state $\*s_{u+n}$ again for the first time after $n$ steps, starting from the same pattern $\*s_u=\*s_{u+n}$ (Supplementary Note~7).
For zero temperature ($\beta\to\infty$) and synchronous updates, $\Omega^{(n)}$ describes the probability of observing a periodic pattern of length $n$ since transitions become deterministic and can result in periodic patterns. 
In general, the distribution of these patterns depends on the higher-order correlations between spins across time steps. However, we observe that for the disordered phase ($m=0$) as well as in the \emph{deep} ordered phase ($m \sim 1$), these correlations disappear for the configurational average in the asynchronous model or in the synchronous model for large $n$ (see Fig.~S5).
In these regions of the phase diagram, we can approximate the probability of observing a pattern $n$ as 
$\br{\Omega^{(n)}}  \approx \exp\br{(1-n)\lambda}$ with $\lambda=\pr{\frac{1+m}{2}}^{N}$ (see Supplementary Note~7).
The expected length until a repeated pattern is observed is then
\begin{align}
   \sum_n n \br{\Omega^{(n)}} 
   \approx \frac{1}{\lambda} = \pr{\frac{2}{1+m}}^{N}.
    \label{eq:average-length}
\end{align}

At the disordered phase ($m=0$), the average length of observed patterns exhibits a maximum value, growing exponentially at a rate of $2^N$. In contrast, when the system enters the ordered phase ($m \sim 1)$, the growth rate decreases as $m$ increases. In the limit $m=1$, the system reaches a static equilibrium of average length $1$, where the same pattern is repeated indefinitely. These results are consistent with expected dynamics under order-disorder phase transitions. Thus, bringing a quasi-deterministic system to a more stochastic regime by decreasing $\beta$ to the critical value $\beta_c$ (with $\Delta J$ smaller than $\Delta J^{c}(\infty)$) increases the diversity of irreversible patterns and hence entropy production. However, further reduction of $\beta$ makes the system more random (i.e., less irreversible transitions), leading to a decrease in entropy production. In contrast, the large entropy production found at the disordered phase at large $\beta$ and $\Delta J$ (wider than $\Delta J^{c}(\infty)$) is caused by diverse oscillatory dynamics whose average pattern length is $2^N$ as in the random dynamics ($\beta=0$). Adding stochasticity to the dynamics by decreasing $\beta$ in this regime reduces the entropy production monotonically.

\subsection*{The SK model with uniformly distributed external fields}
Next, we apply non-zero external fields to the spins. For simplicity, we will consider the synchronous Ising system with unchanging fields $H_{i,u}=H_i$, assuming  a uniform distribution $H_{i} \sim \mathcal{U}\pr{-\Delta H, \Delta H}$. 
Fig.~\ref{fig:order-parameters-DeltaH}(a) and \ref{fig:order-parameters-DeltaH}(b) show the $\beta-\Delta J$ phase diagram for the order parameters. With this change, we observe non-zero correlation $q$ in the area where we previously saw the disordered states ($m=0$ and $q=0$, Fig.~\ref{fig:order-parameters-DeltaJ}(b)).
Fig.~\ref{fig:order-parameters-DeltaH}(c) and \ref{fig:order-parameters-DeltaH}(d) display the order parameters as a function of the inverse temperature and $\Delta H$, where we examine the effect of heterogeneity in the external fields while fixing the coupling variability, $\Delta J=0.2$. The critical line of $\Delta H^c(\beta)$ is obtained in this case as a solution of the following self-consistent equation (Eq.~S3.9):
\begin{align}
    \frac{\Delta H}{J_0} = \int \mathrm{D}z \, \tanh\br{\beta \pr{\Delta H  + \Delta J z }}.
\end{align}
Again, as studied in Supplementary Note~3, this critical phase transition corresponds to the mean-field universality class.
Since the right-hand side term is less than or equal to $1$ regardless of $\beta$ and $\Delta J$, the phase transition occurs only when $\Delta H<J_0$ is satisfied. Intuitively, there is a competition between the dispersion induced by the field diversities $\Delta H$ and  the cohesion induced by the mean coupling strength $J_0$. The ordered phase takes place only if $J_0$ counteracts the dispersion induced by the heterogeneity of external fields. More precisely, the critical $\Delta H^c(\beta)$ at the low temperature limit ($\beta \rightarrow \infty$) is obtained by solving
$\Delta H/J_0 = \int \mathrm{D}z \, \mathrm{sign}\br{\Delta H  + \Delta J z}$. Here we have $\Delta H^c(\infty)=1$. We observe the phase transition by varying $\beta$ if $\Delta H<\Delta H^c(\infty)$, and no phase transition if $\Delta H>\Delta H^c(\infty)$. Note that $q$ increases monotonically with $\beta$ even for $\Delta H>\Delta H^c(\infty)$ when the distributed fields are introduced.

\begin{figure*}
\begin{center}
\includegraphics[width=15.cm]{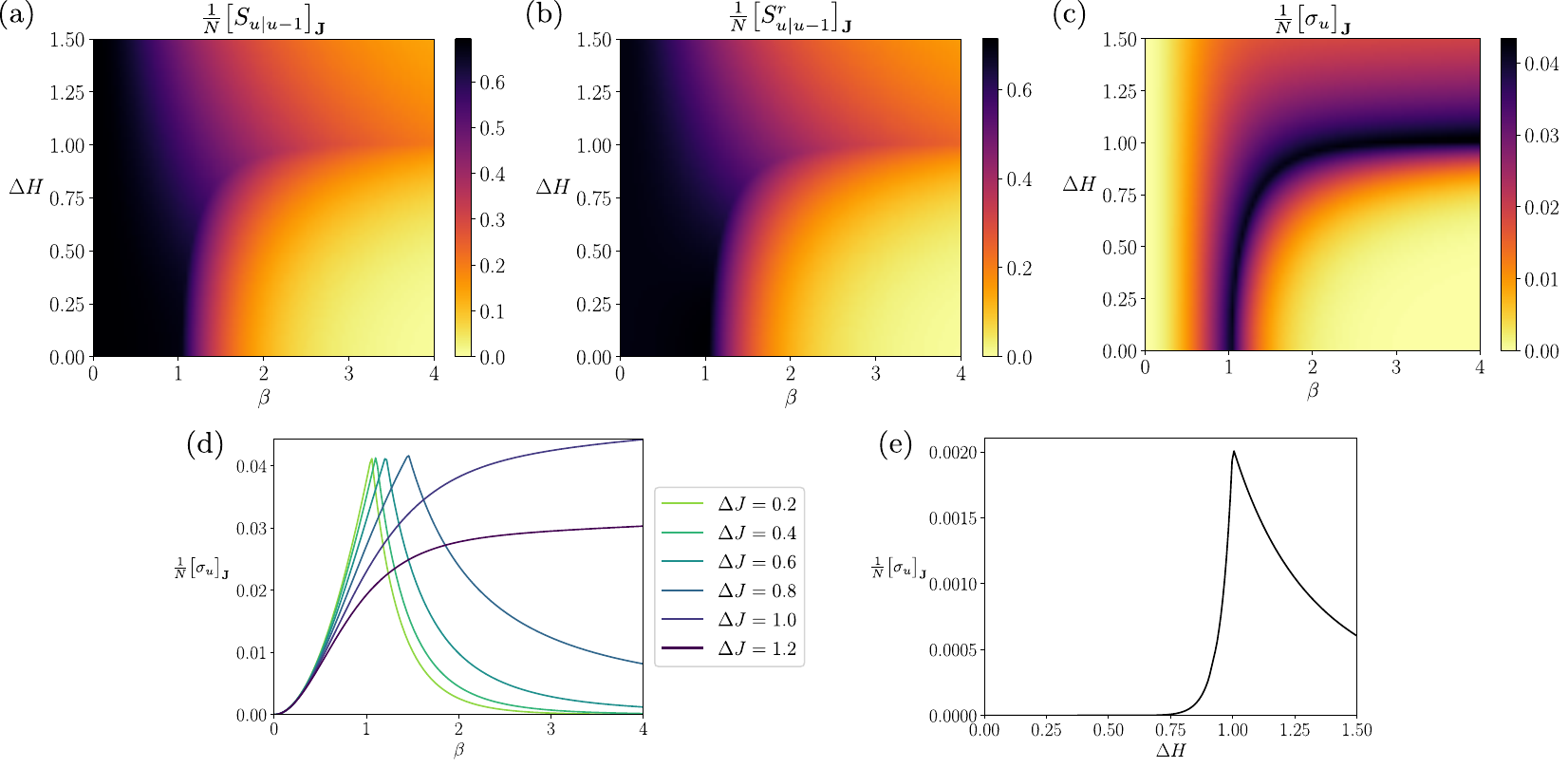}
\end{center}
\caption{\textbf{Conditional entropies and entropy production of the asymmetric SK model with heterogeneous fields under synchronous updates}. (a) The normalized conditional entropy $S_{u|u-1}$ (equivalent to the entropy rate under a steady state). (b) the normalized conditional entropy of the reverse dynamics $S_{u|u-1}^r$. (c) The normalized entropy production at a steady state. (d)  Horizontal sections of the entropy production ($\Delta H=0.2, 0.4, 0.6, 0.8, 1.0$, and $1.2$). It peaks at the critical line. (e) A vertical section of the entropy production at zero temperature ($\beta\to\infty$). All plots are based on a model with fixed parameters $J_0=1$, $\Delta J=0.2$ and variable  $\Delta H$ and $\beta$.
} 
\label{fig:entropy-DeltaH}
\end{figure*}

We now examine the conditional entropy, its reverse, and entropy production for the synchronous system  with distributed fields, using the $\beta$ vs. $\Delta H$ phase diagram. Similarly to the observation in the model without fields, the conditional entropy decreases with higher $\beta$ (it becomes more deterministic processes, see Fig.~\ref{fig:entropy-DeltaH}(a)). The time-reversed conditional entropy also decreases with increasing $\beta$ for all $\Delta H$, indicating that the reverse process is more and more likely to happen regardless of $\Delta H$ (Fig.~\ref{fig:entropy-DeltaH}(b)). As seen previously, the time-reversed conditional entropy diminishes under the ferromagnetic-like states ($\Delta H<\Delta H^c(\infty)$). In contrast, we also observe the reduction of the time-reversed conditional entropy at higher $\beta$ for $\Delta H>\Delta H^c(\infty)$. Note that we observed increased correlations $q$ at higher $\beta$ for $\Delta H>\Delta H^c(\infty)$ when we introduced the non-zero external fields (see Fig.~\ref{fig:order-parameters-DeltaH}(d)), which resulted in the reduction of the reversed entropy similarly to ferromagnetic-like states. Both conditional and time-reversed conditional entropies decrease much slower along the critical line than in other regions, although with different magnitudes. As a result, we see the maximization of the entropy production around critical points more clearly than the $\beta$--$\Delta J$ phase diagram (Fig.~\ref{fig:entropy-DeltaH}(c) and ~\ref{fig:entropy-DeltaH}(d)). 
Finally, at the zero temperature limit ($\beta \rightarrow \infty$), the entropy production peaks at $\Delta H^c(\infty)=1$ (Fig.~\ref{fig:entropy-DeltaH}(e)).

\section*{Discussion}
\label{sec:discussion}

In this paper, we studied in detail the nonequilibrium thermodynamics of the kinetic, asymmetric SK model for both synchronous and asynchronous dynamics. As expected, the order parameters reveal that the model exhibits order-disorder nonequilibrium phase transitions analogous to the paramagnetic-ferromagnetic phase transitions in the equilibrium Ising model. There are, however, no phase transition akin to a spin-glass (which does not emerge due to coupling asymmetry, as previously reported for continuous-time asymmetric SK models in \cite{crisanti1988dynamics}). In addition, we show that the steady-state entropy production is maximized near nonequilibrium phase transition points, being its first derivative discontinuous at these points (Fig.~\ref{fig:entropy-DeltaJ}(d)). 
This result is similar to previously reported critical behavior of the entropy production caused by external stimulation or inertial dynamics (via self-coupling) in homogeneous systems with asynchronous updates by means of naive mean-field approximations or numerical simulations \cite{zhangCriticalBehaviorEntropy2016a,noaEntropyProductionTool2019a,crochik2005entropy}. Nevertheless, our result is novel in that it provides the critical behavior of the entropy production caused by asymmetric, heterogeneous couplings using an exact analytical solution for such complex systems. In addition, the studied model displays a region with disordered oscillations in its phase diagram, where the entropy production takes even larger values than in the critical regime. This phase takes place for disordered systems with low entropy rates, i.e., the heterogeneous connections are strong enough to make the dynamics disordered but quasi-deterministic (Fig.~\ref{fig:entropy-DeltaJ}(c), top-right). In contrast, the entropy production does not increase when we increase the heterogeneity of external fields (Fig.~\ref{fig:entropy-DeltaH}(c)).

Taken together, our results indicate that the behavior of entropy production peaking at a critical point is more general than the simple mean-field, homogeneous models, therefore a non-smooth change of the steady-state entropy production (or entropy dissipated to an external reservoir) can be a useful indicator of a number of nonequilibrium phase transitions. At the same time, our results demonstrate that an increase in entropy production does not necessarily mean that the system is approaching a phase transition point. Instead, combining the order parameters, entropy rate, and entropy production yields a more precise picture of the complex systems and their phase transitions.

Typically, solutions of the symmetric (equilibrium) SK model involve the replica trick to calculate the configurational average of the logarithm of the partition function \cite{nishimori_statistical_2001}. This method introduces an integer number of replicas of a system for averaging the disorder and then recovers the solution using a continuous number of replicas in the zero limit under the replica symmetry or replica-symmetry breaking ansatz. This treatment forces researchers to check the validity of solutions before reaching correct solutions \cite{parisi1979toward,parisi1980sequence}. As an alternative to the replica methods, the path integral methods have been widely used in analyzing the symmetric SK model \cite{sompolinsky1981dynamic,sompolinsky1982relaxational}. However, for partially- or fully-symmetric SK models, the path integral method does not give a definite analytical solution but needs to be computed with Monte Carlo approaches \cite{eissfeller1994mean}. Fortunately, the path integral method derives an exact analytical solution for the case of the fully asymmetric nonequilibrium SK model \cite{crisanti1988dynamics}, which we extended to cover synchronous and asynchronous updates, and theoretically underpinned their nonequilibrium properties by deriving the exact solution of the steady-state entropy production and entropy rates of the system.

Nonequilibrium properties of biological and adaptive systems have received the attention of neuroscience and biological science communities. For example, increased entropy production in macroscopic neural activity was suggested as a signature of physically and cognitively demanding tasks \cite{lynn2021broken}, conscious activity \cite{perl2021nonequilibrium,delafuente2022temporal} or neuropsychiatric diseases like schizophrenia, bipolar disorders, and ADHD \cite{deco2021deep}. While it is not easy to contrast their findings based on the coarse-grained analysis of ECoG or fMRI data with the present results, our precise characterization of the entropy production of the prototypical system sheds light on what kind of behaviors we might expect from these complicated systems. Most importantly, our results indicate two scenarios to increase entropy production by controlling the connection heterogeneity ($\Delta J$) and neuron's nonlinearity ($\beta$). These global changes in the model parameters can be achieved in the brain as gain modulation often mediated by neuromodulators \cite{eldar2013effects}. One scenario to increase entropy production is that the system approaches a critical state as seen in the low $\Delta J$ in Fig.~\ref{fig:order-parameters-DeltaJ} or Fig.~\ref{fig:order-parameters-DeltaH}. The other scenario is to make the system more heterogeneous and sensitive by increasing $\Delta J$ and $\beta$. A significant difference is that the former process maintains stochastic nature while the latter yields quasi-deterministic disorder, as indicated by the high or reduced entropy rate. Therefore, the results suggest that it is crucial to investigate the multiple possibilities of nonequilibrium states to underpin the unconscious (sleep or anesthesia), awake, and engaged states more precisely.

Neuroscientists have often discussed the role of temporal patterns in spiking activities of neurons in computation or in memory consolidation and retrieval. One central topic of the debate is whether neurons, e.g., cortical or hippocampal ones, exhibit precise sequential patterns in a repeated manner \cite{abeles1991corticonics,diesmann1999stable,ikegaya2004synfire,izhikevich2006polychronization}. Such precise sequences should result in a large entropy production similar to the low-temperature disordered phase of the kinetic Ising system. Alternatively, one may explain a broad range of irreversible temporal patterns, including avalanche dynamics \cite{beggs2003neuronal,friedman2012universal}, by the dynamics near the nonequilibrium phase transitions without the precise sequential structure. As suggested by Eq.~\ref{eq:average-length}, the same network that shows simple periodic patterns at zero temperature can retrieve the diverse patterns yielding large entropy production by being poised near the critical phase transition point. The current study highlights the need to dissociate the two scenarios, characterizing different temporally irreversible spiking patterns to understand the distinct roles in neural computation using multiple thermodynamic quantities.

Finally, our analytical solutions offer a benchmark for --  the aforementioned and other -- methods for estimating thermodynamic quantities. For example, characterizing entropy production from brain imaging data requires methods for coarse-graining the phase diagram \cite{lynn2021broken, perl2021nonequilibrium}. The kinetic SK model can serve as a test bench for such methods as it is an analytically tractable system with a well-known phase diagram. Moreover, we can use them to examine both established and novel mean-field theories in estimating the thermodynamic properties of large-scale systems. For example, one can directly fit the Ising model to neuronal spiking data using mean-field methods for finite-size networks, from which one can estimate various thermodynamic quantities of the system \cite{aguilera2021unifying,donner2017approximate}. Accurately estimating these quantities in large networks gives deeper insights into the nonlinear computations of cortical circuitries. The exact solutions provided here can serve to evaluate the accuracy of these approximation methods applied to large-scale networks and provide a benchmark of the thermodynamics quantities of infinitely large networks.

\section*{Data and code availability}

The datasets and code generated in the current study are available in the GitHub repository, \url{https://github.com/MiguelAguilera/asymmetric-SK-model}.


\bibliographystyle{naturemag}
\bibliography{references}

\begin{acknowledgments}
MA was partially funded by the European Union’s Horizon 2020 research and innovation programme under the Marie Skłodowska-Curie grant agreement No 892715 and a Junior Leader fellowship from ``la Caixa'' Foundation (ID 100010434,  code LCF/BQ/PI23/11970024).
HS is supported by JSPS KAKENHI Grant Number JP 20K11709, 21H05246, the National Institutes of Natural Sciences (NINS Program No. 01112005, 01112102), and New Energy and Industrial Technology Development Organization (NEDO), Japan.
\end{acknowledgments}

\section*{Contributions}
M.A. and H.S. designed and reviewed the research; M.A. contributed the analytical results. M.A. and M.I. contributed the numerical results; M.A., M.I. and H.S. contributed the theoretical review and wrote the paper.

\clearpage

\appendix
\onecolumngrid


\begin{center}

{\large Nonequilibrium thermodynamics of the asymmetric Sherrington-Kirkpatrick model }
\vspace{1em} \\
{\large \textbf{Supplementary Information}}
\vspace{1em} \\
Miguel Aguilera \\
\textit{BCAM – Basque Center for Applied Mathematics, 
        Bilbao, Spain \\ 
        IKERBASQUE, Basque Foundation for Science, Bilbao, Spain.\\ 
        School of Engineering and Informatics, 
  University of Sussex, \\
  Falmer, Brighton. United Kingdom}
\vspace{1em} \\
Masanao Igarashi \\
    \textit{Graduate School of Engineering, 
    Hokkaido University,  Sapporo, Japan}
\vspace{1em} \\
Hideaki Shimazaki \\
    \textit{Graduate School of Informatics, Kyoto University, Kyoto, Japan \\
    Center for Human Nature, Artificial Intelligence, and Neuroscience (CHAIN),  \\
Hokkaido University,
        Sapporo, Japan}
\end{center}

\renewcommand\appendixname{Supplementary Note}
\setcounter{figure}{0}
\renewcommand{\thefigure}{S\arabic{figure}}
\renewcommand{\thesection}{\arabic{section}}
\renewcommand{\thesubsection}{S\arabic{section}.\arabic{subsection}}
\renewcommand{\theequation}{S\thesection.\arabic{equation}}
\setcounter{page}{1}.


\section{A path integral approach to the asymmetric SK model}
\label{app:path-integral}

In this Supplementary Note, we compute the generating functional of the asymmetric kinetic Ising model averaged over the quenched couplings with Gaussian distributions, known as the configurational average.

The probability density of a specific trajectory of the kinetic Ising model, $\*s_{0:t}=\{\*s_{0},\*s_{1},\ldots,\*s_{t}\}$, is defined as
\begin{align}
    p(\*s_{0:t}) =& \prod_{u=1}^{t} p(\*s_u\,|\,\*s_{u-1}) p(\*s_0)
    \nonumber\\ =& \exp \Bigg[ \beta\sum_{i,u}  s_{i,u} h_{i,u}
     - \sum_{i,u} \log\br{ 2 \cosh \br{ \beta  h_{i,u}}} \Bigg] p(\*s_0),
\end{align}
where
\begin{align}
h_{i,u}=& H_{i,u}+\sum_j J_{ij} s_{j,u-1}.
\end{align}
Here the summation over $u$ is taken for $u=1,\ldots,t$. For simplicity, we will assume that $p(\*s_0)$ only contains one possible value, that is, the initial distribution is a Kronecker delta $p_0(\*s)=\prod_i \delta\br{s_i,s_{i,0}}$. Although this enables us to ignore the term, the next steps are generalizable to any initial distributions.

The dynamics above  straightforwardly describes a synchronous kinetic Ising model under time-dependent fields $H_{i,u}$ and asymmetric couplings $J_{ij}$. In Supplementary Note \ref{app:async_model_solution}, we will show how this model and its solution can be generalized to cover both synchronous and asynchronous dynamics.

In equilibrium systems, the partition function provides the statistical moments of the equilibrium distribution. Here, to find the statistical properties expected from ensemble trajectories of the asymmetric SK model as well as its steady-state entropy production  (Eq.~\ref{eq:entropy_productionII}), we introduce the following generating functional or dynamical partition function:
\begin{align}
    Z_t(\*g) =& \sum_{\*s_{0:t}} p(\*s_{0:t}) \exp \Bigg[ \sum_{i,u} g_{i,u} s_{i,u} 
    \nonumber\\  & + \sum_{i,u}  g_{u}^{S}   \pr{ \beta s_{i,u} h_{i,u}  - \log\br{2 \cosh\br{\beta h_{i,u}}} } +  \sum_{i,u}  g_{u}^{S^r}   \pr{ 
\beta  s_{i,u-1}  h_{i,u}^r- \log\br{ 2 \cosh\br{\beta h_{i,u}^r}}  } \Bigg]
    \nonumber\\ =&
    \sum_{\*s_{0:t}} \exp \Bigg[ \sum_{i,u} s_{i,u} \beta h_{i,u} - \log\br{2 \cosh\br{\beta h_{i,u}}} + \sum_{i,u} g_{i,u} s_{i,u} 
    \nonumber\\  & + \sum_{i,u}  g_{u}^{S}   \pr{  s_{i,u} \beta h_{i,u}  - \log\br{2 \cosh\br{\beta h_{i,u}}}} +  \sum_{i,u}  g_{u}^{S^r}   \pr{ 
 s_{i,u-1} \beta h_{i,u}^r- \log\br{2 \cosh\br{\beta h_{i,u}^r} } } \Bigg],
    \label{eq:SK-partition-function-II}
\end{align}
where $h_{i,u}^r= H_{i,u}+\sum_j J_{ij} s_{j,u} = h_{i,u+1} + H_{i,u} - H_{i,u+1}$. Note that $h_{i,u}$ have to be defined up to $t+1$ to recover the backwards trajectory. The terms $g_{i,u}$ are designed to obtain the moments and other statistics of the system, and $g_{u}^{S}$ and $g_{u}^{S^r}$ are for its conditional and reversed conditional entropy terms at time $u$.

We will use this generating functional to calculate the statistical moments of various random variables. We will denote them using an average function described as:
\begin{align}
    \ang{f(\cdot)}_{\*g} =& \sum_{\*s_{0:t}} p(\*s_{0:t})  f(\cdot) \exp \Bigg[ \sum_{i,u} g_{i,u} s_{i,u} 
    \nonumber\\  & + \sum_{i,u}  g_{u}^{S}   \pr{ \beta s_{i,u} h_{i,u}  - \log\br{2 \cosh\br{\beta h_{i,u}}} } +  \sum_{i,u}  g_{u}^{S^r}   \pr{ 
\beta  s_{i,u-1}  h_{i,u}^r- \log\br{ 2 \cosh\br{\beta h_{i,u}^r}}  } \Bigg].
\end{align}
For simplicity, we will denote  $\ang{f(\cdot)}_{\*0} $ simply as $\ang{f(\cdot)}$, recovering the statistical moments of the original system.

The configurational average over Gaussian couplings (Eq.~\ref{eq:Gaussian-coulings}) of the generating functional is computed as
\begin{equation}
    \br{Z_t(\*g)}_{\*J} = \int \prod_{i,j} dJ_{ij} p(J_{ij}) Z_t(\*g).
\end{equation}
The configurational average can be solved using a path integral method. To obtain the path integral form, we first insert an appropriate delta integral for the effective fields of each unit for the time steps $u=1,\ldots,t+1$ to the above equation:
\begin{align}
	1 =& 
	\int  d\^\theta \prod_{i,u} \delta\br{\theta_{i,u} - \beta h_{i,u}} \nonumber \\
	=& \frac{1}{\pr{2\pi}^{N(t+1)}}\int  d\^\theta d\^{\hat\theta} \exp\br{ \sum_{i,u} \iu \hat\theta_{i,u} (\theta_{i,u} - \beta H_{i,u} - \beta \sum_j J_{ij} s_{j.u-1}) },
\end{align}
where $\^\theta$ is the $N(t+1)$-dimensional vector composed of the effective fields $\theta_{i,u}$ ($i=1,\ldots,N$ and $u=1,\ldots, t+1$). $\^{\hat\theta}$ is the $N(t+1)$-dimensional conjugate effective field, and we used a Dirac delta function $\delta\br{x-a}=\frac{1}{2\pi} \int_{-\infty}^\infty e^{\iu \zeta(x-a)}\,d\zeta$.
Note that, from now on, all summations and products involving the conjugate effective field $\hat{\^\theta}$ (as well as the order parameters we introduce later) will be performed over the range $u=1,\dots,t+1$.
Next, we replace $\beta h_{i,u}$ in Eq.~\ref{eq:SK-partition-function-II} with the auxiliary variable $\theta_{i,u}$ as well as $\beta h^{r}_{i,u}$ of the reversed couplings at time $u$ with an auxiliary variable $\vartheta_{i,u} = \theta_{i,u+1} + \beta(H_{i,u}-H_{i,u+1})$, and place them inside the integral with respect to $\^\theta$ (i.e., we perform the operation, $f(a)=\int f(x)\delta\br{x-a} dx$). The configurational average is written as
\begin{align}
    \br{Z_t(\*g)}_{\*J} =& \frac{1}{\pr{2\pi}^{N(t+1)}}\int d\^\theta d\^{\hat\theta} 
    \prod_{i,j} dJ_{ij} p(J_{ij}) \nonumber \\ 
    & \cdot \sum_{\*s_{1:t}} \exp \Bigg[ \sum_{i,u}  s_{i,u} (g_{i,u}+ \theta_{i,u})   - \log\br{ 2 \cosh{\theta_{i,u}}} 
    + \sum_{i,u}  g_{u}^{S}   (s_{i,u} \theta_{i,u}  - \sum_{i,u}  \log\br{2 \cosh \theta_{i,u}})
  \nonumber \\ &  +  \sum_{i,u}  g_{u}^{S^r}   \pr{  s_{i,u-1}  \vartheta_{i,u} - \log\br{2 \cosh{\vartheta_{i,u}}}} +\sum_{i,u} \iu  \hat\theta_{i,u} (\theta_{i,u} - \beta H_{i,u} - \beta \sum_j J_{ij} s_{j,u-1}) 
  \Bigg].
  \label{eq:configurational_average}
\end{align}

Using the Gaussian integral formula $
    \int dx \frac{1}{\sqrt{2 \pi b}}^r \exp\br{ax-\frac{\left(x-c\right)^2}{2b}} 
    =\exp\br{ac+\frac{a^2}{2} b }$,
the expectation of $\exp\br{a J_{ij}}$ is computed as
\begin{align}
    \int dJ_{ij} p(J_{ij}) \exp\br{a J_{ij}}
    =\exp\br{aJ_0/N+\frac{a^2}{2} \Delta J^2/N }.
\end{align}
Hence the integral related to $J_{ij}$ in Eq.~\ref{eq:configurational_average} is computed as
\begin{align}
    &\int \Bigg[ \prod_{i,j} dJ_{ij} p(J_{ij}) \Bigg]
    \exp \Bigg[ - \sum_{i,u} \iu  \hat\theta_{i,u} \beta \sum_j J_{ij} s_{j,u-1} 
    \Bigg] 
    \nonumber\\
    =& \prod_{i,j} \Bigg[ \int dJ_{ij} p(J_{ij})
     \exp \Bigg[ - \beta \pr{\sum_{u}  \iu \hat\theta_{i,u}   s_{j,u-1} } J_{ij}
     \Bigg] \nonumber\\
    =& \prod_{i,j} \exp \Bigg[ -  \beta \Bigg( \sum_{u}  \iu \hat\theta_{i,u}   s_{j,u-1}  \Bigg) \frac{J_0}{N} 
    +  \beta^2 \Bigg( \sum_{u}  \iu \hat\theta_{i,u}   s_{j,u-1}\Bigg)^2 \frac{\Delta J^2}{2N}  \Bigg]
    \nonumber\\=& \prod_{i,j} \exp \Bigg[
    - \frac{\beta J_0}{N} \sum_{u} \iu  \hat\theta_{i,u} s_{j,u-1} 
    + \frac{\beta^2 \Delta J^2}{2N}  \sum_{u,v} \iu\hat\theta_{i,u} s_{j,u-1} \iu\hat\theta_{i,v} s_{j,v-1}  \Bigg].
    \label{ap:integral_configurational_average_SK}
\end{align}
Using this result, the Gaussian integral form of the partition function is given as
\begin{align}
    \br{Z_t(\*g)}_{\*J} =&  \frac{1}{\pr{2\pi}^{N(t+1)}} \int d\^\theta d\^{\hat\theta}   
    \sum_{\*s_{1:t}} \exp \Bigg[ \sum_{i,u}  s_{i,u} (g_{i,u}+ \theta_{i,u}) - \sum_{i,u} \log\br{ 2 \cosh{\theta_{i,u}}}  \nonumber\\
    & 
    + \sum_{i,u}  g_{u}^{S}   \pr{ s_{i,u} \theta_{i,u} - \log\br{2 \cosh{\theta_{i,u}}} } +  \sum_{i,u}  g_{u}^{S^r}   \pr{  s_{i,u-1}  \vartheta_{i,u} - \log\br{2 \cosh{\vartheta_{i,u}}}} \nonumber\\ 
    &
    + \sum_{i,u} \iu \hat\theta_{i,u} (\theta_{i,u} - \beta H_{i,u})
         -  \sum_{u} N \beta J_0 \Big( \frac{1}{N}\sum_i \iu \hat\theta_{i,u} \Big) \Big( \frac{1}{N} \sum_j s_{j,u-1} \Big) 
      \nonumber\\ 
    & + \frac{\beta^2 \Delta J^2 }{2N}  \sum_{i,u}\pr{ \iu \hat\theta_{i,u} }^2 
    + \sum_{u>v} N \beta^2 \Delta J^2 \Big( \frac{1}{N}\sum_i  \iu \hat\theta_{i,u} \iu \hat\theta_{i,v} \Big) \Big( \frac{1}{N}\sum_j s_{j,u-1} s_{j,v-1} \Big) 
    \Bigg].
    \label{eq:generating_functional_weight_integral}
\end{align}
Note that, for the term of summation over $u, v$, we separated the $u=v$ terms from the rest, resulting in elimination of the spin variables because  $s_{j,u-1}s_{j,u-1}=1$.

\subsection{Gaussian integral and saddle node approximation}

We will evaluate the aforementioned expression with a Gaussian integral and a saddle node approximation, and show that the saddle node solutions become order parameters. For this goal, we first give an outline of the derivation, and then apply the steps to the above equation. 

Let $C$ be a real value, and $x$ and $y$ be complex values. Eq.~\ref{eq:generating_functional_weight_integral} contains the term in the form of  $\exp\br{C xy}$.  We can represent this term by a double Gaussian integral (a pair of the Gaussian integral formulas) with the form:
\begin{align}
    \label{eq:double_gaussian_integral}
    \exp\br{C xy} =&  \exp\br{\frac{C}{2}\pr{\frac{1}{2}(x+y)^2 + \frac{1}{2}(\iu(x-y))^2} }
    \nonumber\\ =&  \frac{C}{4\pi} \int  dz_R dz_I \exp\br{\frac{C}{2}\pr{-\frac{1}{2}z_R^2 -\frac{1}{2}z_I^2 + (x+y)z_R + \iu(x-y)z_I}}
    \nonumber\\ =&  \frac{C}{4\pi} \int  dz_R dz_I \exp\br{\frac{C}{2}\pr{-\frac{1}{2}z_R^2  -\frac{1}{2}z_I^2  +x  (z_R + \iu z_I) + y  (z_R - \iu z_I)}}.
\end{align}
Because the integrand is an analytic function, we can change the contour of the path integral in the complex space so that it includes the saddle-point solution. This contour integration produces the original value, $\exp\br{C xy}$. Therefore, $z_R$ and $z_I$ are no longer real values but can be complex values.

When $x$ and $y$ are random variables, we can approximate the expectation of $\exp\br{C xy}$ by the saddle node solutions when the constant $C$ is large:
\begin{align}
  \int p(x,y) \exp\br{C xy} dx dy 
   \approx&
    \exp\br{\frac{C}{2} \brc{-\frac{1}{2} {z_R^*}^2  -\frac{1}{2}{z_I^*}^2  + \log \int p(x,y) \exp\br{x  (z_R^* + \iu z_I^*) + y  (z_R^* - \iu z_I^*)} dx dy}},
    \label{eq:double_gaussian_integral_probabilistic}
\end{align}
where $z_R^*$ and $z_I^*$ are the saddle-point solutions that extremize the contents of the braces $\brc{}$ in Eq.~\ref{eq:double_gaussian_integral_probabilistic}. These solutions are given by the following self-consistent equations:
\begin{align}
    z_R^* =& \ang{x+y}, \label{eq:zr*}
    \\ z_I^* =& \iu \ang{x-y}\label{eq:zi*},
\end{align}
where the bracket $\ang{\cdot}$ represents 
\begin{align}
    \ang{f(x,y)} = \frac{\int p(x,y) \exp\br{x  (z_R^* + \iu z_I^*) + y  (z_R^* - \iu z_I^*)} f(x,y) \,dxdy}
    {\int p(x,y) \exp\br{x  (z_R^* + \iu z_I^*) + y  (z_R^* - \iu z_I^*)} \,dxdy}
\end{align}
We reiterate that for the saddle-point solution $z_I^*$ is derived from substituting the exponent by its Taylor expansion around the minimum, i.e., $f(z_I) = f(z_I^*) + \frac{1}{2}f''(z_I^*)(z_I - \iu(x-y))^2  + \mathcal{O}((z_I - \iu(x-y))^3)$, which in this case has an imaginary value. 

To obtain a more intuitive saddle-point solution, we can perform a change of variables
\begin{align}
    z_1^* =& \frac{1}{2}( z_R^* + \iu z_I^*),& z_2^* =&  \frac{1}{2}( z_R^* - \iu z_I^*)
    \label{eq:variable-change1}
    \\ z_R^* =& z_1^* + z_2^*, &z_I^* =& \iu(z_1^* - z_2^*),
    \label{eq:variable-change2}
\end{align}
resulting in
\begin{align}
  \int p(x,y) \exp\br{C xy} dx dy \approx
    \exp\br{\frac{C}{2} \brc{- z_1^* z_2^* + \log \int p(x,y) \exp\br{x  z_1^* + y  z_2^*}dx dy}},
\end{align}
and
\begin{align}
    z_1^* =& \ang{y} \label{eq:z1*},
    \\ z_2^* =& \ang{x}\label{eq:z2*}
\end{align}

In summary, the process previously described consists of 1) introducing a pair of Gaussian integrals, 2) finding a saddle-point solution, and 3) performing a change of variable to recover a solution in terms of expectations of the original variables. We now repeat the process for the integral of the partition function. 

\textbf{(i) Gaussian integrals.}
\hspace{1em} 
First, we introduce Gaussian integrals by applying  Eq.~\ref{eq:double_gaussian_integral} to the quadratic terms in the partition function. Using $C = N \beta J_0$, $x_{u-1} = \frac{1}{N} \sum_j s_{j,u-1}$ and $y_u= -\frac{1}{N}\sum_{i} \iu \hat{\theta}_{i,u} $, we obtain 
\begin{align}
    &\exp\br{\sum_{u} (- N \beta J_0) \Big( \frac{1}{N}\sum_i \iu \hat\theta_{i,u} \Big) \Big( \frac{1}{N} \sum_j s_{j,u-1}  \Big)} \nonumber\\
    =& \prod_u \exp\br{C x_{u-1} y_{u}} \nonumber\\
    =& \pr{ \frac{C}{4\pi}}^t \int  \prod_u dM^+_u dM^-_u \exp\br{ \frac{C}{2}\pr{-\frac{1}{2}(M^+_u)^2  -\frac{1}{2}(M^-_u)^2  +x_{u-1}  (M^+_u + \iu M^-_u) + y_u  (M^+_u - \iu M^-_u)}},
\end{align}
where $M^+_{u}$ and $M^-_{u}$ are real-valued integral variables. Similarly, using  $C = \frac{1}{2} N \beta^2 \Delta J^2$, $x_{u-1,v-1} = \frac{1}{N} \sum_j s_{j,u-1} s_{j,v-1}$, $y_{u,v} = \frac{1}{N}\sum_{i}\hat{\theta}_{i,u} \hat{\theta}_{i,v}$, we have
\begin{align}
    &\exp\br{\sum_{u,v} N \frac{\beta^2 \Delta J^2}{2} \Big( \frac{1}{N}\sum_i  \iu \hat\theta_{i,u} \iu \hat\theta_{i,v} \Big) \Big( \frac{1}{N}\sum_j s_{j,u-1} s_{j,v-1} \Big)} \nonumber\\
    =&\prod_{u>v} \exp\br{C x_{u-1,v-1} y_{u,v}} \nonumber\\
    =&   \pr{ \frac{C}{4\pi}}^{t(t-1)/2} \int  \prod_{u>v} dQ^+_{u,v} dQ^-_{u,v} \exp\br{ \frac{C}{2}\pr{-\frac{1}{2}(Q^+_{u,v})^2  -\frac{1}{2}(Q^-_{u,v})^2  +x_{u-1,v-1}  (Q^+_{u,v} + \iu Q^-_{u,v}) + y_{u,v}  (Q^+_{u,v} - \iu Q^-_{u,v})}},
\end{align}
where $Q^+_{u,v}$ and $Q^-_{u,v}$ are real values. Note that the products over $u$ and $v$ are performed over the range $1,\dots,t+1$. 

With these double Gaussian integrals and defining $d\*M=\prod_u dM^+_u dM^-_u$ and $d\*Q=\prod_{u,v} dQ^+_{u,v} dQ^-_{u,v}$ we can rewrite the partition function as
\begin{align}
    \br{Z_t(\*g)}_{\*J} =&  \frac{(N\beta J_0)^t (N\beta^2 \Delta J^2)^{t(t-1)/2}}{ (4\pi)^{t(t+1)/2}} \int  d\*M  d\*Q
    \exp  \Bigg[
    - N \beta J_0\sum_{u}  \frac{(M_u^+)^2+(M_u^-)^2}{4} 
    \nonumber\\ & - N \beta^2 \Delta J^2  \sum_{u>v}  \frac{(Q_{u,v}^+)^2+(Q_{u,v}^-)^2}{4} 
    + \log \sum_{\*s_{1:t}}\int d\^\theta d\^{\hat \theta} \mathrm{e}^{ \Phi(\*s_{0:t}, \^\theta,\*g) + \Omega(\^{\hat \theta}, \^\theta,\*g)} \Bigg], 
    \label{eq:generating_functional_Gaussian_integral}
\end{align}
where the remaining terms related to the random variable $\*s$ and $\^{\theta}, \^{\hat \theta}$ from the Gaussian integral can be separated into the terms
\begin{align}
	\Phi(\*s_{0:t}, \^\theta,\*g) =&  \sum_{i,u}  \pr{g_{i,u} +  \theta_{i,u}} s_{i,u}
	- \sum_{i,u} \log\br{2 \cosh\br{\theta_{i,u}}}
	\nonumber\\& +\sum_{i,u}  g_{u}^{S}   \pr{ s_{i,u} \theta_{i,u} - \log\br{2 \cosh{\theta_{i,u}} }} +  \sum_{i,u}  g_{u}^{S^r}   \pr{  s_{i,u-1}  \vartheta_{i,u} - \log\br{ 2 \cosh{\vartheta_{i,u}}}}
	\nonumber\\& + \sum_{i,u} \beta J_0 \frac{M_u^+ + \iu M_u^-}{2}   s_{i,u-1}
	+ \sum_{i,u>v} \beta^2 \Delta J^2 \frac{Q_{u,v}^+ + \iu Q_{u,v}^-}{2} s_{i,u-1} s_{i,v-1},
\\ \Omega(\^{\hat \theta}, \^\theta,\*g) =&  \sum_{i,u}  (\theta_{i,u} - \beta H_{i,u} - \beta J_0 \frac{M_u^+ - \iu M_u^-}{2} ) \iu  \hat\theta_{i,u} 
\nonumber\\ & +  \frac{\beta^2 \Delta J^2 }{2}  \sum_{i,u}\pr{ \iu \hat\theta_{i,u} }^2 
	+ \beta^2 \Delta J^2  \sum_{i,u>v} \frac{Q_{u,v}^+ - \iu Q_{u,v}^-}{2} \iu \hat\theta_{i,u}  \iu \hat\theta_{i,v} 
	 - N(t+1) \log\br{2\pi}. 
	 \label{eq:gamma-Gaussian-integral}
\end{align}

Now, the next two steps for solving the integral is to find a saddle-point solution and  perform a change of variables. In the next section, we find that the solutions result in the order parameters of the system.

\textbf{(ii) saddle-point integral solution.} \hspace{1em} 
The exponent of the  integrand above is proportional to $N$, making it possible to evaluate the integral by steepest descent, giving the saddle-point solution as
\begin{align}
    \br{Z_t(\*g)}_{\*J} =& 
    \exp  \Bigg[ \Bigg\{  - N \beta J_0\sum_{u}  \frac{(M_u^{+}(\*g))^2+(M_u^{-}(\*g))^2}{4}
     - N \beta^2 \Delta J^2  \sum_{u>v}  \frac{(Q_{u,v}^{+}(\*g))^2+(Q_{u,v}^{-}(\*g))^2}{4} 
     \label{eq:saddle_node_generating_functional1}
     \nonumber\\ &+ \log \sum_{\*s_{1:t}}\int d\^\theta d\^{\hat \theta} \mathrm{e}^{ \Phi(\*s_{0:t}, \^\theta,\*g) + \Omega(\^{\hat \theta}, \^\theta,\*g)} \Bigg\} \Bigg],
\end{align}
where the optimal values $\*M(\*g), \*Q(\*g)$ are chosen to extremize (maximize or minimize) the quantity between the braces $\{\}$. We introduced the dependence with $\*g$ to denote the optimal values as the solution of $\*M, \*Q$ will be different for different values of $\*g$. 
As in the solutions in Eqs.~\ref{eq:zr*} and \ref{eq:zi*}, the solutions $M_u^{+}(\*g),Q_{u,v}^{+}(\*g)$ and  $M_u^{-}(\*g),Q_{u,v}^{-}(\*g)$ are combinations of the average statistics of the variables of interest (multiplied by the imaginary unit for the latter). 

\textbf{(iii) Change of the variables.} \hspace{1em} 
Having order parameters in this form can be cumbersome. We simplify them to directly capture the average statistics of the system by performing a change of variables at the saddle-point solution as exemplified in Eq.~\ref{eq:variable-change1}, resulting in:
\begin{align}
    \mu_u(\*g) =&  \frac{M_u^+(\*g) + \iu M_u^-(\*g)}{2}, &  m_{u-1}(\*g) =& \frac{M_u^+(\*g) - \iu M_u^-(\*g)}{2},
    \\ \rho_{u,v}(\*g)  =&  \frac{Q_{u,v}^+(\*g) + \iu Q_{u,v}^-(\*g)}{2},  & q_{u-1,v-1}(\*g) =& \frac{Q_{u,v}^+(\*g) - \iu Q_{u,v}^-(\*g)}{2},
\end{align}
or equivalently
\begin{align}
    M_u^+(\*g) =&  m_{u-1}(\*g) + \mu_u(\*g), & M_u^-(\*g) =& \iu (m_{u-1}(\*g) - \mu_u(\*g)),
    \\   Q_{u,v}^+(\*g) =& q_{u-1,v-1}(\*g) + \rho_{u,v}(\*g),  & Q_{u,v}^-(\*g) =& \iu(q_{u-1,v-1}(\*g) - \rho_{u,v}(\*g)),
\end{align}
where now we expect all $\*m(\*g), \^\mu(\*g), \*q(\*g), \^\rho(\*g)$ to be real-valued.

This results in
\begin{align}
    \br{Z_t(\*g)}_{\*J} =& 
    \exp  \Bigg[ \Bigg\{ - N \beta J_0\sum_{u}  \mu_{u}(\*g) m_{u-1}(\*g)
    - N \beta^2 \Delta J^2 \sum_{u>v} \rho_{u,v}(\*g) q_{u-1,v-1}(\*g)
    \label{eq:saddle_node_generating_functional2}
    \nonumber\\&+ \log \sum_{\*s_{1:t}}\int d\^\theta d\^{\hat \theta} \mathrm{e}^{ \Phi(\*s_{0:t}, \^\theta,\*g) + \Omega(\^{\hat \theta}, \^\theta,\*g)} \Bigg\}  \Bigg],
\end{align}
where now $\*m(\*g), \^\mu(\*g), \*q(\*g), \^\rho(\*g)$ are chosen to extremize the quantity between the braces. 
Also, we define the terms
\begin{align}
	\Phi(\*s_{0:t}, \^\theta,\*g) =&  \sum_{i,u}  \pr{g_{i,u} +  \theta_{i,u}} s_{i,u}
	- \sum_{i,u} \log\br{2 \cosh \theta_{i,u}}
	+ \sum_{i,u}  g_{u}^{S}   \pr{ s_{i,u} \theta_{i,u} - \log\br{ 2 \cosh \theta_{i,u}} }
	\nonumber\\& +  \sum_{i,u}  g_{u}^{S^r}   \pr{  s_{i,u-1}  \vartheta_{i,u} - \log\br{ 2 \cosh\vartheta_{i,u}}}
	+ \sum_{i,u} \beta J_0 \mu_{u}(\*g)  s_{i,u-1}
	+ \sum_{i,u>v} \beta^2 \Delta J^2 \rho_{u,v}(\*g) s_{i,u-1} s_{i,v-1},
\\ \Omega(\^{\hat \theta}, \^\theta,\*g) =&  \sum_{i,u}  (\theta_{i,u} - \beta H_{i,u} - \beta J_0 m_{u-1}(\*g)) \iu  \hat\theta_{i,u} 
	\nonumber\\ &	+  \frac{\beta^2 \Delta J^2 }{2}  \sum_{i,u}\pr{ \iu \hat\theta_{i,u} }^2 
	+ \beta^2 \Delta J^2   \sum_{i,u>v} q_{u-1,v-1}(\*g) \iu \hat\theta_{i,u}  \iu \hat\theta_{i,v} 
	- N(t+1) \log\br{2\pi}. 
\end{align}
Note that the summation of $u$ or $v$ related to the order parameters are performed over the range $1,\ldots,t+1$. Also note that integration over disordered connections has removed couplings between units and replaced them with same-unit temporal couplings $\^\rho(\*g)$ and varying effective fields, which are also independent between units, resulting in a mean-field solution where the activity of different spins is independent.

In the next section, we specify the conditions of the extrema, from which we find that some of the extrema are the order parameters.

\subsection{Introduction of the order parameters}
\label{app:order_parameters}

To obtain the values of the order parameters, we extremize the contents of the braces, finding
\begin{align}
	\frac{ \partial \log \br{Z_t(\*g)}_{\*J} }{\partial \mu_{u+1}(\*g)} 
	=& 
    \beta J_0 \pr{ \sum_i \ang{s_{i,u}}_{*,\*g} - N m_u(\*g) }=0; 
    \qquad  m_u(\*g)   =  \frac{1}{N}\sum_i \ang{s_{i,u}}_{*,\*g} ,
	\\ \frac{ \partial \log \br{Z_t(\*g)}_{\*J} }{\partial m_{u-1}(\*g)}
	=&  \beta J_0 \pr{- \sum_i \ang{\iu \hat\theta_{i,u}}_{*,\*g} - N \mu_u(\*g)  }=0;
	\qquad  \mu_u(\*g)  = -\frac{1}{N} \sum_i \ang{\iu \hat \theta_{i,u}}_{*,\*g},
	\\  \frac{ \partial \log \br{Z_t(\*g)}_{\*J} }{\partial \rho_{u+1,v+1}(\*g)} 
	=&  \beta^2 \Delta J^2 \pr{ \sum_i \ang{s_{i,u} s_{i,v}}_{*,\*g} - N q_{u,v}(\*g)  }=0; 
	\qquad q_{u,v}(\*g)  = \frac{1}{N} \sum_i \ang{s_{i,u}s_{i,v}}_{*,\*g} , \label{eq:saddle_node_quv}
	\\  \frac{ \partial \log \br{Z_t(\*g)}_{\*J} }{\partial q_{u-1,v-1}(\*g)} 
	=& \beta^2 \Delta J^2 \pr{ \sum_i \ang{\iu \hat \theta_{i,u} \iu \hat \theta_{i,v}}_{*,\*g} - N \rho_{u,v}(\*g)  }=0; \qquad\rho_{u,v}(\*g) =  \frac{1}{N} \sum_i \ang{\iu \hat \theta_{i,u} \iu \hat \theta_{i,v}}_{*,\*g},
\end{align}
where we define
\begin{align}
     \ang{f(\cdot)}_{*,\*g} =& \frac{ \sum_{\*s_{1:t}}\int d\^\theta d\^{\hat \theta} f(\cdot)\mathrm{e}^{ \Phi(\*s_{0:t}, \^\theta,\*g) + \Omega(\^{\hat \theta}, \^\theta,\*g)} }{ \sum_{\*s_{1:t}}\int d\^\theta d\^{\hat \theta} \mathrm{e}^{ \Phi(\*s_{0:t}, \^\theta,\*g) + \Omega(\^{\hat \theta}, \^\theta,\*g)} } \nonumber \\ &=\br{Z_t(\*g)}_{\*J}^{-1}  \sum_{\*s_{1:t}}\int d\^\theta d\^{\hat \theta} f(\cdot)\mathrm{e}^{ \Phi(\*s_{0:t}, \^\theta,\*g) + \Omega(\^{\hat \theta}, \^\theta,\*g) - N \beta J_0\sum_{u}  \mu_{u}(\*g) m_{u-1}(\*g)
    - N \beta^2 \Delta J^2 \sum_{u>v} \rho_{u,v}(\*g) q_{u-1,v-1}(\*g)}.
\end{align}
These equations are in concordance with Eqs.~\ref{eq:z1*}, \ref{eq:z2*}. 
Here, $\*m(\*g) $, $\^\mu(\*g) $, $\*q(\*g) $, and $ \^\rho(\*g) $ provide the saddle-point solution of the configurational average integral. 

On the other hand, the configurational average of the generating functional holds relations similar to Eqs.~\ref{eq:m_iu}, \ref{eq:R_ij_uv}:
\begin{align}
     \frac{ \partial \br{Z_t(\*g)}_{\*J} }{\partial g_{i,u}} 
    =& \br{\ang{s_{i,u}}_{\*g}}_{\*J},
    \label{eq:dZt_dgiu}
    \\
    \frac{ \partial^2 \br{Z_t(\*g)}_{\*J} }{\partial g_{i,u} \partial g_{j,v}} 
     =&   \br{\ang{s_{i,u} s_{j,v}}_{\*g}}_{\*J}  ,
     \label{eq:dZt_dgiudgiv}
\end{align}
where
\begin{align}
    \ang{f(\cdot)}_{\*g} =& \sum_{\*s_{0:t}}f(\cdot) p(\*s_{0:t}) \exp \Bigg[ \sum_{i,u} g_{i,u} s_{i,u} 
    \nonumber\\  & + \sum_{i,u}  g_{u}^{S}   \pr{ \beta s_{i,u} h_{i,u}  - \log\br{2 \cosh\br{\beta h_{i,u}}} } +  \sum_{i,u}  g_{u}^{S^r}   \pr{ 
\beta  s_{i,u-1}  h_{i,u}^r- \log\br{ 2 \cosh\br{\beta h_{i,u}^r}}  } \Bigg].
\end{align}

In addition, we have the following identities that will be helpful in eliminating spurious solutions
\begin{align}
	 \frac{\partial \br{Z_t(\*g)}_{\*J}}{\partial H_{i,u}} =& \beta\pr{\br{\ang{s_{i,u}}_{\*g}}_{\*J}  - \br{\ang{\tanh\br{\beta h_{i,u}}}_{\*g}}_{\*J} } ,
	\label{eq:spurious_solutionI}
	\\ \frac{\partial^2 \br{Z_t(\*g)}_{\*J}}{\partial H_{i,u}\partial H_{j,v}} =& \beta\pr{ \frac{\partial \br{\ang{s_{i,u}}_{\*g}}_{\*J} }{\partial H_{j,v}} - \frac{\partial \br{\ang{\tanh\br{\beta h_{i,u}}}_{\*g}}_{\*J} }{\partial H_{j,v}}} .
	\label{eq:spurious_solutionII}
\end{align}
Note that the equations above are equal to zero for $\*g=\*0$.

To derive the order parameters, we calculate the same partial derivatives using Eq.~\ref{eq:saddle_node_generating_functional2}. 
The order parameter of the system given by Eq.~\ref{eq:dZt_dgiu} is calculated directly as
\begin{align}
	\frac{ \partial \br{Z_t(\*g)}_{\*J} }{\partial g_{i,u}} 
	=&   \vast(  \ang{s_{i,u}}_{*,\*g}   + \beta J_0\sum_v\frac{ \partial \mu_{v}(\*g) }{\partial g_{i,u}} \pr{N m_{v-1}(\*g) - \sum_j \ang{s_{j,v-1}}_{*,\*g}  } 
	\nonumber\\  &+ \beta J_0\sum_v\frac{ \partial m_{v-1}(\*g) }{\partial g_{i,u}} \pr{N \mu_{v}(\*g) - \sum_j \ang{\theta_{j,v}}_{*,\*g}  } 
	\nonumber\\  & + \beta^2 \Delta J^2\sum_v\frac{ \partial \rho_{v,w}(\*g)}{\partial g_{i,u}} \pr{N q_{v-1,w-1}(\*g) - \sum_j \ang{s_{j,v-1}s_{j,w-1}}_{*,\*g}  } 
	\nonumber\\  &+  \beta^2 \Delta J^2\sum_v\frac{ \partial q_{v-1,w-1}(\*g) }{\partial g_{i,u}} \pr{N \rho_{v,w}(\*g) - \sum_j \ang{\theta_{j,v}\theta_{j,w}}_{*,\*g}  } \vast)\br{Z_t(\*g)}_{\*J}
	\nonumber\\ =&  \ang{s_{i,u}}_{*,\*g} \br{Z_t(\*g)}_{\*J}.  
\end{align}

Similarly, we obtain
\begin{align}
	\frac{ \partial \br{Z_t(\*g)}_{\*J} }{\partial H_{i,u}} 
	=& - \beta \ang{ \iu \hat\theta_{i,u} }_{*,\*g}  \br{Z_t(\*g)}_{\*J}  
    \\
	\frac{ \partial^2 \br{Z_t(\*g)}_{\*J} }{\partial g_{i,u} \partial g_{j,v}} 
	=&   \ang{s_{i,u}s_{j,v}}_{*,\*g} \br{Z_t(\*g)}_{\*J}  
	\\ 
	\frac{ \partial^2 \br{Z_t(\*g)}_{\*J} }{\partial H_{i,u} \partial H_{j,v}}
	=& \beta^2 \ang{ \iu \hat\theta_{i,u}\iu \hat\theta_{j,v} }_{*,\*g}  \br{Z_t(\*g)}_{\*J}  
\end{align}
Here we should note that, as there is no coupling between units, for $i\neq j$ we have a factorized solution $\br{\ang{s_{i,u} s_{j,v}}} = \ang{s_{i,u} s_{j,v}}_{*,\*g} =  \ang{s_{i,u}}_{*,\*g}\ang{ s_{j,v}}_{*,\*g}$. 

Finally, by comparing the above derivatives with Eqs.~\ref{eq:dZt_dgiu}, \ref{eq:dZt_dgiudgiv}, \ref{eq:spurious_solutionI}, and \ref{eq:spurious_solutionII} we obtain the order parameters:
\begin{align}
    m_u(\*g) =&\br{Z_t(\*g)}_{\*J}^{-1} \sum_i \br{\ang{s_{i,u}}_{\*g}}_{\*J},
    \\ q_{u,v}(\*g) =&\br{Z_t(\*g)}_{\*J}^{-1} \sum_i \br{\ang{s_{i,u}s_{i,v}}_{\*g}}_{\*J},
    \\    \mu_u(\*g) =&\br{Z_t(\*g)}_{\*J}^{-1} \sum_i \pr{\br{\ang{s_{i,u}}_{\*g}}_{\*J} - \br{\ang{\tanh\br{\beta h_{i,u}}}_{\*g}}_{\*J}},
    \\    \rho_{u,v}(\*g) =&\br{Z_t(\*g)}_{\*J}^{-1} \sum_i\beta^{-1}\pr{ \frac{\partial \br{\ang{s_{i,u}}_{\*g}}_{\*J} }{\partial H_{j,v}} - \frac{\partial \br{\ang{\tanh\br{\beta h_{i,u}}}_{\*g}}_{\*J} }{\partial H_{j,v}}}. 
\end{align}
Note that, at $\*g=\*0$, $  \mu_u(\*0) =   \rho_u(\*0) = 0$. As well, notice that $\br{Z_t(\*0)}_{\*J}=1$, retrieving activation rates and delayed self-correlations as the order parameters of the system $m_u(\*0), q_{u,v}(\*0)$. Below, we will drop the parenthesis for the order parameters at  $\*g=\*0$, referring to these quantities as $m_u, q_{u,v}$.


Similarly, the forward and reverse entropy rates are calculated from the functions
\begin{align}
	\frac{ \partial \br{Z_t(\*g)}_{\*J} }{\partial g_{u}^{S}} 
	=& 
	\sum_i \ang{ s_{i,u} \theta_{i,u} - \log\br{ 2 \cosh \theta_{i,u}}}_{*,\*g} \br{Z_t(\*g)}_{\*J}  \nonumber\\ =& \sum_{i}\br{\ang{  \pr{ s_{i,u} \theta_{i,u} - \log\br{ 2 \cosh \theta_{i,u}} }}_{\*g}}_{\*J},
	\\ 
	\frac{ \partial \br{Z_t(\*g)}_{\*J} }{\partial  g_{u}^{S^r} } 
	=&\sum_{i}\ang{  \pr{  s_{i,u-1}  \vartheta_{i,u} - \log\br{ 2 \cosh\vartheta_{i,u}}} }_{*,\*g}  \br{Z_t(\*g)}_{\*J}   \nonumber\\ =& \sum_{i} \br{\ang{  \pr{  s_{i,u-1}  \vartheta_{i,u} - \log\br{ 2 \cosh\vartheta_{i,u}}}}_{\*g}}_{\*J},
\end{align}
evaluated at $\*g=\*0$.

\subsection{Mean-field solutions}

After solving the saddle-point integral, we have the following expression for computing relevant  quantities in the system
\begin{align}
	\br{Z_t(\*g)}_{\*J}
	 =& \sum_{\*s_{1:t}}\int d\^\theta d\^{\hat \theta}  \mathrm{e}^{ \Phi(\*s_{0:t}, \^\theta,\*g) + \Omega(\^{\hat \theta}, \^\theta,\*g)}.
	\label{eq:saddle_node_generating_functional3}
\end{align}

At this point, we want to remove the effective fields $\^{\theta}$ and effective conjugate fields $\^{\hat \theta}$. For this goal, (i) we first remove the effective conjugate fields $\^{\hat \theta}$ by recovering the delta functions from their integral forms. Then, (ii) we revert the effective fields $\^{\theta}$ by removing the delta function. This results in the mean-field (factorized) generating functional, from which we obtain the mean-field solutions of order parameters, conditional entropy, or entropy production. 

\textbf{(i) Removing effective conjugate fields}
\hspace{1em} 
We first remove the conjugate fields by recovering a delta function. We rewrite
\begin{align}
	\mathrm{e}^{\Omega(\^{\hat \theta}, \^\theta,\*g)} =& \prod_i\frac{1}{(2\pi)^{N(t+1)}}\exp\br{ \sum_{u}(\theta_{i,u} - \beta H_{i,u} - \beta J_0 m_{u-1}(\*g))   \iu  \hat\theta_{i,u}
	 +   \frac{\beta^2 \Delta J^2 }{2} \sum_{u,v}q_{u-1,v-1}(\*g)  \iu \hat\theta_{i,u} \iu \hat\theta_{i,v} },
\end{align}
defining $q_{u-1,u-1}=1$ and $q_{u-1,v-1}=q_{v-1,u-1}$ to obtain a symmetric matrix. Note that the saddle-node solution Eq.~\ref{eq:saddle_node_quv} was defined only for $u>v$.

We can remove the quadratic terms of $\^{\theta}$ by applying $N(t+1)$-dimensional multivariate Gaussian integrals of the form
\begin{align}
	\mathrm{e}^{\frac{1}{2}\sum_{u,v} K_{u,v} x_u x_v} = \frac{1}{\sqrt{(2\pi)^t |K^{-1}|}}\int d \*z \mathrm{e}^{-\frac{1}{2}\sum_{u,v} K_{u,v} z_u z_v +\sum_{u,v} K_{u,v} x_u z_v , }
\end{align}
for $x_u = -\beta \Delta J \iu \hat\theta_{i,u}$ and $K_{u,v} = q_{u-1,v-1}(\*g)$.
Similarly, we can remove  the quadratic terms of $\^{\hat \vartheta}$ by applying $N$ univariate Gaussian integrals, obtaining
\begin{align}
	  \int d\^{\hat\theta}  \mathrm{e}^{\Omega(\^{\hat \theta}, \^\theta, \*0)} 
	  =&  \frac{1}{(2\pi)^{N(t+1)}} \prod_i \int d\^{\hat\theta}  d \*z   p(\*z)
	  \label{eq:Gaussian_theta_equivalence}
	   \nonumber\\ &\cdot \exp\Bigg[
  \sum_{u}  \iu  \hat\theta_{i,u} (\theta_{i,u} - \beta H_{i,u} - \beta J_0 m_{u-1})  
	-    \beta \Delta J \sum_{u,v} q_{u-1,v-1}(\*g) \iu \hat\theta_{i,u}   z_{v} 
	    \Bigg]
\nonumber	\\ =&  \prod_{i}  \int d \*z   p(\*z) \prod_{u} \delta\br{\theta_{i,u} - \beta\overline h_{i,u}(\*z)}, 
\end{align}
where $\*z = (z_{1},\ldots,z_{t+1})$, and the distribution $p(\*z)=\mathcal{N}(0,\*\Sigma)$ is a multivariate Gaussian with zero mean and inverse covariance 
 $\*\Sigma^{-1} = \*q(\*g)$, and
\begin{align}
	\overline h_{i,u}(\*z) =& 
	H_{i,u} + J_0 m_{u-1}(\*g)
	 + \Delta J  \sum_v z_{v} q_{u-1,v-1}(\*g),
\end{align}

We can simplify the expressions above into 
\begin{align}
	\overline h_{i,u}(\xi_{u}) =& 
	H_{i,u} + J_0 m_{u-1}(\*g) 
	 + \Delta J  \xi_{u},
  \label{app_eq:effective_field}
\end{align}
with $\xi_{u}=\sum_v z_{v} q_{u-1,v-1}$. Let $\^{\xi}=(\xi_{1},\ldots,\xi_{t+1})$, then it follows $p(\^{\xi})=\mathcal{N}(0,\*q)$.
Similarly, we can derive
\begin{align}
	\overline h^r_{i,u}(\xi_{u+1})=& \overline h_{i,u}(\xi_{u}) + H_{i,u} - H_{i,u+1} =
	H_{i,u} + J_0 m_{u}  + \Delta J \xi_{u+1}. 
\end{align}

\textbf{(ii) Removing effective fields}
\hspace{1em} 
We now revert the effective fields $\theta_{i,u}$ to $\beta\overline h_{i,u}(\*z)$ by removing the delta function, which replaces the original $\beta h_{i,u}$ with the mean-field equivalent. 
 
Introducing the equivalences in the previous sections, we have
\begin{align}
	\mathrm{e}^{\Phi(\*s_{0:t}, \^\theta,\*g)} =& \prod_i \exp\Bigg[ \sum_{u}  s_{i,u} \pr{g_{i,u}+  \theta_{i,u}}
	-  \sum_{u} \log\br{2 \cosh\theta_{i,u}} + \sum_{u}  \beta  s_{i,u-1}\widetilde h_{i,u-1}
	\nonumber\\ &+ \sum_{u}  g_{u}^{S}   \pr{ s_{i,u} \theta_{i,u} - \log\br{ 2 \cosh\theta_{i,u}} } 
	+ \sum_{u}  g_{u}^{S^r}   \pr{  s_{i,u-1}  \vartheta_{i,u} - \log\br{2 \cosh\vartheta_{i,u}}} \Bigg],
\end{align}
with
\begin{align}
    \widetilde h_{i,u-1} =  \sum_{u} J_0 \mu_{u}(\*g)  
	+ \sum_{u>v} \beta \Delta J^2 \rho_{u,v}(\*g) s_{i,v-1}.
\end{align}
Note that for $\*g=\*0$, $\widetilde h_{i,u-1}$ terms disappear.

This leads us to the mean-field solution of the configurational average of the generating functional
\begin{align}
     \br{Z_t(\*g)}_{\*J}
	=&   \int d\^\theta  \sum_{\*s_{1:t}} \mathrm{e}^{\Phi(\*s_{0:t}, \^\theta,\*g)} \prod_{i}  \int d\^{\xi}  p(\^{\xi}) \prod_{u} \delta\br{ \theta_{i,u} - \beta\overline h_{i,u}(\xi_{u})}
	\nonumber\\ =& \prod_i \sum_{\*s_{i,1:t}} \int  d\^{\xi}  p(\^{\xi})
	 \exp\Bigg[ \sum_{u} s_{i,u} \pr{g_{i,u} +  \beta\overline h_{i,u}(\xi_{u})} 
	- \sum_{u} \log  2 \cosh\br{ \beta\overline h_{i,u}(\xi_{u})}
	\nonumber\\& + \sum_{u}  \beta  s_{i,u-1}\widetilde h_{i,u-1} + \sum_{u}  \beta(g_{u}^{S} s_{i,u} \overline h_{i,u}(\xi_{u}) + g_{u}^{S^r} s_{i,u-1}\overline h^r_{i,u}(\xi_{u+1})) 
	\nonumber\\&- \sum_{u}\pr{g_{u}^{S} \log\br{ 2 \cosh \br{\beta\overline h_{i,t}(\xi_{t})}} - g_{u}^{S^r}\log\br{2 \cosh \br{\beta\overline h_{i,u}^r(\xi_{u+1})}}} \Bigg].
	\label{eq:saddle_node_generating_functional4}
\end{align}
The summation over $u$ is taken for the range from $1$ to $t$. With $\^\xi$ defined in the range $u=1,\dots,t+1$, we can recover the values of $\overline h_{i,u}$ and $\overline h^r_{i,u}$ for all time steps.

Similarly, statistical moments are obtained as
\begin{align}
     \ang{f(\cdot)}_{*,\*0}
	=&   \prod_i \sum_{\*s_{i,1:t}} \int  d\^{\xi}  p(\^{\xi})
	 f(\cdot) \exp\Bigg[ \sum_{u} s_{i,u} \pr{g_{i,u} +  \beta\overline h_{i,u}(\xi_{u})} 
	- \sum_{u} \log  2 \cosh\br{ \beta\overline h_{i,u}(\xi_{u})}
	\nonumber\\& + \sum_{u}  \beta  s_{i,u-1}\widetilde h_{i,u-1} + \sum_{u}  \beta(g_{u}^{S} s_{i,u} \overline h_{i,u}(\xi_{u}) + g_{u}^{S^r} s_{i,u-1}\overline h^r_{i,u}(\xi_{u+1})) 
	\nonumber\\&- \sum_{u}\pr{g_{u}^{S} \log\br{ 2 \cosh \br{\beta\overline h_{i,t}(\xi_{t})}} - g_{u}^{S^r}\log\br{2 \cosh \br{\beta\overline h_{i,u}^r(\xi_{u+1})}}} \Bigg].
\end{align}

From this equation, we can derive the mean activation rate and the equal-time correlation of the $i$th unit. We note that the diagonal of the covariance matrix of $\^{\xi}$ is equal to 1, hence we arrive at
\begin{align}
	\br{\ang{s_{i,u}}}_{\*J} =& 
	\ang{s_{i,u}}_{*,\*0}  = 
	\int \mathrm{D}z \tanh\br{\beta\overline h_{i,u}(z)},
	\\  \br{\ang{s_{i,u}s_{i,v}}}_{\*J}  =& 
	\ang{s_{i,u}s_{i,v}}_{*,\*0} = 
	\int \mathrm{D}xy^{(q_{u-1,v-1})} \tanh\br{\beta\overline h_{i,u}(x)} \tanh\br{\beta\overline h_{i,v}(y)},
\end{align}
where 
\begin{align}
	\mathrm{D}z =&  \frac{1}{\sqrt{2\pi}} \mathrm{e}^{-\frac{1}{2}z^2},
	\\ \mathrm{D}xy^{(q)} =& \frac{1}{2\pi\sqrt{1-q^2}}\mathrm{e}^{-\frac{x^{2}+y^{2}-2q xy}{2(1-q^2)}}.
\end{align}
Finally, we obtain order parameters
\begin{align}
	m_{u} =& \frac{1}{N}\sum_i \br{\ang{s_{i,u}}}_{\*J}   =\frac{1}{N}\sum_i \int \mathrm{D}z \tanh\br{\beta \pr{H_{i,u} + J_0 m_{u-1} + \Delta J  z}},
	\label{eq-app:order_parameter_m}
	\\  q_{u,v} =&  \frac{1}{N}\sum_i \br{\ang{s_{i,u}s_{i,v}}}_{\*J}   = \frac{1}{N}\sum_i \int \mathrm{D}xy^{(q_{u-1,v-1})} \tanh\br{\beta \pr{H_{i,u} + J_0 m_{u-1} 
	 + \Delta J  x}} \tanh\br{\beta \pr{H_{i,v} + J_0 m_{v-1} 
	 + \Delta J y}}
	 \label{eq-app:order_parameter_q}.
\end{align}
Note that magnetizations $m_u$ are independent of $q_{u,v}$. This is consistent with findings of the asymmetric SK model lacking a spin-glass phase \cite{crisanti1988dynamics}.

The conditional entropy results in
\begin{align}
	\br{S_{u|u-1}}_{\*J}  =& 
	\sum_i \ang{ s_{i,u} \theta_{i,u} - \log\br{ 2 \cosh \theta_{i,u}}}_{*,\*0} 
	\nonumber\\  =& \sum_i  \int   p({\^\xi}) \Big(\tanh\br{\beta\overline h_{i,u}(\xi_{u}) } \beta\overline h_{i,u}(\xi_{u}) 
	 - \log\br{ 2 \cosh \br{\beta \overline h_{i,u}(\xi_{u})}}   \Big)
  	\nonumber\\  =& -\sum_i \int  p({\^\xi})\Big(\tanh \br{\beta \pr{H_{i,u} + J_0 m_{u-1} + \Delta J  \xi_{u}}} \beta \pr{H_{i,u} + J_0 m_{u-1} 
	 + \Delta J  \xi_{u}}
	\nonumber\\ & - \log\br{ 2 \cosh \br{ \beta \pr{H_{i,u} + J_0 m_{u-1}
	 + \Delta J  \xi_{u}} }}\Big)
	\nonumber\\  =& - \sum_i   \int  \mathrm{D}z \Big( \beta \pr{H_{i,u} + J_0 m_{u-1}} \tanh\br{\beta \pr{H_{i,u} + J_0 m_{u-1}  + \Delta J  z}})
	\nonumber\\  &+\beta^2 \Delta J^2 (1-\tanh^2\br{\beta \pr{H_{i,u} + J_0 m_{u-1}  + \Delta J  z}}) 
	\nonumber\\ & - \log\br{2 \cosh \br{\beta \pr{H_{i,u} + J_0 m_{u-1}  + \Delta J  z}}} \Big).
	 \label{eq-app:entropy-rate}
\end{align}

Similarly, the reversed conditional entropy results in
\begin{align}
	\br{S_{u|u-1}^r}_{\*J}  =&
 \sum_{i}\ang{  \pr{  s_{i,u-1}  \vartheta_{i,u} - \log\br{ 2 \cosh\vartheta_{i,u}}} }_{*,\*0}
	\nonumber\\  =&	\sum_i  \int   p({\^\xi}) \Big(\tanh\br{\beta \overline h_{i,u-1}(\xi_{u}) }\beta \overline h^r_{i,u}(\xi_{u+1}))
	- \log\br{2 \cosh \br{\beta\overline h^r_{i,u}(\xi_{u+1}) }}  \Big)
  	\nonumber\\  =& - \sum_i \int  p({\^\xi})\Big( \tanh\br{\beta \pr{H_{i,u-1} + J_0 m_{u-2}  + \Delta J \xi_{u-1}}} (\beta \pr{H_{i,u} + J_0 m_{u}  + \Delta J \xi_{u+1}}) \Big)
	\nonumber\\ & -  \log\br{ 2 \cosh \br{\beta \pr{H_{i,u} + J_0 m_{u}  + \Delta J \xi_{u+1}}}}\Big)
	\nonumber\\  =&  - \sum_i   \int  \mathrm{D}z \Big( (\beta H_{i,u} + \beta J_0 m_{u}) \tanh\br{\beta \pr{H_{i,u-1} + J_0 m_{u-2}  + \Delta J  z}})
	\nonumber\\  &+\beta^2 \Delta J^2  q_{u,u-2} (1-\tanh^2\br{\beta \pr{H_{i,u-1} + J_0 m_{u-2}  + \Delta J  z}})
	\nonumber\\ & -\log\br{2 \cosh \br{\beta \pr{H_{i,u} + J_0 m_{u}  + \Delta J  z}}}   \Big).
	 \label{eq-app:entropy-rate-reverse}
\end{align}
where $\xi_{u+1}$ is decomposed as a conditional Gaussian distribution for a given $\xi_{u-1}$ as $\xi_{u+1}= q_{u,u-2} \xi_{u-1} + \sqrt{1-q_{u,u-2}^2} \zeta_{u+1}$ where $\zeta_{u+1}$ is a normalized Gaussian independent of $\xi_{u-1}$ and the term $ q_{u,u-2}$ the covariance between variables.

Note that if the fields are constant $H_{i,u} = \Theta_i$, the system will converge to a steady state with $m_u=m$ and $q_{u,v}=q$ the entropy production simplifies to
\begin{align}
	\br{\sigma_{u}}_{\*J}  =&\beta^2 \Delta J^2 (1-q ) \sum_i \int  \mathrm{D}z(1-\tanh^2\br{\beta \pr{\Theta_i + J_0 m  + \Delta J  z}}),
	 \label{eq-app:entropy-production}
\end{align}
where $m$ and $q$ are the steady-state solutions of Eqs.~\ref{eq-app:order_parameter_m} and \ref{eq-app:order_parameter_q} respectively. Note that, since every $q_{u,v}$ depends only on $m_u$ and $m_v$, all $q_{u,v}$ converges to the same solution $q$ under the steady state.

\newpage
\section{Solution of the asynchronous asymmetric SK models}
\label{app:async_model_solution}

The kinetic Ising system is sometimes written in the form of a continuous-time master equation with an asynchronous, stochastic update of spins \cite{sompolinsky1981dynamic}. 
Here we extend the discrete-time kinetic system with the synchronous updates in Supplementary Note \ref{app:path-integral} to the continuous-time model with asynchronous updates. We find the dynamical mean-field equation for the order parameters is effectively unchanged in the limit of long-term correlations, but short-term correlations decay slowly and the steady-state entropy production rate requires slight modification from the discrete counterpart.

First, we show that a system with (partially) asynchronous updates is equivalent to defining the time-dependent fields as a doubly stochastic process using independent Bernoulli random variables, $\tau_{i,u} \in \{0,1\}$, which restrict updates of each spin $i$ only to the times in which $\tau_{i,u} =1$. Namely, we define a synchronous, discrete-time master equation
\begin{align}
    p(\*s_u) = \sum_{\*s_{u-1}}p(\*s_{u}\,|\,\*s_{u-1}) p(\*s_{u-1}) 
\end{align}
driven by the following transition probabilities with time-dependent stochastic fields: 
\begin{align}
     p(\*s_{u}\,|\,\*s_{u-1}) =& \prod_i \frac{\exp\br{\beta s_{i,u} h_{i,u}}}{2 \cosh \br{\beta h_{i,u}}},
    \\ h_{i,u} =& H_{i,u} + \sum_j J_{ij} s_{j,u-1},
    \\H_{i,u} =&  \Theta_{i,u} + (1-\tau_{i,u})K s_{i,u-1}.
    \label{eq-app:external_field_stochastic}
\end{align}
where $\tau_{i,u}$ are binary random variables independently following a Bernoulli distribution with rate $\alpha$,  i.e., $p(\tau_{i,u}=1) = \alpha$. When $\tau_{i,u}=1$, the spin updates normally with the field $\Theta_{i,u}$. When $\tau_{i,u}=0$, the field is $K s_{i,u-1}$, i.e., coupled to the previous spin value with a strength $K$. If $K$ is large, the current state is tightly coupled with the previous state. This means that the spin state is unchanged from the previous state in the limit $K\to\infty$. In this limit, the transition probability results in
\begin{align}
    p(s_{i,u}\,|\,\*s_{u-1}) =& 
    \tau_{i,u} w(s_{i,u}\,|\,\*s_{u-1}) 
    + (1-\tau_{i,u}) \delta\br{s_{i,u},s_{i,u-1}}.
\end{align}
where $w(s_{i,u}\,|\,\*s_{u-1})$ is a transition rate given by
\begin{align}
    w(s_{i,u}\,|\,\*s_{u-1}) =& \frac{\exp\br{\beta s_{i,u} h^1_{i,u}}}{2 \cosh \br{\beta h^1_{i,u}}}.
    \\ h_{i,u}^1 =& \Theta_{i,u} + \sum_j J_{ij} s_{j,u-1} .
\end{align}
We note that $\alpha=1$ recovers the synchronous update in Supplementary Note \ref{app:path-integral}. However, for a small $ \alpha$ (such that $N\alpha \ll 1$), only one spin can update at a time and the system becomes equivalent to a single-spin update dynamics. Thus, we can model both the synchronous and asynchronous updates by these doubly-stochastic time-dependent fields.

Since only one spin can be updated at each time-step for a small probability $\alpha$ with the transfer rate  $w(s_{i,u}\,|\,\*s_{u-1})$, the master equation (equivalent to Eq.~\ref{eq:master-equation} in this condition) expected over $\tau_{i,u}$ is described as
\begin{align}
    p(\*s_u)=& \sum_{\*s_{u-1}}p(\*s_u\,|\,\*s_{u-1})p(\*s_{u-1})
     \nonumber\\=&p_{u-1}(\*s_{u}) +  \sum_{\*s_{u-1}} \pr{p(\*s_u\,|\,\*s_{u-1})p(\*s_{u-1}) - p(\*s_{u-1}\,|\,\*s_{u})p_{u-1}(\*s_{u})}
     \nonumber\\ =&  p_{u-1}(\*s_u) + \alpha\sum_i  \pr{w(s_{i,u}\,|\,\*s_{u}^{[i]})p_{u-1}(\*s_{u}^{[i]}) - w(-s_{i,u}\,|\,\*s_{u})p_{u-1}(\*s_{u})}.
     \label{eq:master-equation-async}
\end{align}
where the $^{[i]}$ operator flips the sign of the $i$-th spin (therefore, $\*s_{u-1}=\*s_{u}^{[i]}$), and the second term in the equation stands for the system's probability flow. We now represent the discrete time steps $u=1,2,\ldots$ on the real-valued line of the continuous time. We set the Bernoulli probability $\alpha$ ($\tau_{i,u}$ being $1$) to be identical to the time resolution of the discrete-time step in the continuous time. The starting time of the $u$-th bin is given by $t=(u-1)\alpha$. This condition means that the time $t$ is equivalent to the expected number of the spin updates up to the $u$-th bin. Note that $t$ stands for the continuous time and should not be taken as the last discrete-time step that we use in the subscript of the variables. In the limit $\alpha \to 0$, the system is equivalent to a continuous time description for the new time variable $t$:
\begin{align}
    \frac{d p(\*s,t)}{dt} =& \sum_i  \pr{w(s_{i}\,|\,\*s^{[i]})p(\*s^{[i]},t) - w(-s_{i}\,|\,\*s)p(\*s,t)}.
     \label{eq:master-equation-cont}
\end{align}

This formulation leads to the definition of the entropy change as
\begin{align}
    \frac{d\sigma^\mathrm{sys}(t)}{dt} =& - \frac{d }{dt} \sum_{\*s}  p(\*s,t)\log p(\*s,t)
    \nonumber\\ =& - \sum_{\*s} \frac{d p(\*s,t)}{dt} \log p(\*s,t) - \sum_{\*s} \frac{d p(\*s,t)}{dt}
    \nonumber\\ =& - \sum_{\*s} \frac{d p(\*s,t)}{dt} \log p(\*s,t)
\end{align}
Using Eq.~\ref{eq:master-equation-cont}, the entropy change is further written as
\begin{align}
    \frac{d\sigma^\mathrm{sys}(t)}{dt}
    =&  - \sum_{\*s}\sum_i\pr{w(s_{i}\,|\,\*s^{[i]})p(\*s^{[i]},t) - w(-s_{i}\,|\,\*s)p(\*s,t)}  \log p(\*s,t)
    \nonumber\\ =&  - \frac{1}{2}\sum_{\*s}\sum_i\pr{w(s_{i}\,|\,\*s^{[i]})p(\*s^{[i]},t) - w(-s_{i}\,|\,\*s)p(\*s,t)}  \log p(\*s,t) 
    \nonumber\\ &+ \frac{1}{2}\sum_{\*s}\sum_i\pr{w(-s_{i}\,|\,\*s)p(\*s,t) -w(s_{i}\,|\,\*s^{[i]})p(\*s^{[i]},t) }  \log p(\*s,t)
    \nonumber\\ =&  - \frac{1}{2}\sum_{\*s}\sum_i\pr{w(s_{i}\,|\,\*s^{[i]})p(\*s^{[i]},t) - w(-s_{i}\,|\,\*s)p(\*s,t)}  \log p(\*s,t) 
    \nonumber\\ &+ \frac{1}{2}\sum_i\sum_{\*s^{[i]}}\pr{w(s_{i}\,|\,\*s^{[i]})p(\*s^{[i]},t) - w(-s_{i}\,|\,\*s)p(\*s,t)}  \log p(\*s^{[i]},t)
   \nonumber\\ =& \frac{1}{2}\sum_{\*s}\sum_i\pr{w(s_{i}\,|\,\*s^{[i]})p(\*s^{[i]},t) - w(-s_{i}\,|\,\*s)p(\*s,t)} \log \frac{p(\*s^{[i]},t) }{p(\*s,t) }.
\end{align}
To obtain the third equality, we applied the change of variables from $\*s$ to $\*s^{[i]}$ in the second term, which makes $s_i$ into $-s_i$. We then reverted the summation over $i$ and $\*s^{[i]}$ to $\*s$ and $i$ since both sum all the states and flipping to obtain the fourth equality. 
From this form, the entropy change can be decomposed into an entropy production rate and entropy flow rate terms as follows \cite{van2015ensemble}:
\begin{align}
    \frac{d\sigma^\mathrm{sys}(t)}{dt} =& 
    \underbrace{\frac{1}{2}\sum_{\*s}\sum_i\pr{w(s_{i}\,|\,\*s^{[i]})p(\*s^{[i]},t) - w(-s_{i}\,|\,\*s)p(\*s,t)}  \log  \frac{w(s_{i}\,|\,\*s^{[i]})p(\*s^{[i]},t)}{w(-s_{i}\,|\,\*s)p(\*s,t)}}_{\mathrm{Entropy\,production\,rate}}
    \nonumber\\& + \underbrace{\frac{1}{2}\sum_{\*s}\sum_i\pr{w(s_{i}\,|\,\*s^{[i]})p(\*s^{[i]},t) - w(-s_{i}\,|\,\*s)p(\*s,t)}  \log  \frac{w(-s_{i}\,|\,\*s)}{w(s_{i}\,|\,\*s^{[i]})}}_{\mathrm{Entropy\,flow}}.
\end{align}
Thus, we define the continuous-time steady-state entropy production rate as:
\begin{align}
    \frac{d\sigma(t)}{dt} =&   \frac{1}{2}\sum_{\*s}  \sum_i \pr{w(s_{i}\,|\,\*s^{[i]})p(\*s^{[i]},t) - w(-s_{i}\,|\,\*s)p(\*s,t)} \log \frac{w(s_{i}\,|\,\*s^{[i]})p(\*s^{[i]},t)}{w(-s_{i}\,|\,\*s)p(\*s,t)}
    \nonumber \\=&  \sum_{\*s}  \sum_i w(s_{i}\,|\,\*s^{[i]})p(\*s^{[i]},t)  \log \frac{w(s_{i}\,|\,\*s^{[i]})p(\*s^{[i]},t)}{w(-s_{i}\,|\,\*s)p(\*s,t)}.
\end{align}

\subsection{Generating functional}

Knowing that we obtain the continuous-time asynchronous updates in the limit of $\alpha \to 0$ and $K \to \infty$, in what follows, we will derive the order parameters and entropy production rate with the asynchronous updates in continuous-time by first augmenting the generating functional defined in the discrete-time steps using the doubly stochastic fields and then applying these limits.

Given the augmented fields (Eq.~\ref{eq-app:external_field_stochastic}), the effective field (Eq.~\ref{app_eq:effective_field}) in the thermodynamic limit is decomposed as
\begin{align}
    \overline h_{i,u}(\xi_{u}) = \tau_{i,u} \overline h_{i,u}^{(1)} (\xi_{u})+ (1-\tau_{i,u})\overline h_{i,u}^{(0)} (\xi_{t}), 
\end{align}
where 
\begin{align}
	\overline h_{i,u}^{(1)} (\xi_{u}) =& 
	\Theta_{i,u} + J_0 m_{u-1} 
	 + \Delta J  \xi_{u}, 
	\\ \overline h_{i,u}^{(0)} (\xi_{u}) =& 
	K \pr{s_{i,u-1}+  \frac{1}{K} \pr{\Theta_{i,u} + J_0 m_{u-1} 
	 + \Delta J \xi_{u}}}.
\end{align}
Given that $K$ is large, we can approximate $h_{i,u}^{(0)} (\xi_{u}) \approx K s_{i,u-1}$ because the remaining two terms become negligible for calculating the spin update. Thus we have
\begin{align}
    \overline h_{i,u}(\xi_{u}) \approx \tau_{i,u} \overline h_{i,u}^{(1)} (\xi_{u})+ (1-\tau_{i,u}) K s_{i,u-1}.
\end{align}
for large $K$.

Using the aforementioned $\overline h_{i,u}(\xi_{u})$, the system averaged over variables $\tau_{i,u}$ is now described by the following configurational average of a generating functional:
\begin{align}
     [Z_t(\*g)]_{\*J,\^\tau}
 =& \prod_i \sum_{\*s_{i,1:t}} \sum_{\^\tau_{i,1:t}}\int  d\^{\xi}  p(\^\tau_{i,1:t})
	 \exp\Bigg[ \sum_{u} s_{i,u} \pr{g_{i,u} +  \beta\overline h_{i,u}(\xi_{u})} 
	- \sum_{u} \log  2 \cosh\br{ \beta\overline h_{i,u}(\xi_{u})}
	\nonumber\\&+ \sum_{u}  \beta(g_{u}^{S} s_{i,u} \overline h_{i,u}(\xi_{u}) + g_{u}^{S^r} s_{i,u-1}\overline h^r_{i,u}(\xi_{u+1})) 
	\nonumber\\&- \sum_{u}\pr{g_{u}^{S} \log\br{ 2 \cosh \br{\beta\overline h_{i,t}(\xi_{t})}} - g_{u}^{S^r}\log\br{2 \cosh \br{\beta\overline h_{i,u}^r(\xi_{u+1})}}} \Bigg].
	\label{eq:saddle_node_generating_functional_async}
\end{align}

The order parameters will be obtained similarly as in the general solution with additional averaging over the Bernoulli random variables. To simplify further steps, we will consider the configurational average of the mean and delayed correlation of $N$ individual spins:
\begin{align}
	m_{i,u} =& \br{\ang{s_{i,u}}}_{\*J,\^\tau} = \lim_{\*g\to\*0} \frac{\partial [Z_t(\*g)]_{\*J,\^\tau}}{\partial g_{i,u}} = \sum_{\^\tau_{i,1:t}} p(\^\tau_{i,1:t}) \int \mathrm{D}z \tanh\br{\beta\overline h_{i,u}(z)},
	\\ q_{i,u,v} =&  \br{\ang{s_{i,u}s_{i,v}}}_{\*J,\^\tau}  = \lim_{\*g\to\*0} \frac{\partial^2 [Z_t(\*g)]_{\*J,\^\tau}}{\partial g_{i,u} \partial g_{i,v}} = \sum_{\^\tau_{i,1:t}} p(\^\tau_{i,1:t})  \int \mathrm{D}xy^{(q_{u-1,v-1})} \tanh\br{\beta\overline h_{i,u}(x)} \tanh\br{\beta\overline h_{i,v}(y)},\label{eq-app:q_i_u_v}
\end{align}
which constitute the order parameters
\begin{align}
    m_u =& \frac{1}{N}\sum_i m_{i,u},
    \\q_{u,v} =& \frac{1}{N}\sum_i q_{i,u,v}.
\end{align}

More specifically, since the order parameters $m_u$ and $q_{u,v}$ are given by the expectations over the independent random variables $\tau_{i,u}\sim \mathrm{Bernoulli}(\alpha)$, we further separate $m_{i,u}, q_{i,u,v}$ into the cases where $\tau_{i,u},\tau_{i,v}$ are updated or not:
\begin{align}
    m_{i,u} =&  \sum_{\tau_{i,u}} p(\tau_{i,u})  m_{i,u}^{\tau_{i,u}}
    \nonumber\\=&\sum_{\tau_{i,u}} p(\tau_{i,u}) \int \mathrm{D}z \tanh\br{\beta 	\overline h_{i,u}^{(\tau_{i,u})} (\xi_{u}) },
    \\ q_{i,u,v} =& \sum_{\tau_{i,u},\tau_{i,v}}  p(\tau_{i,u})p(\tau_{i,v})q_{i,u,v}^{\tau_{i,u},\tau_{i,v}}
    \nonumber\\=& \sum_{\tau_{i,u},\tau_{i,v}} p(\tau_{i,u})p(\tau_{i,v}) \int \mathrm{D}xy^{(q_{u-1,v-1})} \tanh\br{\beta\overline h_{i,u}^{(\tau_{i,u})}(x)} \tanh\br{\beta\overline h_{i,v}^{(\tau_{i,v})}(y)},
\end{align}
where $m_{i,u}^0 = m_{i,u-1}$ and $q_{i,u,v}^{0,0} =q_{i,u-1,v-1}$. 

\subsection*{Mean activation rate order parameter}

First, we look into the mean activation rate. Since $p(\tau_{i,u}=1) = \alpha$, the expectation of this order parameter by $\tau_i$ is given by
\begin{align}
	m_{u} =&  \alpha \frac{1}{N}\sum_i  \int \mathrm{D}z \tanh\br{\beta \pr{\Theta_{i,u} + J_0 m_{u-1} + \Delta J  z}} + (1-\alpha) m_{u-1},
	\label{eq-app:async_order_parameter_m}
\end{align}
where we assumed $K \to \infty$ to obtain the second term. 
The above equation gives the dynamical mean-field equation in the discrete time steps. We now represent the discrete time steps $u=1,2,\ldots$ on the real-valued line of the continuous time using $t=(u-1)\alpha$. By representing the mean activation rate at the $u-1$th bin by $m(t)$, i.e., $m(t)= m_{u-1}$, the equation above is described as:
\begin{align}
	\frac{m(t+\alpha) - m(t)}{\alpha }=&  \frac{1}{N}\sum_i  \int \mathrm{D}z \tanh\br{\beta \pr{\Theta_{i}(t( + J_0 m(t) + \Delta J  z}} - m(t)
	\label{eq-app:async_order_parameter_m_t}
\end{align}
Hence, we obtain the following differential equation in the limit of $\alpha \to 0$:
\begin{align}
	\frac{d m(t)}{d t} =& - m(t) + \frac{1}{N}\sum_i  \int \mathrm{D}z \tanh\br{\beta \pr{\Theta_{i}(t) + J_0 m(t) + \Delta J  z}}.
	\label{eq-app:async_order_parameter_m_cont}
\end{align}
Note that Eqs.~\ref{eq-app:async_order_parameter_m} converges to the same formula given by Eq.~\ref{eq-app:order_parameter_m} when the magnetic fields are equal to $\Theta_{i,u}$.

\subsection*{Delayed self-correlation order parameter}

Next, we compute the order parameter of the delayed self-correlation. We can simplify the decomposition of $q_{i,u,v}$ in two steps as:
\begin{align}
     q_{i,u,v} =&  \sum_{\tau_{i,v}} q_{i,u,v}^{\tau_{i,v}} p(\tau_{i,v}),
    \\ q_{i,u,v}^{\tau_{i,v}}  =& \sum_{\tau_{i,u}} q_{i,u,v}^{\tau_{i,u},\tau_{i,v}} p(\tau_{i,u}),
    \label{eq-app:update_q_async}
\end{align}
where the variable $q_{i,u,v}^{\tau_{i,v}} $ captures the marginal order parameter when we know $\tau_{i,v}$ , i.e., if the spin at time $v$ has been updated or not, regardless of whether the spin at time $u$ has been updated or not. $q_{i,u,v}^{\tau_{i,u},\tau_{i,v}}$ is the order parameter given that we know both the update occurred at $u$ and $v$.

We can decompose $q_{i,u,v}$ into the two cases, where spin $v$ is or is not updated:
\begin{align}
     q_{i,u,v} =& (1-\alpha)  q_{i,u,v}^{0} + \alpha q_{i,u,v}^{1}, 
  \nonumber \\ 
  =&(1-\alpha) q_{i,u,v-1} + \alpha  q_{i,u,v}^{1}.
	 \label{eq-app:async_order_parameter_q}
\end{align}
Here we used the equivalence $q_{i,u,v}^{0}=q_{i,u,v-1}$ under $K \to \infty$ because, if the spin at time $v$ is not updated, it takes the value of the spin at the previous time $v-1$, at which we do not know if the spin is updated or not. We then calculate the other variable $q_{i,u,v}^{1}$ in terms of the updates of $u$ and $v$:
\begin{align}
	 q_{i,u,v}^{1}   =& (1-\alpha)  q_{i,u,v}^{0,1}  + \alpha q_{i,u,v}^{1,1} 
  \nonumber \\ =& (1-\alpha)  q_{i,u-1,v}^{1} + \alpha \int \mathrm{D}xy^{(q_{u-1,v-1})} \tanh\br{\beta \pr{\Theta_{i,u} + J_0 m_{u-1} 
	 + \Delta J  x}} \tanh\br{\beta \pr{\Theta_{i,v}  + J_0 m_{v-1} 
	 + \Delta J y}}.
	 \label{eq-app:async_order_parameter_q*}
\end{align}
Here we used $ q_{i,u,v}^{0,1}=q_{i,u-1,v}^{1}$ for the same reason previously mentioned, applied at time $u$. $q_{i,u,v}^{1,1}$ is the self-delayed correlation $q_{i,u,v}$ as in Eq.~\ref{eq-app:q_i_u_v} with $H_{i,u}=\Theta_{i,u}$ and $H_{i,v}=\Theta_{i,v}$ because the spin was updated at these time steps with correlation between effective fields $q_{u-1,v-1}$, and there are no constraints from the previous spins. 
	 
Similarly to the mean activation rate, in the small $\alpha\to 0$ limit, with  $t' \equiv \alpha (u-1)$, $t \equiv \alpha (v-1)$, the previous equations lead to
\begin{align}
	\frac{d q_{i}(t',t)}{d t} =& - q_i(t',t) +  q_{i}^{1}(t',t),
	 \label{eq-app:async_order_parameter_q_cont1}
	 \\ \frac{d q_{i}^{1}(t',t)}{dt'}=& - q_{i}^{1}(t',t) +   \int \mathrm{D}xy^{(q(t',t))} \tanh\br{\beta \pr{\Theta_{i}(t') + J_0 m(t') 
	 + \Delta J  x}} \tanh\br{\beta \pr{\Theta_{i}(t)  + J_0 m(t)
	 + \Delta J y}},
	 \label{eq-app:async_order_parameter_q_cont2}
\end{align}
The equations above are solved iteratively with boundary conditions:
\begin{align}
	q_{i}(t,t) =& q_{i}^{1}(t,t), = 1,
	\nonumber \\ q_{i}(t',t) =& q_{i}(t,t') ,
	\nonumber \\ q_{i}^{1}(t',t) =& q_{i}^{1}(t,t') ,
	\nonumber \\ \quad t'\geq 0, \, & t\geq 0.
\end{align}
Similar boundary conditions apply for the discrete time description in $u,v$.
The delayed self-correlation order parameter $q(t',t)$ is obtained as the average of $q_{i}(t',t)$ over the spins. Similarly to the mean activation rate, the delayed self-correlation converges to Eq.~\ref{eq-app:order_parameter_q} obtained under the synchronous update.

\subsubsection*{Long-range limit}

Assuming a time-independent $\Theta_{i,u} (=\Theta_{i})$, we can calculate the convergence values of $q_{i,u,v}$ in two steps, separating the dynamics in $u$ and the dynamics in $v$. First, from Eq.~\ref{eq-app:async_order_parameter_q*}, it is easy to see that, knowing $q_{u-1,v-1}$ for all values of $u$ and a fixed $v$, the convergence point for  $u\gg v$ will be
\begin{align}
    q_{i,\infty, v}^{1} \equiv  & \lim_{u \to\infty} q_{i,u,v}^{1}     \nonumber  \\ =  &\lim_{u\to\infty}   \int \mathrm{D}xy^{(q_{u-1,v-1})} \tanh\br{\beta \pr{\Theta_{i} + J_0 m_{u-1} 
	 + \Delta J  x}} \tanh\br{\beta \pr{\Theta_{i}  + J_0 m_{v-1} 
	 + \Delta J y}},
\end{align}
or, in continuous time
\begin{align}
    q_{i}^{1}(\infty, t) \equiv   &\lim_{t'\to\infty}   \int \mathrm{D}xy^{(q_{t',t})} \tanh\br{\beta \pr{\Theta_{i} + J_0 m(t') 
	 + \Delta J  x}} \tanh\br{\beta \pr{\Theta_{i}  + J_0 m(t) 
	 + \Delta J y}}.
\end{align}

Now we can solve the dynamics in $v$ independently to $u$, writing Eq.~\ref{eq-app:async_order_parameter_q} for $u\gg v$ as
\begin{align}
     q_{i,\infty,v} 
  =&(1-\alpha) q_{i,\infty,v-1} + \alpha  q_{i,\infty,v}^{1},
\end{align}
as well as its continuous-time equivalent
\begin{align}
     \frac{dq_{i}(\infty,t)}{dt} 
  =&- q_{i}(\infty,t) + q_{i}^1(\infty,t).
\end{align}

If we assume a starting point $q_{i,\infty,v}$ for a given $v$, then we find that $q_{i,\infty,v}$ converges for a large $v$ (still under $u\gg v$) to
\begin{align}
    q_{i,\infty,\infty} \equiv& \lim_{v \to \infty}  q_{i,\infty,v}
    = \lim_{v \to \infty}  q_{i,\infty,v}^{1} \nonumber\\
    =& \lim_{v \to \infty} \lim_{u \to \infty}  \int \mathrm{D}xy^{(q_{i,u-1,v-1})} \tanh\br{\beta \pr{\Theta_{i} + J_0 m_{u-1} 
	 + \Delta J  x}} \tanh\br{\beta \pr{\Theta_{i}  + J_0 m_{v-1} 
	 + \Delta J y}}.
	 \label{eq-app:async_order_parameter_q_convergence}
\end{align}
Similarly, we define the continuous-time equivalent function
\begin{align}
    q_{i}(\infty,\infty)
    \equiv&  \lim_{t \to \infty} q_{i}(\infty, t)
    = \lim_{t \to \infty} q_{i}^{1}(\infty, t) 
     \nonumber\\
    =& \lim_{t \to \infty} \lim_{t' \to \infty}  \int \mathrm{D}xy^{(q_{i}(t',t))} \tanh\br{\beta \pr{\Theta_{i} + J_0 m(t') 
	 + \Delta J  x}} \tanh\br{\beta \pr{\Theta_{i}  + J_0 m(t) 
	 + \Delta J y}}.
	 \label{eq-app:async_order_parameter_q_convergence_cont}
\end{align}
Note that Eqs.~\ref{eq-app:async_order_parameter_m_cont}, \ref{eq-app:async_order_parameter_q_convergence_cont} converge to the same values as Eqs.~\ref{eq-app:order_parameter_m}, \ref{eq-app:order_parameter_q} when the magnetic fields are equal to $\Theta_{i}$, proving that the synchronous and asynchronous asymmetric SK models have identical solutions for their order parameters.

\begin{figure}
    \centering
    \includegraphics[width=9cm]{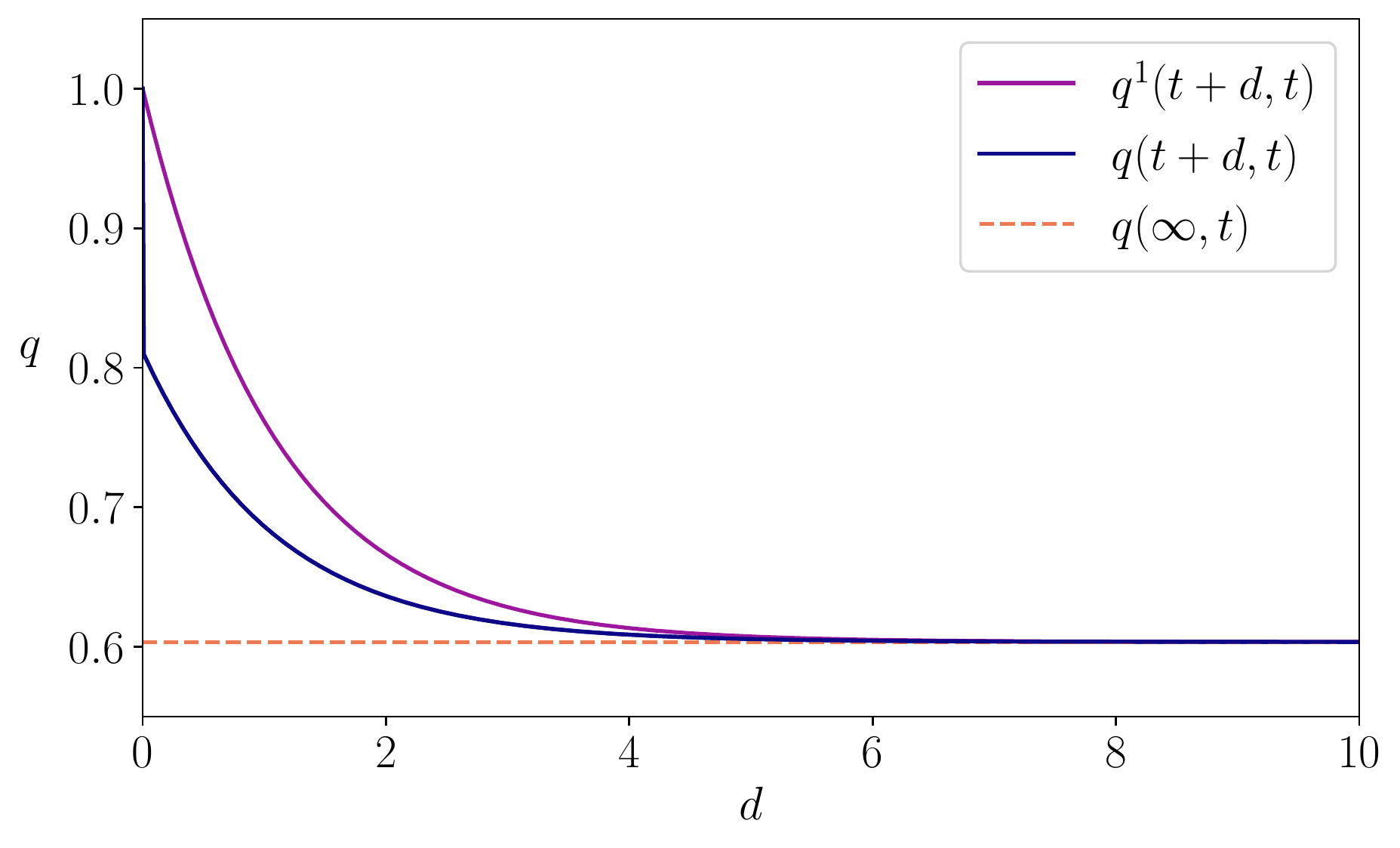}
    \caption{Result of simulating Eq.~\ref{eq-app:async_order_parameter_q_cont1},\ref{eq-app:async_order_parameter_q_cont2} to calculate $q(t+d,t)$ (blue line) and $q^{1}(t+d,t)$ (purple line)  for $0 \leq d \leq 10$ with $\alpha=0.01$, $\beta=2$, $\Theta_{i,u}=0$, $J_0=1$, and $\Delta J = 0.5$. We also calculated the long range limit using Eq.~\ref{eq-app:async_order_parameter_q_convergence_cont} (dashed orange line). We observe that both variables converge for large $d$ to the result given by Eq.~\ref{eq-app:async_order_parameter_q_convergence_cont}. In addition, we find that while $q^{1}(t+d,t)$ decays smoothly, $q(t+d,t)$ has a discontinuity at $d=0$ since $q(t,t)=1$.}
    \label{fig:q_decay}
\end{figure}

In Fig.~\ref{fig:q_decay}, we numerically confirmed our theoretical result by simulating an exemplary dynamics of the self-correlation order parameters in the continuous-time domain. The procedure is as follows. We select an Euler step of $\alpha=0.01$ and an initial values of $q_{i}(t',0)=  \delta\br{d}$. Given $q_{i}^{1}(0,0)=1$, we computed a forward pass of the values of $q_i^{1}(d,0)$ for larger values of $d$ ($d \geq 0$), using Eq.~\ref{eq-app:async_order_parameter_q_cont2}. 
Then, given $t'=t+d$ we calculated one Euler step of Eq.~\ref{eq-app:async_order_parameter_q_cont1} in $t$, updating the value of $q_i(t+d,t)$ for all values of $d$. We iterated the above procedure until the function $q_i(t+d,t)$ and $q_i^{1}(t+d,t)$ converge to fixed values. We confirmed that the convergence value of the process was the same one as directly calculating $q_i(\infty,\infty)$.

\subsection*{Entropy production}

Assume that time-constant fields $\Theta_i$, i.e., the fields $H_{i,u} = \Theta_i + (1-\tau_{i,u})K s_{i,u-1}$ are stochastic processes driven by time-independent $\tau_{i,u}$. In this case the average difference of the forward and reverse entropy rates in Eqs.~\ref{eq-app:entropy-rate}, \ref{eq-app:entropy-rate-reverse} converges to:
\begin{align}
	\br{\sigma_{u}}_{\*J,\^\tau}  =& \sum_{\^\tau_{u},\^\tau_{u-1}} p(\^\tau_{u})p(\^\tau_{u-1}) \pr{S_{u|u-1} -S_{u|u-1}^r }
	\nonumber \\=& \sum_{\^\tau_{u}} p(\^\tau_{u}) \pr{  \beta^2 \Delta J^2 (1-q_{u,u-2}) \sum_i \int  \mathrm{D}z(1-\tanh^2\br{\beta \pr{H_{i,u}+ J_0 m_{u-1}  + \Delta J  z}}) },
\end{align}
for the steady-state values of $m_{u-1}$ and $q_{u,u-2}$ (independent of $u$).

For asynchronous updates in a steady state, non-zero contributions to the entropy production at spin $i$ only occur during spin updates, i.e., $\tau_{i,u} =1$ occurring with a probability $\alpha$ (we can see this in the equation above where the $\tanh^2\br{\beta \pr{H_{i,u} + J_0 m  + \Delta J  z}}$ term becomes equal to 1 when $\tau_{i,u} =0$).
This leads to the steady-state entropy production value of 
\begin{align}
	\br{\sigma_{u}}_{\*J,\^\tau}  =&  \alpha \beta^2 \Delta J^2 (1-q_{u,u-2} ) \sum_i \int  \mathrm{D}z(1-\tanh^2\br{\beta \pr{\Theta_{i} + J_0 m_u  + \Delta J  z}}) 
	 \label{eq-app:entropy-production_async}.
\end{align}
In the continuous-time limit $\alpha \to 0$, with a change of variables $t=(u-1)\alpha$, we have the entropy production rate defined in continuous time: 
\begin{align}
	\br{\frac{d\sigma(t)}{dt}}_{\*J,\^\tau}  =& \lim_{\alpha \to 0} \beta^2 \Delta J^2 (1-q(t+\alpha,t-\alpha) ) \sum_i \int  \mathrm{D}z(1-\tanh^2\br{\beta \pr{\Theta_{i} + J_0 m(t)  + \Delta J  z}}).
\end{align}
Note that due to the discontinuity of $q(t+d,t)$ (or $q(t+d,t-d)$ equivalently), the limit in $d\to0$ is different from $q(t,t)=1$ (see Fig.~\ref{fig:q_decay}). This property, $\lim_{d\to 0} q(t+d,t-d)<1$, guarantees that the entropy production rate can be non-zero for the adequate parameters.

\newpage
\section{Ferromagnetic critical phase transition in the infinite kinetic Ising model with Gaussian couplings and uniform weights}
\label{app:phase_transitions}
We define a kinetic Ising network of infinite size under synchronous updates ($\alpha=1$), with random fields $H_{i,u}=H_i$, where $H_i$ are uniformly distributed following $\mathcal{U}(-\Delta H ,\Delta H)$, and couplings $J_{ij}$ follow a Gaussian distribution $\mathcal{N}(\frac{1}{N}, \frac{\Delta J^2}{N})$.

As we have found that the asymmetric SK model with arbitrary fields follows a mean-field solution, calculating the effects of disorder in the fields becomes easier, as we can approximate the update equations of the order parameters in the thermodynamic limit $N\to\infty$ as an integral with a large number of units:
\begin{align}
    m_u = \frac{1}{N}\sum_i  m_{i,u} =& \frac{1}{2\Delta H}\int_{-\Delta H}^{\Delta H} dh \int \mathrm{D}z \tanh\br{\beta\pr{h  + J_0  m_{u-1} + \Delta J z }} 
  \nonumber \\ =& \frac{1}{2\beta \Delta H} \int \mathrm{D}z \log \frac{\cosh\br{\beta\pr{\Delta H + J_0  m_{u-1} + \Delta J z} }}{\cosh\br{\beta\pr{-\Delta H + J_0  m_{u-1} + \Delta J z} }}.
    \label{eq:mean-activation-order-parameter}
\end{align}

Similarly, the delayed self-correlation parameter:
\begin{align}
    q_{u,v} = \frac{1}{N} \sum_i R_{ii,u,v} =&  \frac{1}{2\Delta H} \int_{-\Delta H}^{\Delta H} dh  \int \mathrm{D}xy^{(q_{u-1,v-1})} \tanh\br{\beta\pr{h + J_0  m_{u-1} + \Delta J x}} \tanh\br{\beta\pr{h + J_0  m_{v-1} + \Delta J y}}
    \nonumber\\ =&   1 + \frac{1}{2\beta \Delta H}   \int \mathrm{D}xy^{(q_{u-1,v-1})} \dfrac{\left(\mathrm{e}^{2H_{v,y}{\beta}}+\mathrm{e}^{2\beta H_{u,x}}\right)}{\left(\mathrm{e}^{2\beta H_{v,y}}-\mathrm{e}^{2\beta H_{u,x}}\right)} \log\br{
    \frac{\mathrm{e}^{-2\beta \Delta H}+\mathrm{e}^{2\beta H_{u,x}}}{\mathrm{e}^{2\beta \Delta H}+\mathrm{e}^{2\beta H_{u,x}}}  
    \frac{\mathrm{e}^{2{\beta}\Delta H}+\mathrm{e}^{2\beta H_{v,y}}}{\mathrm{e}^{-2{\beta}\Delta H}+\mathrm{e}^{2\beta H_{v,y}}}},
    \label{eq:correlation-order-parameter}
\end{align}
with $H_{u,x} = J_0  m_{u-1} + \Delta J x$ and $H_{v,y}  = J_0  m_{v-1} + \Delta J y$.

In the zero-temperature limit, $\beta\to\infty$, these expressions have the following forms:
\begin{align}
    m_u  =& \frac{1}{2\beta \Delta H} \int \mathrm{D}z \pr{\abs{\Delta H + J_0  m_{u-1} + \Delta J z}  -\abs{-\Delta H + J_0  m_{u-1} + \Delta J z} },
    \label{eq:mean-activation-order-parameter-T0}
\end{align}
\begin{align}
    q_{u,v} =&    1 + \frac{1}{2\beta \Delta H}   \int \mathrm{D}xy^{(q_{u-1,v-1})}  \mathrm{sign}\br{H_{v,y}-H_{u,x}} \beta \Bigg( \abs{\Delta H+ H_{u,x}} -  \abs{\Delta H - H_{u,x}} - \abs{\Delta H+ H_{v,y}} +  \abs{\Delta H - H_{v,y}} \Bigg).
    \label{eq:correlation-order-parameter-T0}
\end{align}

Finally, the normalized conditional entropy in the thermodynamic limit and normalized reversed conditional entropy are given as
\begin{align}
     \br{\frac{1}{N} S_{u|u-1}}_{\*J}  =& \frac{1}{2\Delta H}\int_{-\Delta H}^{\Delta H} dh \int \mathrm{D}z \Big(
     \beta^2 \Delta J^2  (1-\tanh^2\br{\beta\pr{h  + J_0  m_{u-1} + \Delta J z }}) 
	\nonumber\\ & + \beta h \tanh\br{\beta\pr{h  + J_0  m_{u-1} + \Delta J z }}  - \log\br{ 2 \cosh \br{\beta \pr{h  + J_0  m_{u-1} + \Delta J z }}}  \Big) + \beta J_0 m_u m_{u-1}
  \nonumber \\ =& \frac{1}{2 \beta \Delta H} \int \mathrm{D}z  \Big(\beta^2 \Delta J^2\tanh\br{ \beta\pr{\Delta H + J_0  m_{u-1} + \Delta J z}}-\tanh\br{ \beta\pr{-\Delta H + J_0  m_{u-1} + \Delta J z}})
    \nonumber \\ & +\frac{1}{2 \beta \Delta H}\pr{ \varphi\pr{\beta \Delta H,\beta  J_0  m_{u-1} +\beta  \Delta J z }) - \varphi\pr{-\beta \Delta H, \beta J_0  m_{u-1} +\beta  \Delta J z }}  \Big) + \beta J_0 m_u m_{u-1},
    \label{eq:rate-ent-result}
\end{align}
and
\begin{align}
     \frac{1}{N} \br{S_{u|u-1}^r }_{\*J}=& \frac{1}{2\Delta H}\int_{-\Delta H}^{\Delta H} dh \int \mathrm{D}z \Big( q_{u,u-2}  \beta^2 \Delta J^2  (1-\tanh^2\br{\beta \pr{h  + J_0  m_{u-2} + \Delta J z }})
	\nonumber\\ & - \log\br{2 \cosh \br{\beta \pr{h  + J_0  m_{u} + \Delta J z }}}  \Big)
     \Bigg)+ \beta J_0 m_{u-1}m_{u+1}
    \nonumber\\ =& \frac{1}{2 \beta \Delta H} \int \mathrm{D}z  \Big( \beta^2 \Delta J^2 q_{u,u-2}  (\tanh\br{ \beta\pr{\Delta H + J_0  m_{u-2} + \Delta J z}}-\tanh\br{ \beta\pr{-\Delta H + J_0  m_{u-2} + \Delta J z}}) 
     \nonumber\\ &+\frac{1}{2 \beta \Delta H}\pr{  \varphi\pr{\beta \Delta H, \beta J_0  m_{u} + \beta \Delta J z }) -  \varphi\pr{-\beta \Delta H,\beta  J_0  m_{u} + \beta \Delta J z }}  \Big) + \beta J_0 m_{u-1} m_{u+1},
    \label{eq:rate-ent-rev-result}
\end{align}
where we define
\begin{align}
     \varphi\pr{h, w }  = h\log\br{1+\exp\br{2 h+2w}}+ \mathrm{Li}_2\br{-\exp\br{2 h+2w}}+ h w
\end{align}
with $\mathrm{Li}_s\br{x}$ being the polylogarithm function.

\subsection{Critical points}

Assuming a nonequilibrium steady state in which $m_u=m_{u-1}=m$, we obtain the critical point of the system by computing the non-zero solutions of the first order Taylor expansion around $m=0$ of the right-hand part of Eq.~\ref{eq:mean-activation-order-parameter},
\begin{align}
    m \approx& \frac{1}{2\beta \Delta H} \int \mathrm{D}z \log \frac{\cosh\br{\beta\pr{\Delta H + \Delta J z} }}{\cosh\br{\beta\pr{-\Delta H + \Delta J z} }}
    \nonumber\\ &+\frac{1}{2\beta \Delta H} \int \mathrm{D}z \pr{\tanh\br{\beta\pr{\Delta H  + \Delta J z }} - \tanh\br{\beta\pr{-\Delta H  + \Delta J z }}} \beta J_0 m
    \nonumber\\ =& \frac{1}{ \Delta H} \int \mathrm{D}z \tanh\br{\beta\pr{\Delta H  + \Delta J z }}  J_0 m.
\end{align}
This equation yields the self-consistent equation whose solution gives the critical inverse temperature, $\beta_c$:
\begin{align}
    \frac{ \Delta H}{J_0} = \int \mathrm{D}z \tanh\br{\beta \pr{\Delta H  + \Delta J z }}.
    \label{eq:self_consistent_eq_for_beta}
\end{align}

In the special case where $\Delta H=0$, the expansion around $m=0$ results in 
\begin{align}
    m \approx&  \int \mathrm{D}z \tanh\br{\beta\pr{ \Delta J z} }
    +  \int \mathrm{D}z \pr{1-\tanh^2\br{\beta\pr{ \Delta J z }} } \beta J_0 m
  \nonumber \\ =& \int \mathrm{D}z \pr{1-\tanh^2\br{\beta\pr{ \Delta J z }} } \beta J_0 m.
\end{align}
The critical value $\beta_c$ is given by the solution of the equation, 
\begin{align}
    \frac{1}{ \beta J_0} = \int \mathrm{D}z \pr{1-\tanh^2\br{\beta\pr{ \Delta J z }} }.
    \label{eq:self_consistent_eq_for_beta_H=0}
\end{align}

Similarly, we can find the critical value of $\Delta J$ at the limit of zero temperature by solving the equation in the $\beta\to\infty$ limit: 
\begin{align}
    \frac{1}{ \Delta H} \int \mathrm{D}z \, \mathrm{sign}\br{\pr{\Delta H  + \Delta J z }}  J_0 = 1.
    \label{eq:self_consistent_eq_at_zero_temp}
\end{align}

\subsection{Critical exponents}
We can characterize  critical exponents of the system using the normalized inverse temperature $\tau=-\frac{\beta - \beta_c}{\beta_c}$. We first note that the first order Taylor expansion of the following term around the critical $\beta_c$ yields
\begin{align}
    \frac{1}{ \Delta H} \int \mathrm{D}z \tanh\br{\beta\pr{\Delta H  + \Delta J z }}  J_0  \approx& 1 +  \frac{1}{ \Delta H} \int \mathrm{D}z \pr{1 - \tanh^2 \pr{\beta_c\pr{\Delta H  + \Delta J z }} } \pr{\Delta H  + \Delta J z } J_0 \pr{\beta - \beta_c}
    \nonumber\\ =& 1 - K' \pr{\beta - \beta_c}.
\end{align}
We also note that the value of $m$ around $\beta=\beta_c$ with the third order Taylor expansion is given as
\begin{align}
    m \approx &   \frac{1}{\Delta H} \int \mathrm{D}z \tanh\br{\beta\pr{\Delta H  + \Delta J z }} J_0 m 
    \nonumber\\-&  \frac{1}{3 \beta  \Delta H} \int \mathrm{D}z \tanh\br{\beta\pr{\Delta H + \Delta J z }} \brc{1-\tanh^2 \br{\beta\pr{\Delta H  + \Delta J z }}} \pr{\beta J_0 m}^3
  \nonumber \\ =& (1 - K' \pr{\beta - \beta_c})m - K'' m^3,
\end{align}
from which we obtain
\begin{align}
m \propto& \pr{\beta - \beta_c}^{\frac{1}{2}}.
\end{align}
Thus we have a critical exponent $\frac{1}{2}$, which is consistent with the scaling exponent of the order parameter of the mean-field universality class typically denoted by the symbol `$\beta$' in the literature.

Similarly, we can compute the susceptibility to a uniform magnetic field $B$ added on top of $H_i$, having that
\begin{align}
    \frac{\partial m}{\partial B}\Big\rvert_{B=0} =& \frac{1}{2\beta \Delta H} \int \mathrm{D}z \brc{ \tanh\br{\beta\pr{\Delta H  + \Delta J z }} - \tanh\br{\beta\pr{-\Delta H  + \Delta J z }} } \pr{\beta + \beta J_0  \frac{\partial m}{\partial B}\Big\rvert_{B=0}  }
    \nonumber\\ =& \frac{1}{ \Delta H} \int \mathrm{D}z \brc{ \tanh\br{\beta\pr{\Delta H  + \Delta J z }} } \pr{1+ J_0  \frac{\partial m}{\partial B}\Big\rvert_{B=0}  },
\end{align}
which evaluated at the limit $\tau\to 0$ results in
\begin{align}
    \frac{\partial m}{\partial B}\Big\rvert_{B=0}  =& \pr{1 - K' \tau} \pr{\frac{1}{J_0} + \frac{\partial m}{\partial B} },
    \\ \frac{\partial m}{\partial B}\Big\rvert_{B=0}  \propto & \frac{1 - K' \tau}{\tau} \approx \pr{-\tau}^{-1},
\end{align}
retrieving the $\gamma =1$ exponent that is consistent with the mean-field universality class.

Note that these critical exponents, corresponding to the mean-field universality class, are also the same found in the order-disorder phase transition of the equilibrium SK model \cite{nishimori_statistical_2001}. Note that the spin-glass phase, not present for asymmetric couplings, has different exponents and does not correspond to this universality class \cite{oppermann2008universality}).

\newpage
\section{Comparison with the equilibrium SK model}
\label{app:equilibrium-SK}

\begin{figure*}[ht]
\begin{center}
\begin{tabular}{ll}
 (a)&  (b)  \\
  \includegraphics[width=7.5cm]{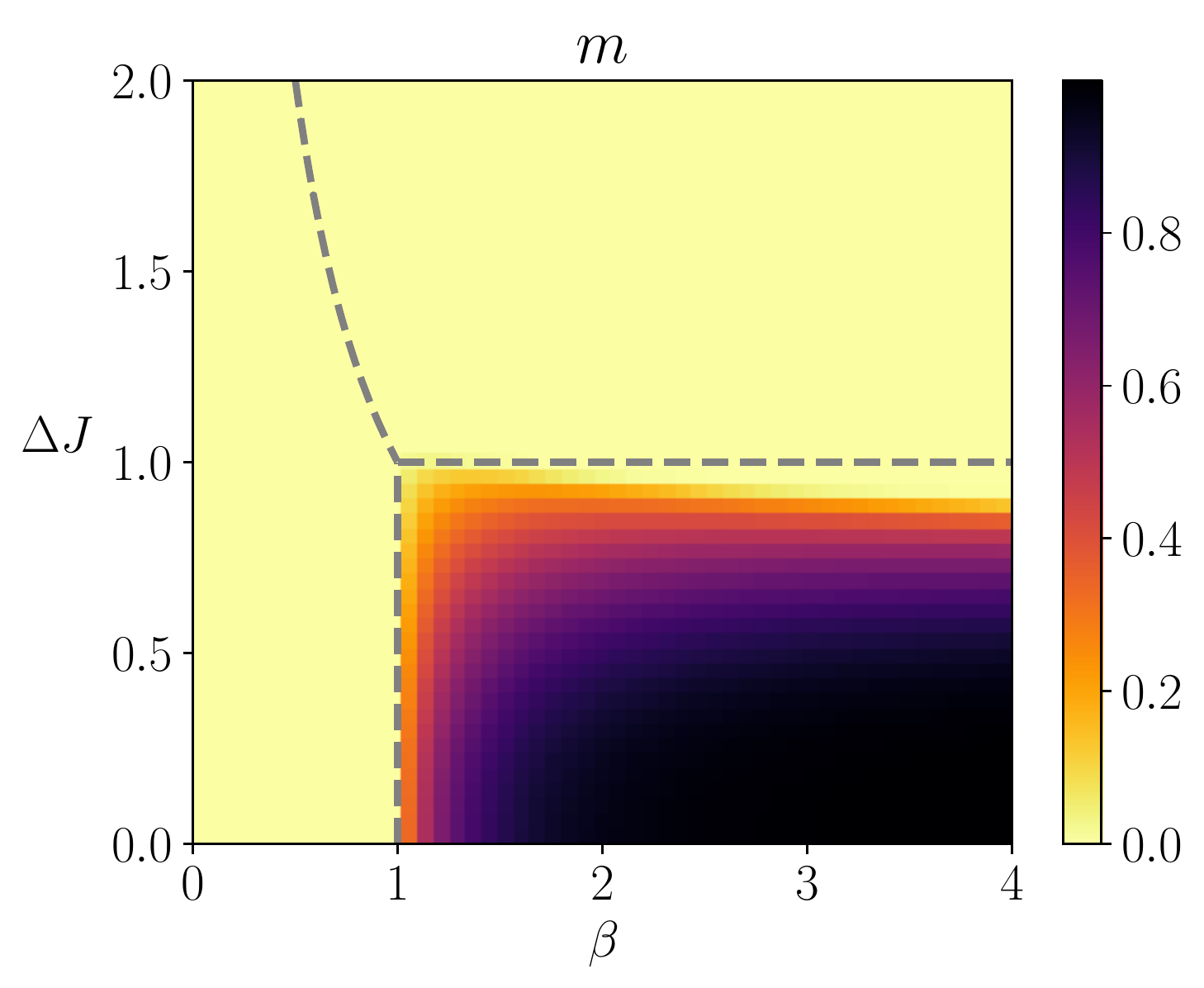} &
  \includegraphics[width=7.5cm]{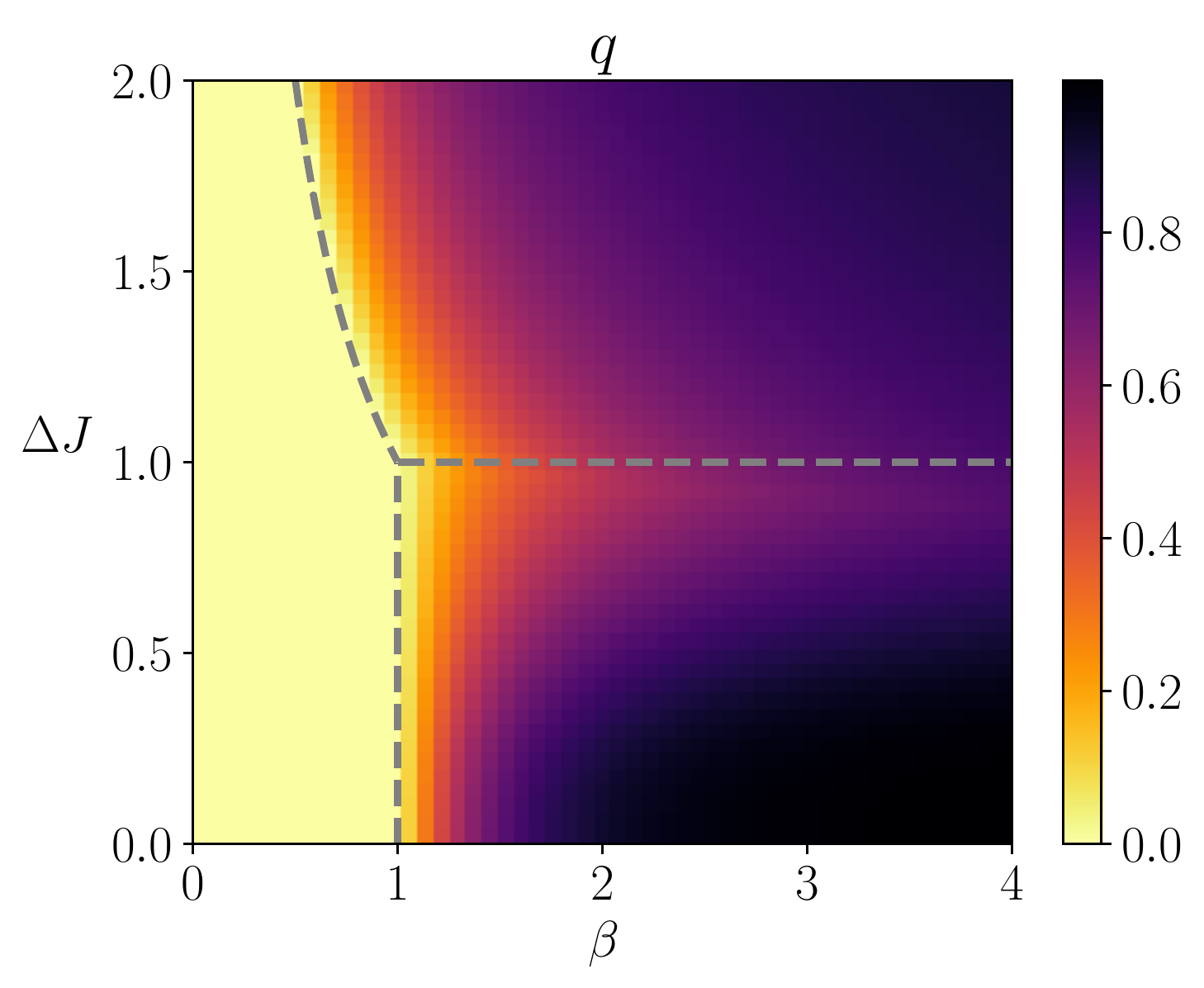} 
\end{tabular}
\end{center}
\caption{\textbf{Order parameters of the equilibrium SK model with zero fields}. An approximate solution of the model with symmetric couplings is calculated under the replica-symmetry assumption \cite{nishimori_statistical_2001}. The average magnetization $m$ and the average delayed self-coupling $q$ are shown for a model with fixed parameters $J_0=1$, $\Delta H=0$ and variable $\Delta J$ and $\beta$. The dashed line represents the critical line separating disordered (left), ordered (bottom-right) and spin-glass (top-right) phases.
} 
\label{fig:order-parameters-equilibrium}
\end{figure*}

To illustrate distinct behaviors between the symmetric and asymmetric SK models, we compare the order parameters of the asymmetric SK model with those of its equilibrium counterparts. We use the replica-symmetric solution of the model \cite{nishimori_statistical_2001}, which becomes unstable for the spin-glass phase but still yields an approximate phase diagram of the system. Figure \ref{fig:order-parameters-equilibrium} displays the phase diagram of the order parameters of an equilibrium SK model, which is equivalent to that  of the nonequilibrium SK model in the main text shown in Fig.~\ref{fig:order-parameters-DeltaJ}. 

\newpage
\section{Convergence times}
\label{app:convergence-time}

\begin{figure}[ht]
\begin{center}
\begin{tabular}{ll}
 (a) & (b)\\
  \includegraphics[width=7.5cm]{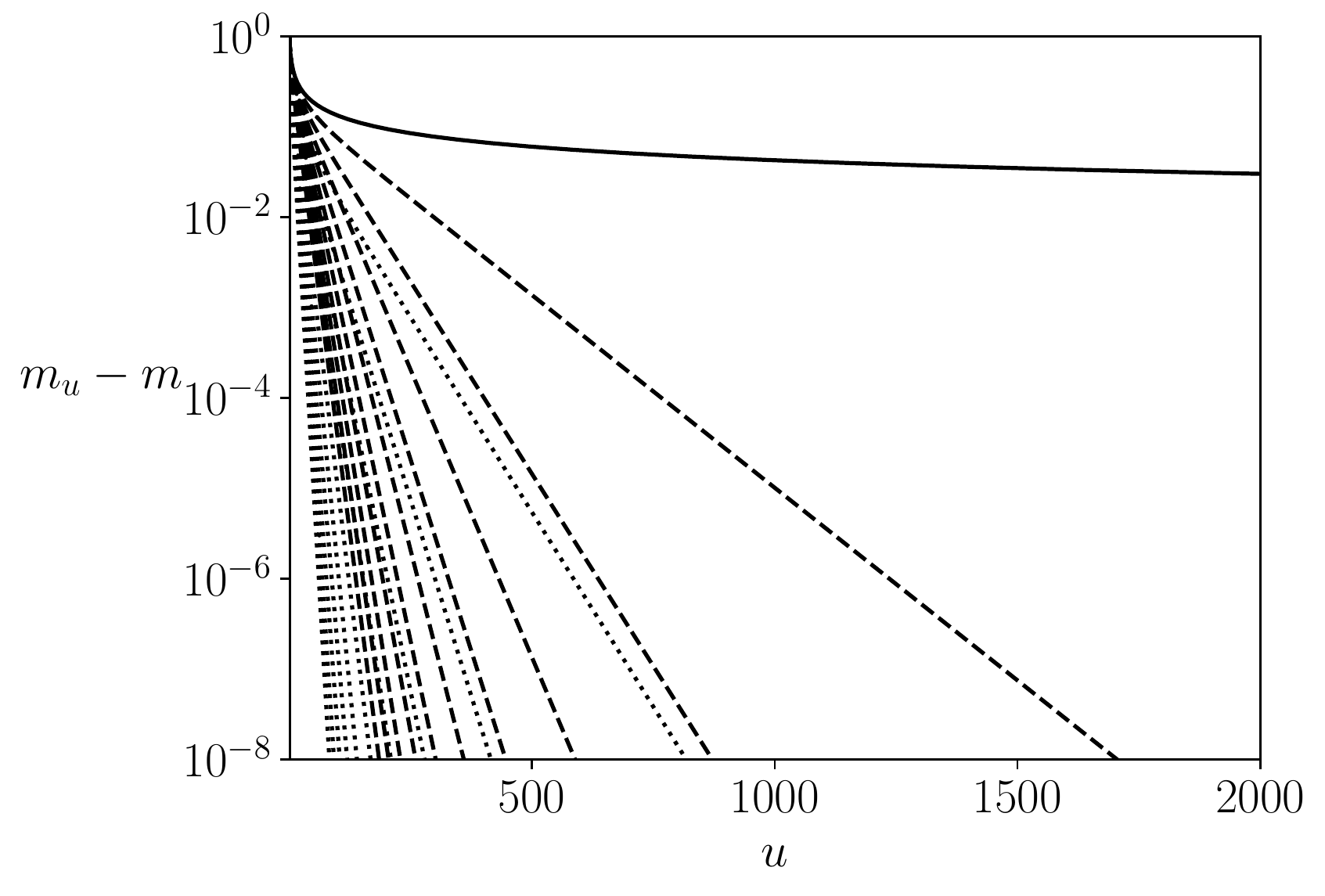}&
  \includegraphics[width=7.5cm]{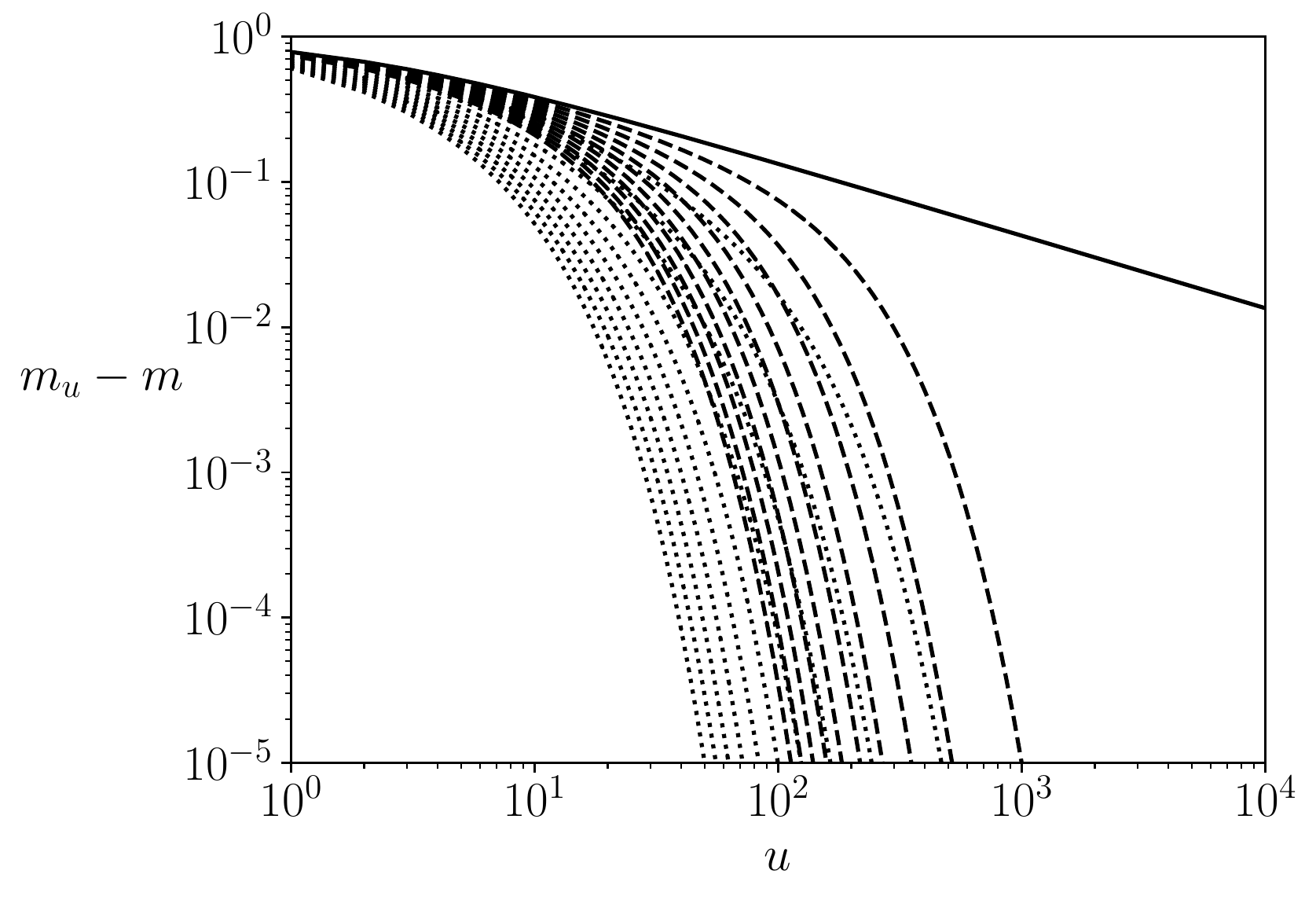}
\end{tabular}
\end{center}
\caption{\textbf{Convergence time}. Convergence times of the average-rate order parameter $m_u$ to its stationary state value $m$ at the critical point (black), ordered phase (dotted line) and disordered phase (dashed line) for a semilog (a) and log-log scales (b). All lines show an exponential decay, except for the system at criticality, which shows a power-law decay. We can know that the disordered phase is not a spin-glass phase due to the exponential decay. 
} 
\label{fig:convergence-time}
\end{figure}

Spin glasses show a particular slow decay functions, which converge non-exponentially following a non-trivial slow function \cite{joy1998relationship}. This finding is replicated in models like the equilibrium SK model \cite{cugliandolo1994evidence,sompolinsky1981dynamic}.
To refute the existence of a spin-glass phase with such slow non-exponential decay, we simulated the convergence of the average magnetization as the dynamics reaches a nonequilibrium steady state. Using the critical inverse temperature $\beta_c$ for $\Delta J=0.2, \Delta H =0$, we use 11 values of $\Delta J$ uniformly distributed in the interval $[0.19,0.21]$. In Fig.~\ref{fig:convergence-time}, we observe, at the critical value ($\Delta J =0.2$, black line), the convergence of magnetization follows a power law, as expected. Conversely, both the ordered and disordered phases ($\Delta J<0.2$, dotted line, $\Delta J>0.2$, dashed line) converge as exponential functions. The figure is calculated for $\alpha=1$, but the behavior is similar for all $\alpha$. These results confirm that the disordered phase is not a spin-glass phase, as spin glasses show a non-exponential slow decay characterized by a non-trivial function.

\newpage

\section{Numerical simulations}

To verify our theoretical solutions of the order parameters and entropy production obtained under the configurational average, we ran numerical simulations of the kinetic Ising systems with synchronous and asynchronous updates for many random realizations of the model parameters. Here we explain how we performed the numerical simulations and calculated the statistics from the sample trajectories.

First, we constructed a fully asymmetric kinetic Ising system of size $N$ with random couplings. The elements $J_{ij}$ of the coupling matrix $\*J$ were randomly sampled from independent Gaussian distributions, $J_{ij} \sim \mathcal{N}\pr{\frac{J_0}{N},\frac{\Delta J}{N}}$. We simulated trajectories of $t=2^7$ steps for synchronous updates and $T=2^7 N$ steps for asynchronous updates (see in the following) for $101$ values of the inverse temperature $\beta$ in the range $[0,4]$. For each system with size $N$, we run the simulations using $R=400,000$ random configurations of the matrix $\*J$, and we further repeated the process for different system sizes.

In the simulation, we updated each spin in accordance to the transition probability:
\begin{align}
    p(s_{i,u}\,|\,\*{s}_{u-1})=&  \frac{\exp\br{ \beta s_{i,u} h_{i,u}}}{2 \cosh\br{\beta h_{i,u}}},
    \\ h_{i,u}=& \Theta_i+\sum_j J_{ij} s_{j,u-1}.
\end{align}
For the Ising systems with synchronous updates, we simultaneously updated all spins at each time step, which is equivalent to setting $\alpha=1$ in Eq.~\ref{eq:Ising2}.
For asynchronous updates, we randomly selected a single spin at each time step and updated the selected spin using the above equation to capture the behavior of the system in the continuous time limit with $\alpha\to 0$ and $K \to \infty$. In this time limit, updates will be infrequent and only one spin is updated at a time. That is, most of the time $\tau_{i,u}=0$. However, for computational efficiency we only simulated the steps where a spin is updated with $\tau_{i,u}=1$ (a random event happening with probability $N\alpha$). Thus, $t$ steps of the system with synchronous updates
in Eq.~\ref{eq:Ising2} with $\alpha \to 0$ corresponds to $T$ steps in our simulation, where $T$ is a stochastic variable corresponding to a Binomial distribution $B(Nt,\alpha)$ (which in the limit $\alpha\to 0$ is equivalent to a Poisson distribution $\mathrm{Pois}(Nt\alpha)$).
To make the behavior of the synchronous and asynchronous system have equivalent speeds, we choose to use $t=2^7$ for the synchronous system, and $T=2^7N$ for the asynchronous systems, approximately corresponding to $t=2^7/\alpha$. This guarantees that we have a total of $2^7N$ individual spin updates for both systems.

Next, we computed the order parameters and entropy production at a steady state from the sample trajectories as follows. We calculated the steady-state average activation rate from the last time step $t$ of the sample trajectories as: 
\begin{align}
    \widehat m = \frac{1}{N}\sum_i \br{\ang{s_{i,t}}}_{\*J,\^\tau}, 
\end{align}
where $\ang{\cdot}_{\*J} $ is an average over the $R$ configurations of $\*J$. The last time step $t$ is $t=2^7$ for the synchronous updates and $\frac{1}{\alpha}2^7$ for asynchronous updates (corresponding to $2^7N$ updates in our simulation) and large enough to make the systems reach the steady state. 

We computed the steady-state delayed self-correlation from the samples in a similar way. For the synchronous Ising systems, it was computed as:
\begin{align}
    \widehat q = \frac{1}{N}\sum_i \ang{s_{i,t}s_{i,t-1}}_{\*J}, 
\end{align}
with $\alpha = 1$ whereas in the asynchronous Ising model we used
\begin{align}
    \widehat q = \frac{1}{N}\sum_i \br{\ang{s_{i,t}s_{i,t-d}}}_{\*J,\^\tau} 
\end{align}
with $d=10/\alpha$ (with $t-d$ corresponding to the point  $T-10N$ in our simulation) 
to obtain the delayed correlation of the spin states between two distant time points (i.e., a long-range correlation).

Finally, we calculated the steady-state entropy production from the samples using
\begin{align}
	\widehat \sigma_{t}
	=& \alpha
	 \br{\ang{ \sum_i\log \frac{p(s_{i,t}\,|\,\*s_{t-1})}{p(s_{i,t-1}\,|\,\*s_{t})}}}_{\*J,\^\tau}
	\nonumber\\
	=&
	\alpha \left[\left\langle \sum_i \Theta_{i}  (s_{i,t} - s_{i,t-1}) 
	+ \sum_{ij} J_{ij} (s_{i,t}s_{j,t-1} - s_{i,t-1}s_{j,t}) \right.\right.
	\nonumber\\ & \left.\left. -  \log (2 \cosh (\Theta_{i}   + \sum_j J_{ij} s_{j,t-1} ))
	+ \log ( 2 \cosh (\Theta_{i}    + \sum_j J_{ij} s_{j,t} ) ) \right\rangle\right]_{\*J,\^\tau}.
	\label{eq:entropy_production_sample}
\end{align}
The steady-state entropy production of the synchronous update is obtained by setting $\alpha = 1$ and effectively removing the average over $\^\tau$. Note in our results we normalize this entropy production by the number of spins $N$ to make it independent of the system size.

\begin{figure*}[t]
\begin{center}
\includegraphics[width=18cm]{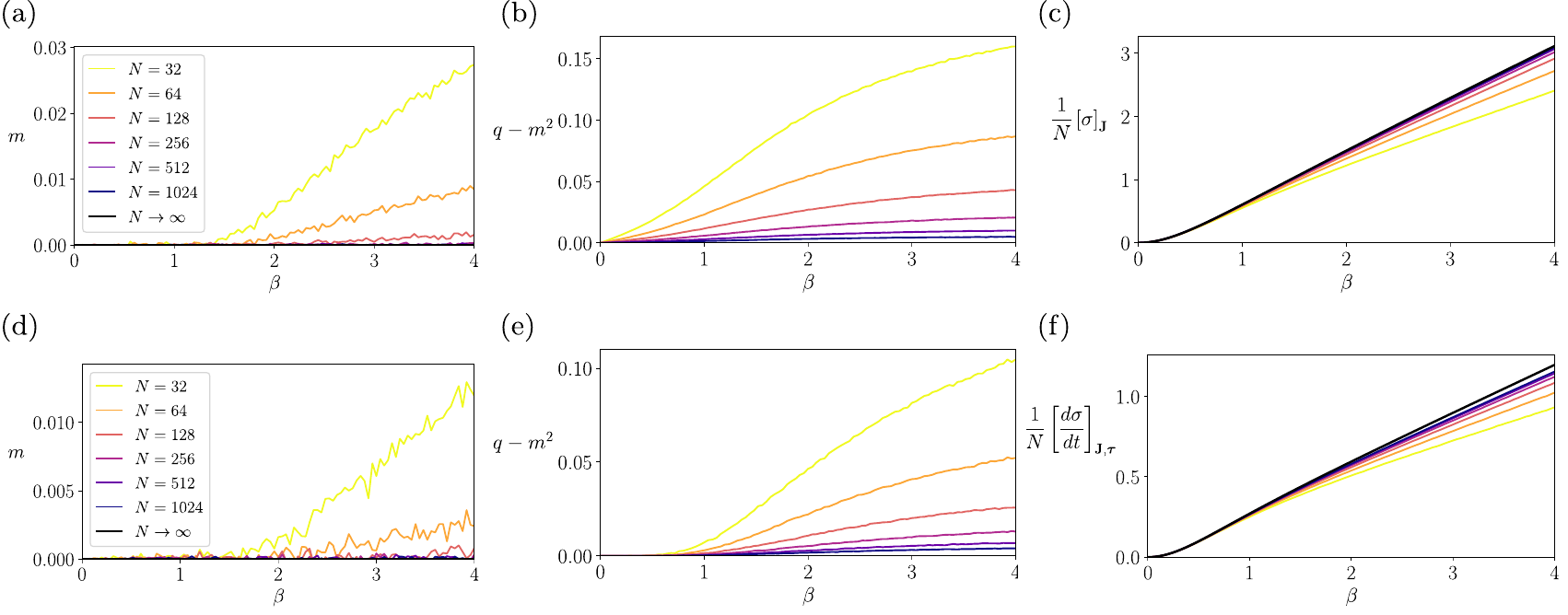}   
\end{center}
\caption{\textbf{Verification of the exact mean-field solutions by simulating the kinetic Ising systems}. We repeated the simulations for systems of size $N = 32, 64, 128, 256, 512, 1024$ with synchronous (top) and asynchronous (bottom) updates with $\Theta_{i,u}=0$ and $\Delta J =1$.
(a,d) Sampling estimation of the mean activation rate $\widehat m$ compared with the theoretical order parameter $m$ (black lines).   
(b,e) Sampling estimation of the average delayed self-correlations $\widehat q - \widehat m^2$ compared with the theoretical order parameter $q-m^2$ (black lines). 
(c,f) Sampling estimation of the entropy production and entropy production rate $\widehat \sigma, \lim_{\alpha\to 0}  \frac{1}{\alpha}\widehat \sigma$ computed from the sample trajectories (Eq.~\ref{eq:entropy_production_sample},~\ref{eq:entropy-production-rate_sample}) compared with its mean-field value at the thermodynamic limit $\frac{1}{N}\br{\sigma}_{\*J},\frac{1}{N}\br{\frac{d\sigma}{dt}}_{\*J,\^\tau}$ (black lines, Eqs.~\ref{eq:sigma_mean-field},\ref{eq:sigma-rate_mean-field}). 
} 
\label{fig:simulation2}
\end{figure*}

The steady-state entropy rate for the asynchronous updates is given by $\lim_{\alpha\to 0} \widehat \sigma_{t} / \alpha$, to make it independent of the update rate $\alpha$:
\begin{align}
	\lim_{\alpha\to 0}  \frac{1}{\alpha}\widehat \sigma_{t}
	=&
	\left[\left\langle \sum_i \Theta_{i}  (s_{i,t} - s_{i,t-1}) 
	+ \sum_{ij} J_{ij} (s_{i,t}s_{j,t-1} - s_{i,t-1}s_{j,t}) \right.\right.
	\nonumber\\ & \left.\left. -  \log (2 \cosh (\Theta_{i}   + \sum_j J_{ij} s_{j,t-1} ))
	+ \log ( 2 \cosh (\Theta_{i}    + \sum_j J_{ij} s_{j,t} ) ) \right\rangle\right]_{\*J,\^\tau}.
	\label{eq:entropy-production-rate_sample}
\end{align}
Similarly to the synchronous case,  we normalize this entropy production rate by the number of spins $N$ to make it independent of the system size.

To verify our exact mean-field solutions, we simulated networks of different sizes with synchronous and asynchronous updates for the parameters $\Theta_{i,u}=0$ and $\Delta J =0.5$ (Fig.~\ref{fig:simulation}) or  $\Delta J =1$ (Fig.~\ref{fig:simulation2}). These results corroborate our theoretical predictions and confirm that the steady-state entropy production peaks at the phase transitions and increases when the significantly heterogeneous system approaches the quasi-deterministic regimes.


\newpage
\section{Dynamic patterns}
\label{app:dynamic-patterns}

To characterize the complexity of the spin kinetics, we count the number of states that returns to themselves after a number of transition steps with different lengths for a large but finite size system under steady state. 
We describe this probability as
\begin{align}
    \Omega^{(n)} = p(\*s_{u}=\*s_{u+n},  \*s_{u} \neq \*s_{u+n-1}, \*s_{u} \neq \*s_{u+n-2},\, \dots).
\end{align}
We will assume the synchronous Ising system with any $n$, or the asynchronous Ising system with large $n$, so the order parameters converge to their long-range values $m,q$.

Given this probability distribution, we can calculate the average transition length as
\begin{align}
    \sum_n n \Omega^{(n)}.
\end{align}
A transition length of 1 results in a static pattern, where the same pattern appears consecutively. A transition length longer than 1 indicates that the dynamics exhibits a cyclic pattern on average. For simplicity, we assume that there are no fields $\Theta_{i,u}=0$. Also, under steady state for the configurational average, the self-correlations are uniform across elements, having $\br{\ang{s_{i,u}s_{i,v}}}  =q$.

\subsection{The proportion of patterns with length $n$} 

The proportion for the patterns with an arbitrary length $n$ is obtained as follows. First, we note that by using a Kronecker delta $\delta[\*s_{u},\*s_{u-k}]$, it can be written as
\begin{align}
    \Omega^{(n)} =&  \sum_{\*s_{0:t}} p(s_{0:t})\delta[\*s_{u},\*s_{u-n}] \prod_{k=1}^{n-1} (1-\delta[\*s_{u},\*s_{u-k}])
    \nonumber\\ =&  \Psi^{(n-1)} - \Psi^{(n)}, 
    \\ \Psi^{(n)} =& \sum_{\*s_{0:t}} p(s_{0:t}) \prod_{k=1}^{n} (1-\delta[\*s_{u},\*s_{u-k}]).
\end{align}
Here, $\Omega^{(n)}$ is the probability of observing $n-1$ patterns different from $\*s_{u}$ during consecutive state updates and finally observing $\*s_{u}$ at the $n$-th step. In turn, $\Psi^{(n)}$ is the probability of just observing $n$ patterns different from $\*s_{u}$ during $n$ state updates. Both $\Omega^{(n)}$ and $\Psi^{(n)}$ are probability distributions that meet $\sum_{n=1}^\infty \Omega^{(n)} =1$ and $\sum_{n=1}^\infty \Psi^{(n)} =1$. This is guaranteed by defining $\Omega^{(1)} =1 - \Psi^{(1)}$, so that $\sum_{n=1}^\infty \Omega^{(n)}  = 1 - \Psi^{(\infty)} = 1 $ (as the probability $\Psi^{(\infty)} $ converges to zero).

We can expand $\Psi^{(n)}$ as:
\begin{align}
    \Psi^{(n)} =&  1 - \sum_k  {\Delta}_k +  \sum_{k<l}  {\Delta}_{k,l} -  \sum_{k<l<m}  {\Delta}_{k,l,m} + \ldots,
    \nonumber\\  {\Delta}_{k,l,m,\dots} =&  \sum_{\*s_{0:t}} p(s_{0:t}) \delta[\*s_{u},\*s_{u-k}]\delta[\*s_{u},\*s_{u-l}]\delta[\*s_{u},\*s_{u-m}] \cdots.
\end{align}

In steady state, the product of $\delta[\*s_{u},\*s_{u-k}]\delta[\*s_{u},\*s_{u-l}]\delta[\*s_{u},\*s_{u-m}] \cdots$ for any set of $k\neq l\neq m \cdots$ results in the same value. This simplifies the previous equation to
\begin{align}
    \Psi^{(n)} =& \sum_{k=0}^n \binom{n}{k} (-1)^{k}  {\Delta}^{(k)},
    \\  {\Delta}^{(k)}  =& \prod_{k'=1}^k \delta[\*s_{u},\*s_{u-k'}].
\end{align}

Using $\delta[\*s_{u},\*s_{u-k}] = \prod_i  \frac{1+s_{i,u}s_{i,u-k}}{2}$, the configurational average of ${\Delta}^{(n)}$ becomes 
\begin{align}
    \br{ {\Delta}^{(n)} }_{\*J,\^\tau}  =& \br{\sum_{\*s_{0:t}} p(s_{0:t}) \prod_{k=1}^n \prod_{i=1}^{N} \frac{1+s_{i,u}s_{i,u-k}}{2}}_{\*J,\^\tau} 
    \nonumber\\ =&  \frac{1}{2^{nN}}  \br{  \left\langle \prod_{i=1}^{N} \prod_{k=1}^n (1+s_{i,u}s_{i,u-k}) \right\rangle}_{\*J,\^\tau}
    \nonumber\\ =&  \frac{1}{2^{nN}}  \br{  \left\langle \prod_{i=1}^{N} (1+s_{i,u}s_{i,u-1})(1+s_{i,u}s_{i,u-2})\cdots(1+s_{i,u}s_{i,u-n}) \right\rangle} _{\*J,\^\tau}
    \nonumber\\ =&  \frac{1}{2^{nN}}  \br{  \left\langle \prod_{i=1}^{N} \sum_{k=0}^{n}\binom{n}{k} (s_{i,u})^k  s_{i,u-i_1}s_{i,u-i_2}\cdots s_{i,u-i_k} \right\rangle}_{\*J,\^\tau},
\end{align}
where $\{i_1, i_2, \ldots, i_k\}$ are the set of $k$ indices chosen form $1,\ldots, n$. Since $s_{i,u}s_{i,u}=1$, $(s_{i,u})^k$ is $1$ when $k$ is even and $s_{i,u}$ when $k$ is an odd number. It can be summarized as
\begin{align}
    (s_{i,u})^k = \frac{1+(-1)^k}{2} + s_{i,u} \frac{1-(-1)^k}{2}.
\end{align}
Using this equation, we obtain
\begin{align}
    \br{ \Delta^{(n)} }_{\*J,\^\tau}
    =& \frac{1}{2^{nN}}  \br{  \left\langle \prod_{i=1}^{N} \sum_{k=0}^{n}\binom{n}{k} \pr{\frac{1+(-1)^k}{2} + s_{i,u} \frac{1-(-1)^k}{2}} \prod_{l=1}^{k} s_{i,u-i_l} \right\rangle}_{\*J,\^\tau}
    \nonumber\\=& \frac{1}{2^{nN}} \prod_{i=1}^{N} \sum_{k=0}^{n}\binom{n}{k} \pr{ \frac{1+(-1)^k}{2} \br{ \left\langle \prod_{l=1}^{k} s_{i,u-i_l} \right\rangle}_{\*J,\^\tau}+  \frac{1-(-1)^k}{2}  \br{ \left\langle s_{i,u} \prod_{l=1}^{k} s_{i,u-i_l} \right\rangle }_{\*J,\^\tau}  }.
    \label{eq:Upsilong_configurational_average}
\end{align}
The expression above is difficult to compute in general without resorting to extra assumptions, but we can illustrate its behavior  for small pattern lengths.

The proportion of static patterns (1-periodic) in the system in steady state (in large $t$ limit) can be calculated as
\begin{align}
    \Omega^{(1)} =&  1 - \Psi^{(1)} = \sum_{\*s_{0:t}} p(s_{0:t}) \delta[\*s_{u},\*s_{u-1}] 
    \nonumber\\ =&  \sum_{\*s_{0:t}} p(s_{0:t}) \prod_i \frac{1+s_{i,u}s_{i,u-1}}{2}.
\end{align}
The configurational average of the proportion is 
\begin{align}
    \br{\Omega^{(1)}}_{\*J,\^\tau} =& \pr{\frac{1+q}{2}}^N.
\end{align}
That is, for $q=1$ (e.g., $\Delta J = 0$ and $\beta\to\infty$), the system will display only static patterns.

The proportion of 2-periodic patterns can be calculated as
\begin{align}
    \Omega^{(2)} =&  \Psi^{(1)} - \Psi^{(2)} =\sum_{\*s_{0:t}} p(s_{0:t}) \delta[\*s_{u},\*s_{u-2}](1-\delta[\*s_{u},\*s_{u-1}] ) 
    \nonumber\\ =&  \sum_{\*s_{0:t}} p(s_{0:t}) \prod_i \frac{1+s_{i,u}s_{i,u-2}}{2} \pr{1-\prod_i \frac{1+s_{i,u}s_{i,u-1}}{2}},
    \\ \br{\Omega^{(2)}}_{\*J,\^\tau} =&  2^{-N} \pr{(1+q)^N  - \pr{\frac{1+3q}{2}}^N} =   \br{\Omega^{(1)}}_{\*J,\^\tau} \pr{1  - \pr{\frac{1+3q}{2+2q}}^N}.
\end{align}

Similarly, the proportion of 3-periodic patterns is
\begin{align}
    \Omega^{(3)} & = \Psi^{(2)} - \Psi^{(3)} =  \sum_{\*s_{0:t}} p(s_{0:t}) \delta[\*s_{u},\*s_{u-3}](1-\delta[\*s_{u},\*s_{u-1}])(1-\delta[\*s_{u},\*s_{u-2}] )) 
    \nonumber\\& = \sum_{\*s_{0:t}} p(s_{0:t}) \prod_i \frac{1+s_{i,u}s_{i,u-3}}{2} \pr{1-\prod_i \frac{1+s_{i,u}s_{i,u-1}}{2}}\pr{1-\prod_i \frac{1+s_{i,u}s_{i,u-2}}{2}  },
    \\ \br{\Omega^{(3)}}_{\*J,\^\tau} & =2^{-N}\pr{ (1+q)^N  - 2 \pr{\frac{1+3q}{2}}^N + \pr{\frac{1+6q+\rho^{(3)}}{4}}^N} 
    \nonumber\\& = \br{\Omega^{(1)}}_{\*J,\^\tau} \pr{1  - 2 \pr{\frac{1+3q}{2+2q}}^N + \pr{\frac{1+6q+\rho^{(3)}}{4+4q}}^N},
\end{align}
where $\rho_t^{(3)}= \br{\ang{s_{i,u}s_{i,u-1}s_{i,u-2}}}$, which can be calculated as a three-dimensional Gaussian integral
\begin{align}
    \rho_{u}^{(3)} =&\frac{1}{N}\sum_i \int p(\xi)
	\prod_{\tau=0}^3\tanh\br{\beta\overline h_{i,u-\tau}(\xi_{u-\tau}) }.
	\label{eq:def_order_parameter_rho}
\end{align}

\subsection{Expected cycle length in the disordered and deep-ordered phases} 
\begin{figure}
    \centering
    \includegraphics[width=6cm]{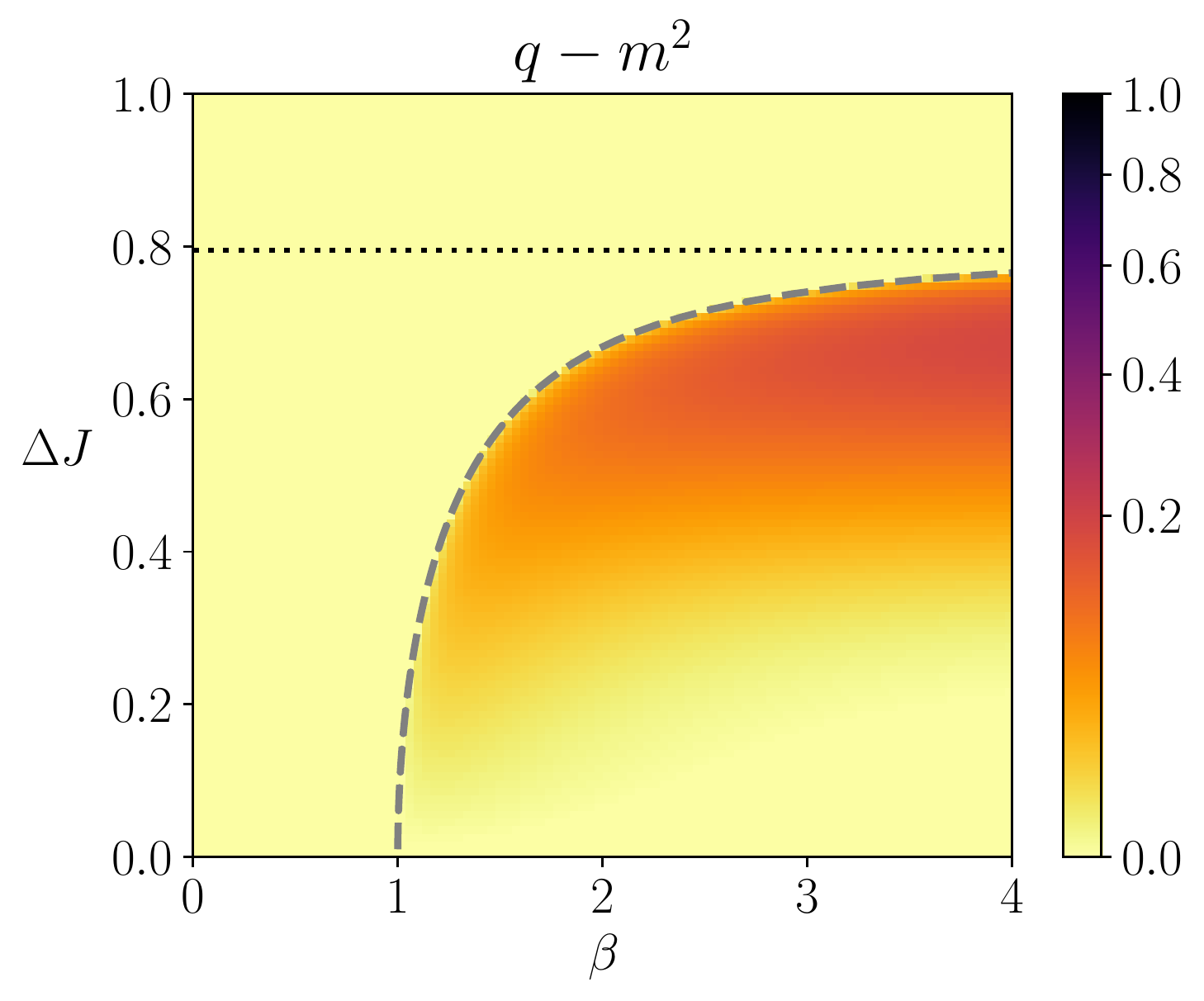}
    \caption{Correlated activity in the configurational average, captured by spin covariances $q-m^2$. We observe that spins become independent both in the disordered phase and in the \emph{deep} ordered phase.}
    \label{fig:q-m2}
\end{figure}

As shown in Fig.~\ref{fig:q-m2}, the system in steady state exhibits uncorrelated dynamics across the time steps ($q-m^2=0$) at the disordered phase ($\Delta J(\beta) > \Delta J^c(\beta)$) and at the limit of $\beta \rightarrow \infty$ in the ordered phase ($\Delta J(\beta) < \Delta J^c(\beta)$). We call the latter the deep ordered phase. For the uncorrelated dynamics, the configurational averages of spins in Eq.~\ref{eq:Upsilong_configurational_average} are given by the product of $m$, which results in
\begin{align}
    \br{ {\Delta}^{(n)} }_{\*J,\^\tau}
    =& \frac{1}{2^{nN}} \pr{ \sum_{k=0}^{n}\binom{n}{k} \pr{\frac{1+(-1)^k}{2} m^k + \frac{1-(-1)^k}{2}  m^{k+1}  } }^N
    \nonumber\\=& \frac{1}{2^{nN}}\pr{\frac{1}{2}((1+m)^n + (1-m)^n + m(1+m)^n - m(1-m)^n)}^N
    \nonumber\\=& \frac{1}{2^{(n+1)N}} \pr{(1+m)^{n+1} +(1-m)^{n+1}}^N.
    \label{eq:Upsilong_configurational_average_uncorrelated}
\end{align}
In the following, we provide the configurational average of the proportion for the disordered and deep-ordered phases.

\subsubsection{Disordered phase}
In the disordered phase, the mean activation rate is $m=0$. This results in $ \br{ {\Delta}^{(n)} } = 2^{-nN}$. Hence we obtain
\begin{align}
  \br{ \Psi^{(n)}}_{\*J,\^\tau}  =&\sum_{k=0}^n \binom{n}{k} (-1)^{k}  2^{-kN} = \pr{1-2^{-N}}^n,
\end{align}
by using the binomial expansion formula. The configurational average of the proportion is computed as 
\begin{align}
  \br{\Omega^{(n)}}_{\*J,\^\tau}  =&2^{-N} \pr{1-2^{-N}}^{n-1}  \approx  \lambda \exp\br{(1-n)\lambda},
\end{align}
where $\lambda = 2^{-N}$. The proportion of the longer cycle length exponentially decays with the rate $2^{-N}$, and the average cycle length is given by $1/\lambda = 2^N$. Namely, 
\begin{align}
  \sum_n n \br{\Omega^{(n)}}_{\*J,\^\tau} = 2^{N},
\end{align}
which is the number of possible patterns.

\subsubsection{Deep ordered phase} 
In the deep ordered phase, $m$ is positive. We expect $(1+m)^{n+1} \gg (1-m)^{n+1}$ for sufficiently large $m$ ($m \sim 1$). Under this condition, Eq.~\ref{eq:Upsilong_configurational_average_uncorrelated} is approximated as 
\begin{align}
    \br{ {\Delta}^{(n)} }_{\*J,\^\tau} =&
     \pr{\frac{1+m}{2}}^{(n+1)N}\pr{1+\pr{\frac{1-m}{1+m}}^{n+1} }^N  \approx \pr{\frac{1+m}{2}}^{(n+1)N}.
\end{align}
Therefore, we obtain
\begin{align}
  \br{ \Psi^{(n)}}_{\*J,\^\tau} \approx& Z^{-1} \pr{\frac{1+m}{2}}^{N}\sum_{k=0}^n \binom{n}{k} (-1)^{k} \pr{\frac{1+m}{2}}^{kN} 
  = 
  \pr{1-\pr{\frac{1+m}{2}}^{N}}^{n},
\end{align}
by the binomial expansion formula. Here, we have introduced $Z=\sum_{k=0}^n \binom{n}{k}\br{ {\Delta}^{(k)} } $ as a normalization factor to compensate for approximation errors and ensure the consistency condition $\sum_n \br{ \Psi^{(n)}}= 1$ (which also guarantees by definition $\sum_n \br{ \Omega^{(n)}}= 1$).
Using this expression, we can compute the probabilities $\br{\Omega^{(n)}}$ of pattern lengths as
\begin{align}
  \br{\Omega^{(n)}}  \approx& 
  \Bigg( \pr{\frac{1+m}{2}}^{N}\pr{1-\pr{\frac{1+m}{2}}^{N}}^{n-1} \approx \lambda \exp\br{(1-n)\lambda},
      \label{eq:Omega-deep-ordered-phase}
\end{align}
where 
\begin{align}
  \lambda =& \pr{\frac{1+m}{2}}^{N}.
\end{align}
The last step is the exponential approximation of the geometric distribution guaranteed for a large $N$. In this case, the average pattern length is 
\begin{align}
  \sum_n n \br{\Omega^{(n)}} \approx &  \frac{1}{\lambda} = \pr{\frac{2}{1+m}}^{N}.
\end{align}
The result reveals that the average pattern length grows exponentially with the system size $N$, and the growth rate becomes slower as $m$ increases.

We note that the result for the deep ordered phase includes the result of disordered phase: inserting $m=0$ yields $\lambda=2^{-N}$ and the average pattern length $1/\lambda=2^N$, which we found at the disordered phase.

\clearpage

\begin{@fileswfalse}
\bibliography{bib}
\end{@fileswfalse}

\end{document}